# Relativistic Six-Fluid Rotating-Plasma Force-Balance Equilibria for p-$^{11}$B Fusion Conditions in Spherical Tori

Y.-K. Martin Peng [a] and Akio Ishida

Beijing ENN Fusion Energy Science and Technology Co., Ltd., Beijing 101111, China

Beijing Key Laboratory of High Magnetic Field Spherical Torus Fusion Energy, Beijing 100871, China

Hebei Key Laboratory of Compact Fusion, Langfang 065001, China

[a] e-mail address: pengyuankai@enn.cn

**Abstract**

We formulate an axisymmetric spherical torus (ST) six-component magnetofluid equilibrium model for p-$^{11}$B plasmas composed of thermal and suprathermal proton, boron, and electron fluids. The model is designed for p-$^{11}$B parameter regimes where ion energies, strong current drive, energetic electron populations, and ion-electron temperature separation can render a single-fluid static-equilibrium description inadequate. It generalizes rotating-plasma equilibrium theory by incorporating relativistic thermal enthalpy corrections to the electron fluids via $h(\mathrm{T}/m_e c^2)$ and by constructing modified flux functions $Y_\alpha$ associated with canonical angular momentum for each component α. The equilibrium is determined by a Grad-Shafranov-like Ampere law for the poloidal magnetic flux $\psi$, algebraic density relations constrained by quasineutrality, and prescribed profile functions $F_\alpha(Y_\alpha)$, $T_\alpha(Y_\alpha)$ and $K_\alpha(Y_\alpha)$. The manuscript includes the relativistic magnetofluid equations, the axisymmetric equilibrium formulation, the six-fluid computational method, and four numerical equilibria: two reactor-relevant ST p-$^{11}$B parameter sets and two EHL-2 ST cases. The examples span toroidal beta $\beta_T$= 0.058-0.253, plasma currents $I_p$= 2.697-12.59 MA, and suprathermal electron temperatures $T_{eh}$= 287 keV-2.37 MeV, with $h$ up to 19.1. Over these parameter ranges, Ampere's law accuracy, electrostatic potential accuracy, and component-resolved radial force balance can have errors ranging from 3% to below 0.1%. These results indicate that high-temperature p-$^{11}$B ST equilibria can differ qualitatively from the single-fluid tokamak equilibria and require analysis of equilibrium component-resolved force balance, relativistic electron enthalpy, and explicit reproducibility diagnostics.

## I. INTRODUCTION

Given the environmentally friendly and cost-effective nature of abundant aneutronic fuel, ENN Science and Technology Development Co., Ltd. (ENN) has decided to pursue proton-boron fusion in spherical-torus (ST) plasmas.[1] The proton-boron fusion reactor requires an ion temperature $T_i$ ~ 150 keV and a hot-ion mode $T_i/T_e > 2$ to reduce electron radiation loss. The drift-velocity difference between protons and boron is also crucial for enhancing the p-$^{11}$B fusion reaction rate. An aspect ratio below 1.8 limits the central solenoid size and makes ECRH current drive inevitable. To achieve the p-$^{11}$B fusion burn condition, ICRH and hydrogen NBI (neutral beam injection) heating can be used. ECRH and ICRH power produce resonant suprathermal and non-resonant thermal plasma components. Suprathermal protons and boron ions at low concentrations and at MeV energies are considered to exploit the double peaks in the fusion cross section near 165 keV and 675 keV.[2,3]

To describe the force-balance equilibrium of a rotating plasma near the proton-boron fusion-burn condition, we adopt a six-fluid model for the bulk thermal and low-density suprathermal species, including protons, boron ions, and electrons. Because suprathermal electron energies can be in the MeV range, a relativistic description is required. In the ECRH current drive experiment in EXL-50/EXL-50U, suprathermal relativistic electrons play a significant role in confinement [4,5]. A detailed comparison of the rapidly rotating multi-magneto-fluid force-balance equilibrium [6] with the static magneto-fluid equilibrium [7,8] is presented in this paper.

Section II presents the equations of relativistic magnetofluid dynamics, Section III gives the axisymmetric equilibrium formulation, and Section IV describes the six-fluid computational method. Comparative numerical results are presented in Section V; the complete inputs, outputs, and diagnostic plots for the following four equilibria are collected in Appendix E:

High $\beta_T$ equilibrium near fusion burn conditions,

Low $\beta_T$ equilibrium near fusion burn conditions,

Low $\beta_T$ EHL-2 equilibrium,

High $\beta_T$ EHL-2 equilibrium,

where EHL-2 is the next-generation ST experiment at ENN. Summary and discussion are presented in Section VI, followed by the conclusion in Section VII. Because the temperature of supra-thermal electrons can reach the MeV range under fusion burn conditions, the function $h$ is used in Appendix A to capture the relativistic effects of thermal motion. Using a relativistic fluid model, the dispersion relation for the ion sound mode is derived in Appendix B. There, by comparing the expressions for the plasma frequency and thermal velocity of the non-relativistic species with those of the relativistic species, the expressions for the plasma frequency and thermal velocity of the relativistic species are clarified, including the ion sound velocity of proton-boron plasma. The numerical accuracy of the equilibrium solution is examined in Appendix C, and the Debye length of the equilibrium plasma is examined in Appendix D.

## II. EQUATIONS OF RELATIVISTIC MAGNETOFLUID DYNAMICS

This section presents the relativistic forms of the continuity equation and the equation of motion for a single species, omitting the species subscript for simplicity. The continuity equation is given by

$$\frac{\partial \gamma n}{\partial t} + \boldsymbol{\nabla} \cdot (\gamma n \boldsymbol{u}) = 0. \tag{1}$$

Here, $n$ is the density in the proper frame in which the fluid volume element concerned is at rest, $\boldsymbol{u}$ is the fluid velocity, and $\gamma \equiv \left(1 - \frac{u^2}{c^2}\right)^{-1/2}$. Note that $\gamma n$ is the density in the laboratory frame in which the fluid volume element concerned moves with $\boldsymbol{u}$, and the factor $\gamma$ results from Lorentz contraction.[9] Suppose that the density is defined as $n = \frac{\Delta N}{\Delta V}$ in the proper frame where $\Delta N$ is the number of particles contained in the volume $\Delta V$. In the laboratory frame where this volume element moves with the velocity $\boldsymbol{u}$, the volume $\Delta V$ shrinks to $\frac{\Delta V}{\gamma}$ via Lorentz contraction. Since $\Delta N$ is not changed, the density in the laboratory frame becomes $\gamma n$.

The equation of fluid motion is then given by

$$\frac{\partial}{\partial t}(mn\gamma^2 h(T^*)\boldsymbol{u}) + \frac{\partial}{\partial x_j}\left(mn\gamma^2 h(T^*)\boldsymbol{u}u_j\right) + \boldsymbol{\nabla}p = q\gamma n\boldsymbol{E} + q\gamma n\boldsymbol{u} \times \boldsymbol{B}, \tag{2}$$

where *m* and *q* are the mass and charge of a particle consisting of the fluid species concerned, *p*=*nT* is the pressure, and ***E*** and ***B*** are the electric and magnetic fields, respectively.[9,10] Here it is understood to take the sum of j=1, 2 and 3 implied for the repeated j, where $x_1 = x$, $x_2 = y$, and $x_3 = z$. The quantity $h(T^*)$ is defined by

$$h(T^*) \equiv \frac{\epsilon + p}{mnc^2} = \frac{K_3(1/T^*)}{K_2(1/T^*)} \ , \tag{3}$$

where $\epsilon$ is the energy density of the fluid, which contains the rest mass energy, $T^* \equiv T/mc^2$, *T* is the temperature of the fluid in energy units, and $K_n$(z) is the modified Bessel function of the n$^{\text{th}}$ kind. Derivation of Eq. (3) is shown in Appendix A of [11] where the function *h* is represented by $g_{ep}$. Note that $\epsilon + p$ is the enthalpy density. The quantity $mn\gamma^2 h\boldsymbol{u}$ should be understood as $\gamma n \times m\gamma\boldsymbol{u} \times h$, where $\gamma n$ is density, $m\gamma\boldsymbol{u}$ is momentum due to macroscopic motion in the laboratory frame, and *h* represents the relativistic effect due to thermal motion. $\gamma^2 h$ becomes unity in the non-relativistic limit.

We describe the equations in the laboratory frame and adopt SI units. Using the vector and scalar potentials, **E** and **B** are expressed as

$$\boldsymbol{E} = -\frac{\partial \boldsymbol{A}}{\partial t} - \boldsymbol{\nabla}V_E \quad \text{and} \quad \boldsymbol{B} = \boldsymbol{\nabla} \times \boldsymbol{A}, \tag{4}$$

where $V_E$ is the electrostatic potential, and ***A*** is the magnetic vector potential. Using the continuity equation and Eq. (4) in Eq. (2) and multiplying 1/*n*, the equation of motion for a "fluid element" is given by

$$\gamma\frac{\partial}{\partial t}\boldsymbol{P} + m\gamma u\boldsymbol{\nabla}(\gamma u h) + \boldsymbol{\nabla}T + \frac{T}{n}\boldsymbol{\nabla}n + q\gamma\boldsymbol{\nabla}V_E = \gamma\boldsymbol{u} \times (\boldsymbol{\nabla} \times \boldsymbol{P}), \tag{5}$$

where $\boldsymbol{P} \equiv m\gamma h\boldsymbol{u} + q\boldsymbol{A}$ is the generalized momentum, and *u* is the magnitude of the velocity $\boldsymbol{u}$. See [11] for the derivation, which is similar to earlier work on two-fluid and three-fluid equilibrium models [12,13].

The macroscopic "fluid element" contains many microscopic particles. As a result, Eq. (5) describing the behavior of the generalized momentum $\boldsymbol{P}$ also accounts for the effects of the pressure gradient, the electric field, the relativistic macroscopic velocity through $\gamma$, and the relativistic velocity of thermal motion through $h(T^*)$. For an axisymmetric plasma, Eq. (5) conserves the z-component of the generalized angular momentum, $\hat{z} \cdot (\boldsymbol{R} \times \boldsymbol{P}) = RP_\phi$, satisfying

$$\frac{\partial RP_\phi}{\partial t} + \boldsymbol{u} \cdot \boldsymbol{\nabla}\left(RP_\phi\right) = 0, \tag{6a}$$

where $RP_\phi = m\gamma hRu_\phi + qRA_\phi$. (6b)

This quantity plays a key role in axisymmetric equilibrium formulation.

## III. AXISYMMETRIC EQUILIBRIUM FORMULATION

For each fluid species, we adopt the following two equilibrium equations from Eqs. (1) and Eq. (5).

$$\boldsymbol{\nabla} \cdot (\gamma n \boldsymbol{u}) = 0, \tag{7}$$

$$m\gamma u \boldsymbol{\nabla}(\gamma u h) + \boldsymbol{\nabla} T + T\boldsymbol{\nabla} \ln n + q\gamma \boldsymbol{\nabla} V_E = q\gamma \boldsymbol{u} \times \boldsymbol{\Omega}, \tag{8}$$

where $\boldsymbol{\Omega} \equiv q^{-1}\boldsymbol{\nabla} \times \boldsymbol{P} = \boldsymbol{B} + \boldsymbol{\nabla} \times (q^{-1} m\gamma h \boldsymbol{u})$. (9)

In this paper, $\boldsymbol{\Omega}$ represents a modified magnetic field. Maxwell's equations then provide

$\boldsymbol{\nabla} \times \boldsymbol{B} = \mu_0 \sum_\alpha \boldsymbol{j}_\alpha$ where $\boldsymbol{j}_\alpha = q_\alpha \gamma_\alpha n_\alpha \boldsymbol{u}_\alpha$, (10)

$$\boldsymbol{\nabla} \cdot \boldsymbol{B} = 0. \tag{11}$$

We adopt the charge neutrality condition instead of the Poisson equation,

$$\sum_\alpha q_\alpha \gamma_\alpha n_\alpha = 0. \tag{12}$$

Hereafter we adopt the right-handed cylindrical coordinates $(R, \phi, Z)$. Since the magnetic field $\boldsymbol{B}$, the modified magnetic field $\boldsymbol{\Omega}$ and the density flux $\gamma n \boldsymbol{u}$ are divergence-free, these can be expressed in terms of flux or stream functions $\psi, Y$ and $\Phi$:

$$\boldsymbol{B} = \boldsymbol{\nabla}\psi(R,Z) \times \boldsymbol{\nabla}\phi + RB_\phi \boldsymbol{\nabla}\phi, \tag{13}$$

$$\boldsymbol{\Omega} = \boldsymbol{\nabla} Y(R,Z) \times \boldsymbol{\nabla}\phi + R\Omega_\phi \boldsymbol{\nabla}\phi, \tag{14}$$

$$n\gamma \boldsymbol{u} = \boldsymbol{\nabla}\Phi(\mathrm{R},\mathrm{Z}) \times \boldsymbol{\nabla}\phi + Rn\gamma \mathrm{u}_\phi \boldsymbol{\nabla}\phi. \tag{15}$$

Combining the generalized angular momentum given by Eq. (6b) with the definition of $\boldsymbol{\Omega}$ by Eq. (9), one arrives at the following relation:

$$Y = RP_{\phi}/q = \psi + \frac{m}{q}\gamma h R u_{\phi}, \text{ and} \tag{16}$$

$$R\Omega_{\phi} = RB_{\phi} - R^2 \boldsymbol{\nabla} \cdot \left(\frac{mh}{qnR^2} \boldsymbol{\nabla}\Phi\right). \tag{17}$$

Since in axisymmetric equilibrium, the L.H.S. of Eq. (8) has no $\phi$-component, the following relation must be satisfied,

$$\boldsymbol{\nabla}\phi \cdot (\boldsymbol{u} \times \boldsymbol{\Omega}) = 0. \tag{18}$$

This gives $\boldsymbol{\nabla}\phi \cdot \left(\boldsymbol{\nabla}\Phi(\mathrm{R}, \mathrm{Z}) \times \boldsymbol{\nabla}Y(R,Z)\right) = 0$ at arbitrary R and Z. The above is satisfied when

$$\Phi = \Phi\left(Y(R,Z)\right), \tag{19}$$

i.e., the function $\Phi$ can be an arbitrary function of $Y(R,Z)$.

We therefore can adopt the assumption,

$$\boldsymbol{\Omega} \cdot \boldsymbol{\nabla}T = 0, \tag{20}$$

the validity of which is discussed in [14] and is a natural generalization of $\boldsymbol{B} \cdot \boldsymbol{\nabla}T = 0$ for a rotating fluid.

It should be noted that 1) the poloidal flow velocity of a species is much smaller than its thermal velocity, and 2) for axisymmetric rotating plasma, Eq. (8), giving $\boldsymbol{\Omega} \equiv q^{-1}\boldsymbol{\nabla} \times \boldsymbol{P}$, is more fundamental than $\boldsymbol{B}$. Applying Eq. (20) then leads to

$$T = T\left(Y(R,Z)\right) \quad \text{and} \quad h = h\left(\frac{T\left(Y(R,Z)\right)}{mc^2}\right), \tag{21}$$

i.e., the function $T$ can also be an arbitrary function of $Y(R,Z)$.

Next, consider the force balance in the $\boldsymbol{\Omega}$-direction, which is given by

$$\boldsymbol{\nabla}\phi \cdot \left(\left\{m\gamma u \boldsymbol{\nabla}(\gamma u h) + \boldsymbol{\nabla}T + \frac{T}{n}\boldsymbol{\nabla}n + q\gamma\boldsymbol{\nabla}V_E\right\} \times \boldsymbol{\nabla}Y\right) = 0. \tag{22}$$

Since $\frac{1}{n}\boldsymbol{\nabla}n = \frac{1}{\bar{n}}\boldsymbol{\nabla}\bar{n} = \boldsymbol{\nabla}\ln\bar{n} = \boldsymbol{\nabla}\left(\ln n - \ln n_{ref}\right) = \boldsymbol{\nabla}\ln n$, one can define $\bar{n} \equiv n/n_{ref}$, where $n_{ref}$ is a suitable reference density to be introduced in Section IV. Using Eq. (20), Eq. (22) can be written as

$$\nabla\phi \cdot \left(\nabla\left\{\tfrac{1}{2}m(\gamma u)^2 h + T(1+\ln n) + qV_E\right\} \times \nabla Y\right) + \nabla\phi \cdot \{(\gamma-1)\mathrm{q}\nabla V_E \times \nabla Y\} = 0. \qquad (23)$$

Since the rotational velocity of the supra-thermal electron fluid is much less than the speed of light, we can neglect the second term in Eq. (23). In the example equilibrium 1 reported in Appendix E, the peak $u_{eh}$ can reach 5900 km/s, giving $\gamma - 1 = 2 \times 10^{-4}$ and justifying this approximation.

We can therefore define a new function *F* from Eq. (23) as

$$F \equiv \tfrac{1}{2}m(\gamma u)^2 h + T(1+\ln n) + qV_E, \qquad (24)$$

which can also be an arbitrary function of *Y*,

$$F = F\big(Y(R,Z)\big). \qquad (25)$$

Finally, using Eq. (25), the force balance equation Eq. (8) can be written as

$$\frac{dF}{dY} - \frac{dT}{dY}\ln n + \frac{1}{2}m(\gamma u)^2\frac{dh}{dT}\frac{dT}{dY} = \frac{q}{R}\left(\gamma u_\phi - \frac{1}{n}\frac{d\Phi}{dY}\Omega_\phi\right). \qquad (26)$$

Solving Eq. (26) for the toroidal rotational velocity, we have for each species $\alpha$

$$\gamma_\alpha u_{\alpha\phi} = q_\alpha^{-1} R\left(F'_\alpha - T'_\alpha \ln n_\alpha + \tfrac{1}{2}m_\alpha(\gamma_\alpha u_\alpha)^2\frac{dh_\alpha}{dT_\alpha}T'_\alpha\right) + \frac{1}{n_\alpha}\Phi'_\alpha\Omega_{\alpha\phi}, \qquad (27)$$

where the prime represents the derivative with respect to $Y_\alpha$. The toroidal current density of each species is therefore given by

$$j_{\alpha\phi} = q_\alpha n_\alpha \gamma_\alpha u_{\alpha\phi} = n_\alpha R\left(F'_\alpha - T'_\alpha \ln n_\alpha + \tfrac{1}{2}m_\alpha(\gamma_\alpha u_\alpha)^2\frac{dh_\alpha}{dT_\alpha}T'_\alpha\right) + q_\alpha\Phi'_\alpha\Omega_{\alpha\phi}. \qquad (28)$$

It is well known that the Grad-Shafranov equilibrium model has two arbitrary functions of the magnetic flux $\psi$. In the present model, each fluid has three arbitrary functions $F(Y)$, $T(Y)$ and $\Phi(Y)$ with *Y* given by Eq. (16). In this paper, we call these functions the profile functions. Note also that $Y = \frac{RP_\phi}{q}$ where $RP_\phi$ is the z-component of the generalized angular momentum given by Eq. (6b).

Considering Ampere's equation and using Eq. (13),

$$\nabla \times \boldsymbol{B} = \nabla(RB_\phi) \times \nabla\phi - R\left\{\frac{\partial}{\partial R}\left(\frac{1}{R}\frac{\partial\psi}{\partial R}\right) + \frac{1}{R}\frac{\partial^2\psi}{\partial Z^2}\right\}\nabla\phi. \qquad (29)$$

The toroidal component of Ampere's equation can be written as

$$\frac{\partial}{\partial R}\left(\frac{1}{R}\frac{\partial\psi}{\partial R}\right)+\frac{1}{R}\frac{\partial^2\psi}{\partial Z^2}=-\mu_0 j_\phi, \tag{30}$$

where $j_\phi=\sum_\alpha j_{\alpha\phi}$. Using Eq. (15), the poloidal component of Ampere's equation can be written as

$$RB_\phi=\sum_\alpha \mu_0 q_\alpha \Phi_\alpha. \tag{31}$$

Explicit expression of the charge neutrality becomes complex as the number of plasma fluids increases. We will discuss its solution in the next section.

## IV. COMPUTATIONAL METHOD FOR THE SIX-FLUID PLASMA MODEL

Here, we study the equilibrium of proton-boron plasmas in fusion- and reactor-relevant ST. For such high-temperature plasma, it may be useful to use electron current drive via ECRH, proton current drive via ICRH, and neutral beam injection for both electrons and protons. These current drive methods produce high-energy, low-density suprathermal components. The non-resonant thermal plasma components remain at lower temperatures and higher densities. The plasma considered here therefore consists of the thermal proton, boron, and electron fluids, represented by subscripts pl, bl, and el, and the suprathermal proton, boron, and electron fluids, represented by subscripts ph, bh, and eh.

In a p-boron fusion plasma, the thermal electron temperature can reach several tens of keV. We apply the relativistic form of equations derived in section III to the el and eh fluids, and for the other four fluids, we apply non-relativistic form of equations, putting $\gamma_\alpha=1$, $h_\alpha=1$ and $\frac{dh_\alpha}{dT_\alpha}=0$ for $\alpha=pl, bl, ph$ and $bh$.

### A. Dimensionless equilibrium equations

For convenience in numerical computation, we adopt dimensionless variables hereafter. The primary scales are (1) $L_{ref}$ a reference length of a plasma, (2) $I_{ref}$ a reference plasma current, and (3) $n_{ref}$ a reference density. These lead to reference variables for the magnetic field $B_{ref}\equiv\frac{\mu_0 I_{ref}}{L_{ref}}$, velocities $u_{ref}\equiv\frac{B_{ref}}{(\mu_0 m_p n_{ref})^{1/2}}$, temperatures $T_{ref}\equiv m_p u_{ref}^2$, pressures $p_{ref}\equiv n_{ref}T_{ref}$, scalar potential $V_{Eref}\equiv\frac{T_{ref}}{e}$, vector potential $A_{ref}\equiv B_{ref}L_{ref}$, poloidal magnetic flux $\psi_{ref}\equiv B_{ref}L_{ref}^2$, current density $j_{ref} \equiv \frac{B_{ref}}{\mu_0 L_{ref}}$, poloidal flow surface function $\Phi_{ref} \equiv n_{ref} u_{ref} L_{ref}^2$, the function $F_{ref} = T_{ref}$, and electric field $E_{ref} \equiv u_{ref} B_{ref}$. We introduce the parameter $\varepsilon \equiv \frac{l_{pR}}{L_{ref}}$, where $l_{pR} \equiv \frac{c}{\sqrt{\frac{e^2 n_{ref}}{\varepsilon_0 m_p}}}$ is the proton inertial length for the reference density $n_{ref}$. Since the three primary scales are chosen arbitrarily, it is useful to also use the following expressions for velocity, temperature, magnetic field, and poloidal magnetic flux, respectively:

$$u_{ref} = \sqrt{\frac{\mu_0}{m_p}} \frac{I_{ref}}{L_{ref}\sqrt{n_{ref}}} \ , \quad T_{ref} = \mu_0 \left(\frac{I_{ref}}{L_{ref}\sqrt{n_{ref}}}\right)^2 \ , \ B_{ref} \equiv \frac{\mu_0 I_{ref}}{L_{ref}} \ , \quad \psi_{ref} = \mu_0 L_{ref} I_{ref}.$$

For simplicity, we neglect symbols for dimensionless variables unless otherwise noted. The following dimensionless equations are then obtained:

$$F_{pl} = \frac{1}{2} u_{pl}^2 + T_{pl}\left(1 + \ln n_{pl}\right) + V_E = F_{pl}\left(Y_{pl}\right), \tag{32a}$$

$$F_{ph} = \frac{1}{2} u_{ph}^2 + T_{ph}\left(1 + \ln n_{ph}\right) + V_E = F_{ph}\left(Y_{ph}\right), \tag{32b}$$

$$F_{bl} = \frac{1}{2}\frac{m_b}{m_p} u_{bl}^2 + T_{bl}\left(1 + \ln n_{pb}\right) + Z_b V_E = F_{bl}(Y_{bl}), \tag{32c}$$

$$F_{bh} = \frac{1}{2}\frac{m_b}{m_p} u_{bh}^2 + T_{bh}(1 + \ln n_{bh}) + Z_b V_E = F_{bh}(Y_{bh}), \tag{32d}$$

$$F_{el} = \frac{1}{2}\frac{m_e}{m_p} u_{el}^2 h(T_{el}^*) + T_{el}(1 + \ln n_{el}) - V_E = F_{el}(Y_{el}), \tag{32e}$$

$$F_{eh} = \frac{1}{2}\frac{m_e}{m_p} u_{eh}^2 {\gamma_{eh}}^2 h(T_{eh}^*) + T_{eh}(1 + \ln n_{eh}) - V_E = F_{eh}(Y_{eh}), \tag{32f}$$

where $\gamma_{eh}(u_{eh}) = \left(1 - \frac{u_{eh}^2}{\bar{c}^2}\right)^{-1/2}$, $T_{eh}^* = T_{eh}/\left(\frac{m_e}{m_p}\bar{c}^2\right)$, and $\bar{c} \equiv \frac{c}{u_{ref}}$. By setting $\gamma_{el}(u_{el}) = 1$ we obtain

$$Y_{pl} = \psi + \varepsilon R u_{pl\phi}, \tag{33a}$$

$$Y_{ph} = \psi + \varepsilon R u_{ph\phi}, \tag{33b}$$

$$Y_{bl} = \psi + \varepsilon\left(\frac{m_b}{m_p Z_b}\right) R u_{bl\phi}, \tag{33c}$$

$$Y_{bh} = \psi + \varepsilon\left(\frac{m_b}{m_p Z_b}\right) R u_{bh\phi}, \tag{33d}$$

$$Y_{el} = \psi - \varepsilon\left(\frac{m_e}{m_p}\right) h(T_{el}^*) R u_{el\phi}, \text{ and} \tag{33e}$$

$$Y_{eh} = \psi - \varepsilon\left(\frac{m_e}{m_p}\right) \gamma_{eh} h(T_{eh}^*) R u_{eh\phi}. \tag{33f}$$

To select the external toroidal field, we introduce a new profile function for the poloidal mass density flux,

$$\Phi_\alpha(Y_\alpha) = -\varepsilon K_\alpha\big(Y_\alpha(R,Z)\big) \quad \text{for } \alpha = pl, ph, bl, bh, el, eh. \tag{34}$$

The toroidal rotational velocities can then be written as

$$u_{pl\phi} = \varepsilon R\big(F_{pl}' - T_{pl}' \ln n_{pl}\big) - \frac{\varepsilon}{n_{pl}} K_{pl}' \Omega_{pl\phi}, \tag{35a}$$

$$u_{ph\phi} = \varepsilon R\big(F_{ph}' - T_{ph}' \ln n_{ph}\big) - \frac{\varepsilon}{n_{ph}} K_{ph}' \Omega_{ph\phi}, \tag{35b}$$

$$u_{bl\phi} = \frac{\varepsilon}{Z_b} R(F_{bl}' - T_{bl}' \ln n_{bl}) - \frac{\varepsilon}{n_{bl}} K_{bl}' \Omega_{bl\phi}, \tag{35c}$$

$$u_{bh\phi} = \frac{\varepsilon}{Z_b} R(F_{bh}' - T_{bh}' \ln n_{bh}) - \frac{\varepsilon}{n_{bh}} K_{bh}' \Omega_{bh\phi}, \tag{35d}$$

$$u_{el\phi} = -\varepsilon R\left(F_{el}' - T_{el}' \ln n_{el} + \frac{1}{2}\frac{m_e}{m_p} u_{el}^2 \frac{dh(T_{el}^*)}{dT_{el}} T_{el}'\right) - \frac{\varepsilon}{n_{el}} K_{el}' \Omega_{el\phi}, \text{ and} \tag{35e}$$

$$\gamma_{eh} u_{eh\phi} = -\varepsilon R\left(F_{eh}' - T_{eh}' \ln n_{eh} + \frac{1}{2}\frac{m_e}{m_p} (\gamma_{eh} u_{eh})^2 \frac{dh(T_{eh}^*)}{dT_{eh}} T_{eh}'\right) - \frac{\varepsilon}{n_{eh}} K_{eh}' \Omega_{eh\phi}, \tag{35f}$$

where

$$\Omega_{\alpha\phi} = B_\phi + \varepsilon^2 \frac{e}{q_\alpha}\frac{m_\alpha}{m_p} R \nabla \cdot \left(\frac{K_\alpha'}{n_\alpha R^2} \nabla Y_\alpha\right) \quad \text{for } \alpha = pl, ph, bl, bh, \text{ and} \tag{36a}$$

$$\Omega_{\alpha\phi} = B_\phi - \varepsilon^2 \frac{m_e}{m_p} R \nabla \cdot \left(\frac{h_\alpha K_\alpha'}{n_\alpha R^2} \nabla Y_\alpha\right) \quad \text{for } \alpha = el, eh. \tag{36b}$$

The toroidal current densities are then written as,

$$j_{pl\phi} = \varepsilon^{-1} n_{pl} u_{pl\phi}, \tag{37a}$$

$$j_{ph\phi} = \varepsilon^{-1} n_{ph} u_{ph\phi}, \tag{37b}$$

$$j_{bl\phi} = \varepsilon^{-1} Z_b n_{bl} u_{bl\phi}, \tag{37c}$$

$$j_{bh\phi} = \varepsilon^{-1} Z_b n_{bh} u_{bh\phi}, \tag{37d}$$

$$j_{el\phi} = -\varepsilon^{-1} n_{el} u_{el\phi}, \text{ and} \tag{37e}$$

$$j_{eh\phi} = -\varepsilon^{-1} n_{eh} \gamma_{eh} u_{eh\phi}. \tag{37f}$$

The toroidal component of Ampere's equation is then written as

$$R \frac{\partial}{\partial R}\left(\frac{1}{R}\frac{\partial \psi}{\partial R}\right) + \frac{\partial^2 \psi}{\partial Z^2} = -R j_{\phi}, \tag{38}$$

where $j_{\phi} = j_{pl\phi} + j_{ph\phi} + j_{bl\phi} + j_{bh\phi} + j_{el\phi} + j_{eh\phi}$. (39)

The poloidal component of Ampere's equation is written as

$$RB_{\phi} = \mathrm{K}_{el} + \mathrm{K}_{eh} - \mathrm{K}_{pl} - \mathrm{K}_{ph} - Z_b \mathrm{K}_{bl} - Z_b \mathrm{K}_{bh}. \tag{40}$$

Solving Eq. (32) for $V_E$, the electrostatic potential $V_E$ must then satisfy

$$V_E = \tilde{F}_{pl} - T_{pl}(1 + \ln n_{pl}) \tag{41a}$$

$$V_E = \tilde{F}_{ph} - T_{ph}\left(1 + \ln n_{ph}\right), \tag{41b}$$

$$V_E = \frac{1}{Z_b}\{\tilde{F}_{bl} - T_{bl}(1 + \ln n_{bl})\}, \tag{41c}$$

$$V_E = \frac{1}{Z_b}\{\tilde{F}_{bh} - T_{bh}(1 + \ln n_{bh})\}, \tag{41d}$$

$$V_E = -\{\tilde{F}_{el} - T_{el}(1 + \ln n_{el})\}, \tag{41e}$$

$$V_E = -\{\tilde{F}_{eh} - T_{eh}(1 + \ln n_{eh})\}. \tag{41f}$$

Here, $\tilde{F}_{pl} = F_{pl} - \frac{1}{2}u_{pl}^2$, $\tilde{F}_{ph} = F_{ph} - \frac{1}{2}u_{ph}^2$, $\breve{F}_{bl} = F_{bl} - \frac{1}{2}\frac{m_b}{m_p}u_{bl}^2$, $\tilde{F}_{bh} = F_{bh} - \frac{1}{2}\frac{m_b}{m_p}u_{bh}^2$,

$\tilde{F}_{el} = F_{el} - \frac{1}{2}\frac{m_e}{m_p}u_{el}^2 h(T_{el}^*)$, and $\tilde{F}_{eh} = F_{eh} - \frac{1}{2}\frac{m_e}{m_p}u_{eh}^2 h(T_{eh}^*)$.

**B. Densities consistent with electrostatic potential**

To derive densities consistent with the electrostatic potential $V_E$, we use simultaneous equations of density ratio $n_\alpha/n_{el}$ for $\alpha = el, eh, pl, ph, bl$ and $bh$. To do so, eliminate the electrostatic potential $V_E$ and the quantity $(1 + \ln n_{el})$ from Eq. (41) by using the relation $1 + \ln n_\alpha = 1 + \ln n_{el} + \ln(n_\alpha/n_{el})$.

Eliminating $V_E$ from Eq.(41e) and (41a),

$$-\tilde{F}_{el} + T_{el}(1 + \ln n_{el}) = \tilde{F}_{pl} - T_{pl}\left(1 + \ln n_{el} + \ln\frac{n_{pl}}{n_{el}}\right)$$

Also eliminating $V_E$ from Eq.(41e) and (41f),

$$\tilde{F}_{el} - T_{el}(1 + \ln n_{el}) = \tilde{F}_{eh} - T_{eh}\left(1 + \ln n_{el} + \ln\frac{n_{eh}}{n_{el}}\right)$$

Eliminating $(1 + \ln n_{el})$ from the above two equations, we have Eq.(42a).

$$\frac{n_{eh}}{n_{el}} = \exp\left\{\frac{T_{eh}-T_{el}}{T_{eh}}\left(\frac{\tilde{F}_{eh}-\tilde{F}_{el}}{T_{eh}-T_{el}} - \frac{\tilde{F}_{pl}+\tilde{F}_{el}}{T_{pl}+T_{el}} + \frac{T_{pl}}{T_{pl}+T_{el}}\ln\frac{n_{pl}}{n_{el}}\right)\right\}, \tag{42a}$$

Applying the same technique, we obtain the following three equations.

$$\frac{n_{ph}}{n_{el}} = \exp\left\{\frac{T_{ph}+T_{el}}{T_{ph}}\left(\frac{\tilde{F}_{ph}+\tilde{F}_{el}}{T_{ph}+T_{el}} - \frac{\tilde{F}_{eh}-\tilde{F}_{el}}{T_{eh}-T_{el}} + \frac{T_{eh}}{T_{eh}-T_{el}}\ln\frac{n_{eh}}{n_{el}}\right)\right\}, \tag{42b}$$

$$\frac{n_{bh}}{n_{el}} = \exp\left\{\frac{\frac{T_{bh}}{Z_b}+T_{el}}{\frac{T_{bh}}{Z_b}}\left(\frac{\frac{\tilde{F}_{bh}}{Z_b}+\tilde{F}_{el}}{\frac{T_{bh}}{Z_b}+T_{el}} - \frac{\tilde{F}_{ph}+\tilde{F}_{el}}{T_{ph}+T_{el}} + \frac{T_{ph}}{T_{ph}+T_{el}}\ln\frac{n_{ph}}{n_{el}}\right)\right\}, \tag{42c}$$

$$\frac{n_{bl}}{n_{el}} = \frac{1}{Z_b}\left(1 + \frac{n_{eh}}{n_{el}} - \frac{n_{pl}}{n_{el}} - \frac{n_{ph}}{n_{el}} - Z_b\frac{n_{bh}}{n_{el}}\right), \text{ and} \tag{42d}$$

$$\frac{n_{pl}}{n_{el}} = \exp\left\{\frac{T_{pl}+T_{el}}{T_{pl}}\left(\frac{\tilde{F}_{pl}+\tilde{F}_{el}}{T_{pl}+T_{el}} - \frac{\frac{\tilde{F}_{bl}}{Z_b}+\tilde{F}_{el}}{\frac{T_{bl}}{Z_b}+T_{el}} + \frac{\frac{T_{bl}}{Z_b}}{\frac{T_{bl}}{Z_b}+T_{el}}\ln\frac{n_{bl}}{n_{el}}\right)\right\}. \tag{42e}$$

Here, Eq. (42d) represents the electric charge-neutrality condition, which is examined in Appendix C. The charge-neutrality condition is well satisfied in the four example equilibria reported in Appendix E.

The density of each fluid can be computed using Eqs. (43) and (44):

$$n_{el} = \exp\left(\frac{\tilde{F}_{pl}+\tilde{F}_{el}}{T_{pl}+T_{el}} - 1 - \frac{T_{pl}}{T_{pl}+T_{el}} \ln\frac{n_{pl}}{n_{el}}\right), \text{ and} \tag{43}$$

$$n_\alpha = \frac{n_\alpha}{n_{el}} \times n_{el} \quad \text{for } \alpha = pl, ph, bl, bh, eh. \tag{44}$$

**C. Computational method**

The magnitude of the 2nd term of $\Omega_{\alpha\phi}$ in Eq. (36) is estimated to be much less than that of the 1st term $B_\phi$. Therefore, we treat the 2nd term as a correction term to be verified afterward. The partial differential equation that must be solved is Eq. (38), while the others are algebraic equations. The computational flowchart follows Step 11; Fig. 29 in Appendix E compares the initial model of the toroidal plasma current density and the poloidal magnetic flux over the mid-plane radius *R*.

**Step 1**. Select the current of the poloidal magnetic field coils. Using the Biot-Savart law, calculate the poloidal magnetic flux, $\psi_{pfcoil}$. See Tables III - XIV for the poloidal field coil locations and currents, the coefficients of the input surface functions, and the output plasma parameters used in the Example equilibria presented in Appendix E.

**Side step 1.1**. Make the initial $\psi_{plasma-model}$. Applying the Biot-Savart law to a suitable toroidal current model, $j_{\phi-model}$, such as Maekawa's model [15], calculate the poloidal magnetic flux due to this model plasma current, $\psi_{plasma-model}$.

**Side step 1.2**. Make model poloidal magnetic flux, $\psi_{model} = \psi_{pfcoil} + \psi_{plasma-model}$. Since the experimental data near p-$^{11}$B fusion burn condition is not available at the present time, we must use the boundary value of $\psi_{model}$ to solve Ampere's equation Eq. (38) initially. Later, we iterate the boundary condition in **Step 11**.

**Step 2**. Select 18 suitable profile functions, $F_\alpha(Y_\alpha), T_\alpha(Y_\alpha)$ and $K_\alpha(Y_\alpha)$ for $\alpha = pl, ph, bl, bh, el, eh.$

We adopt $K_\alpha(Y_\alpha) = 0$ for $\alpha = pl, ph, bl, bh, eh$ and $K_{el}(Y_{el}) = CK_{el}0 = const$ which gives the external toroidal magnetic field. (See Eq. (40).) No poloidal flow is assumed in the present stage of equilibrium computation.

About $F_\alpha(Y_\alpha), T_\alpha(Y_\alpha)$, we adopt two types of functional forms for the thermal components and the supra-thermal components.

For the thermal components, i.e. $\alpha = pl, bl, el$

$$T_\alpha\big(Y_\alpha(R,Z)\big) = CT_\alpha 0 + CT_\alpha 1 \times gfun_{C1} \times \left[\tanh\left(\frac{X1}{gfun_{C21}}\right)\right]^{gfun_{nt}} \quad \text{for } X1 \geq 0, \tag{45a}$$

$$T_\alpha\big(Y_\alpha(R,Z)\big) = CT_\alpha 0 \quad \text{for } X1 < 0,$$

$$F_\alpha\big(Y_\alpha(R,Z)\big) = CF_\alpha 0 + CF_\alpha 1 \times gfun_{C1} \times \left[\tanh\left(\frac{X1}{gfun_{C21}}\right)\right]^{gfun_{nf}} \quad \text{for } X1 \geq 0, \tag{45b}$$

$$F_\alpha\big(Y_\alpha(R,Z)\big) = CF_\alpha 0 \quad \text{for } X1 < 0,$$

where $X1 \equiv -(Y_\alpha(R.Z) - Psi_{crit1})$.

For the suprathermal components, i.e. $\alpha = ph, bh, eh$

$$T_\alpha\big(Y_\alpha(R,Z)\big) = CT_\alpha 0 + CT_\alpha 1 \times gfun_{C1} \times \left[\tanh\left(\frac{X2}{gfun_{C2}}\right)\right]^{gfun_{nf}} \quad \text{for } X2 \geq 0, \tag{46a}$$

$$T_\alpha\big(Y_\alpha(R,Z)\big) = CT_\alpha 0 \quad \text{for } X2 < 0,$$

$$F_\alpha\big(Y_\alpha(R,Z)\big) = CF_\alpha 0 + CF_\alpha 1 \times gfun_{C1} \times \left[\tanh\left(\frac{X2}{gfun_{C2}}\right)\right]^{gfun_{nt}} \quad \text{for } X2 \geq 0, \tag{46b}$$

$$F_\alpha\big(Y_\alpha(R,Z)\big) = CF_\alpha 0 \quad \text{for } X2 < 0,$$

where $X2 \equiv -(Y_\alpha(R.Z) - Psi_{crit2})$.

Here, $CT_\alpha 0, CT_\alpha 1, CF_\alpha 0, CF_\alpha 1$ are constants. $gfu_{C2}$, $gfun_{C21}$ are constant parameters to determine the degree of hollowness of current density and flow. $gfun_{nt}$ and $gfun_{nf}$ are constant parameters to control the profile shape of density and temperature. Parameter $gfun_{C1}$ is adjusted to achieve convergence of Ampere's equation.

Note that the boundaries of the flow and the current density of the thermal and supra-thermal fluids are determined by conditions X1 = 0 and X2 = 0. Therefore, constant parameters $Psi_{crit1}$ and $Psi_{crit2}$ determine the boundaries of the flow and the current density of the thermal and suprathermal fluids. The computational boundary is rectangular with $R_0 \leq R \leq R_f$ and $Z_0 \leq Z \leq Z_f$. The outer boundary refers to the right vertical computational boundary at $R = R_f$. Selecting $Psi_{crit1}$ and $Psi_{crit2}$, it is possible that there is no flow or current density of each fluid along the outer boundary. Parameters $CT_\alpha 0$ and $CF_\alpha 0$ for every species fluid can be determined to satisfy $V_E = 0$ for the prescribed densities along the outer boundary.

Since the rotational energy and the electrostatic potential energy is small compared to the thermal energy, the density is mainly given by Exp ($F_\alpha$ / $T_\alpha - 1$), i.e. Exp ($CF_\alpha 1/CT_\alpha 1 - 1$).

**Step 3**. Solve Eq. (45) to find the 0th order poloidal magnetic flux function $\psi$, which satisfies the boundary condition,

$$R\frac{\partial}{\partial R}\left(\frac{1}{R}\frac{\partial\psi}{\partial R}\right)+\frac{\partial^2\psi}{\partial Z^2}=-Rj_{\phi-model}. \quad (47)$$

Put $Y_\alpha = \psi$, find the 0th-order densities, compute the toroidal velocities using Eq. (33), update $Y_\alpha$ and compute the current densities using Eq. (37).

**Step 4**. Update $\psi$, solving Eq. (38) for updated $Q \equiv -Rj_\phi$.

**Step 5**. Update $Y_\alpha$, $B_\phi$, $\Omega_{\alpha\phi}$ and $u_{\alpha\phi}$.

**Step 6**. Solve Eq. (42) for density ratio $n_{eh}/n_p$ iteratively given the other quantities.

**Step 7**. Update densities $n_\alpha$ for $\alpha = pl, ph, bl, bh, el, eh$.

**Step 8**. If the convergence of Ampere's law is not sufficient, go back to Step 4.

**Step 9**. If the convergence of Ampere's law is sufficient, verify that the potential $V_E$ is solved accurately.

**Step 10**. If $V_E$ is solved well, check if $(\psi - \psi_{plasma})$ is consistent with $\psi_{pfcoil}$.

**Step 11**. If so, stop computation. If not, go back to Step 4 using the present $\psi$ as the new boundary value of $\psi$, i.e., iteration of the boundary value of $\psi$.

To solve Eq. (38) together with Eqs. (45) and (46), we use the second-order finite difference method with equal mesh intervals in the (R, Z) directions. To speed up computation, we combine successive over-relaxation (SOR) with a progressive multigrid refinement scheme with three grid levels. Example 1 equilibrium is first computed using the $100 \text{ in } R \times 100 \text{ in } Z$ mesh numbers on the finest grid. To check convergence with respect to grid size, the Example 1 equilibrium is also computed using the $200 \times 100 \text{ in } R \times Z$ mesh numbers. Since the error of this check is 1% or less, Examples 2, 3, and 4 are computed using only $100 \text{ in } R \times 100 \text{ in } Z$ mesh numbers.

Flow chart to solve the 6-fluid axisymmetric force-balance equilibrium

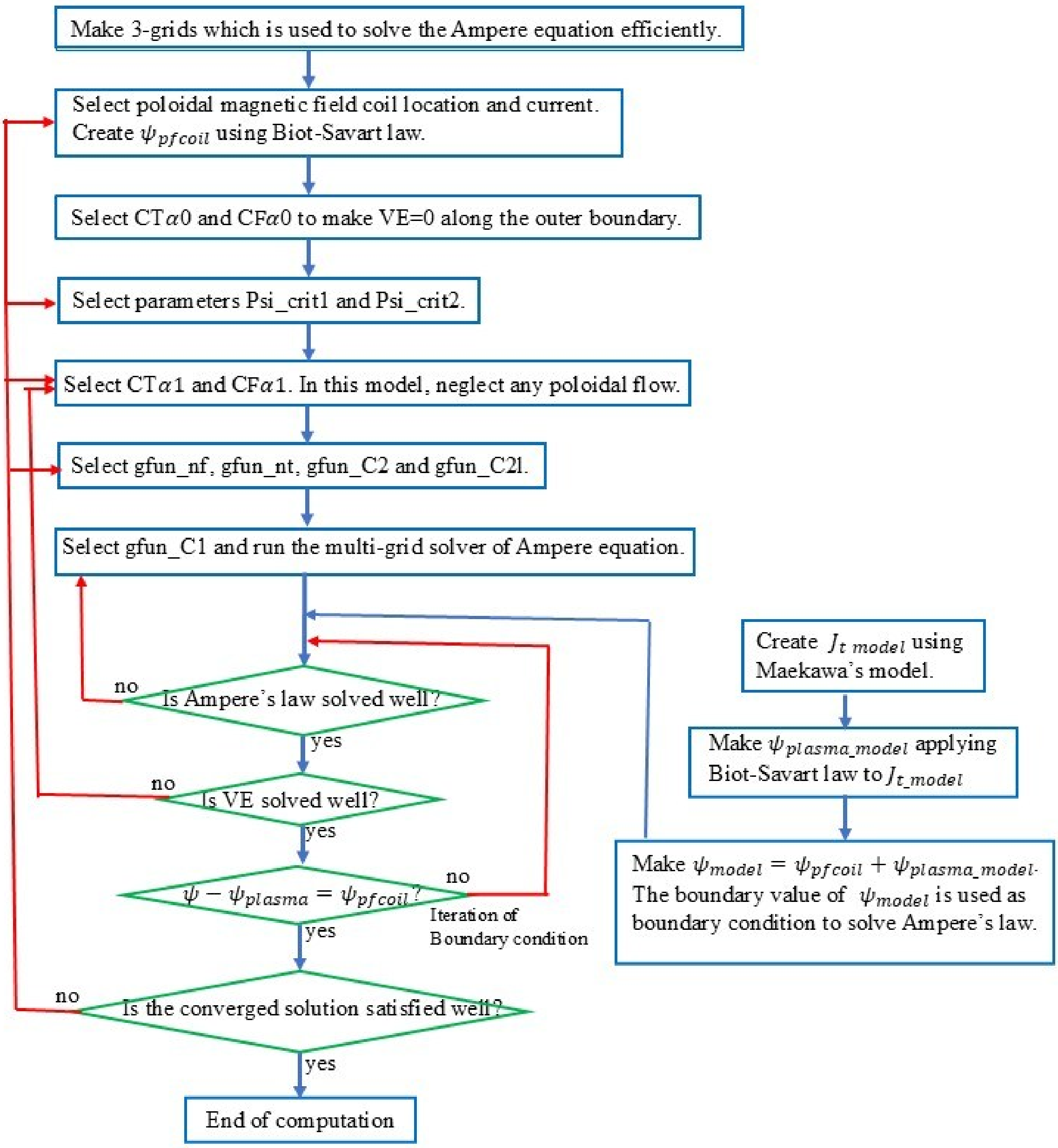

## V. COMPARATIVE NUMERICAL RESULTS

The four equilibrium calculations cover two driven-burn conditions at 100 keV–1 MeV and two sub-burn conditions for the upcoming EHL-2 experiment at 20 keV–300 keV.[1,5] Detailed poloidal-field-coil currents, profile-function inputs, complete field, density, temperature, and plasma-flow profiles, and component-resolved diagnostics are provided in Appendix E. We compare the principal global quantities of the equilibria here to assess their parameter ranges, including the roles of relativistic electrons.

TABLE I. Global parameters of the four six-fluid equilibria.

| Parameter | Equilibrium 1 | Equilibrium 2 | Equilibrium 3 | Equilibrium 4 |
|---|---|---|---|---|
| **Regime** | High-$\beta_T$ fusion | Low-$\beta_T$ fusion | Low-$\beta_T$ EHL-2 | High-$\beta_T$ EHL-2 |
| **Major radius, Rmajor (m)** | 1.426 | 1.482 | 0.979 | 0.967 |
| **Minor radius, a (m)** | 0.862 | 0.911 | 0.548 | 0.535 |
| **Plasma current, Ip (MA)** | 12.59 | 6.36 | 2.697 | 4.459 |
| **Field at major radius (T)** | 3.383 | 2.971 | 2.722 | 2.267 |
| **Toroidal beta $\beta_T$ within LCFS** | 0.253 | 0.095 | 0.058 | 0.236 |
| **Safety factor q95** | 4.28 | 6.31 | 6.69 | 3.26 |
| **Thermal-electron temperature on axis (keV)** | 66.7 | 22.4 | 15.6 | 17.1 |
| **Suprathermal electron temperature on axis (keV)** | 2373.80 | 294.66 | 287.42 | 436.48 |
| **Peak suprathermal electron *h*** | 19.1 | 2.824 | 2.77 | 3.93 |

Equilibrium 1 has the largest toroidal beta, plasma current, suprathermal electron temperature, and relativistic thermal enthalpy correction. Equilibrium 4 demonstrates that a high $\beta_T$ EHL-2 configuration can also be represented, while Equilibria 2 and 3 provide lower $\beta_T$ comparisons.

TABLE II. Relative numerical residuals reported for the four equilibria.

| **Diagnostic** | **Equilibrium 1** | **Equilibrium 2** | **Equilibrium 3** | **Equilibrium 4** |
|---|---|---|---|---|
| **Ampere's law** | 0.000614 | 0.00098 | 0.00106 | 0.00106 |
| **Electron potential $V_E$** | 0.029 | 0.00168 | 0.02036 | 0.00772 |
| **pl force balance** | 0.00189 | 0.0083 | 0.0022 | 0.0039 |
| **bl force balance** | 0.00213 | 0.0102 | 0.0025 | 0.0026 |
| **el force balance** | 0.00223 | 0.0096 | 0.0030 | 0.0049 |
| **ph force balance** | 0.00157 | 0.0058 | 0.0021 | 0.0050 |
| **bh force balance** | 0.00164 | 0.0057 | 0.0030 | 0.0044 |
| **eh force balance** | 0.00163 | 0.0046 | 0.0025 | 0.0052 |

Across the four cases, the reported Ampere-law residuals range from 0.000614 to 0.00106, the electron-potential residuals range from 0.00168 to 0.029, and the component force-balance residuals range from 0.00157 to 0.0102. Appendix E provides the full radial force-balance plots and case-specific validation figures.

## VI. DISCUSSION

### A. Physical significance of the six-fluid equilibrium

A key physical implication is that high-temperature p-$^{11}$B spherical-torus equilibria cannot be specified adequately by the pressure and current profiles alone. Instead, they require six component force-balance conditions, six modified fluxes $Y_\alpha$, six density fields, and eighteen prescribed profile functions $F_\alpha(Y_\alpha)$, $T_\alpha(Y_\alpha)$, and $K_\alpha(Y_\alpha)$. This added structure is required because protons and boron ions have different inertia per unit charge; suprathermal and thermal populations can have different flows and temperatures; and the suprathermal electron fluid can strongly affect the toroidal current and electrostatic potential.

The modified flux $Y_\alpha$ is physically meaningful. As the toroidal canonical momentum is conserved under axisymmetry, strong plasma flows displace the component flow flux distribution away from ψ, as given by Eq. (16). The thermal electron flux may depart from ψ when $h$(T*) is large, even when $\gamma_{eh}$ is close to unity. The boron ion flux may depart from ψ because $m_b/q_b$ is large. The sign of $q_\alpha$ implies opposite shifts in the electron and ion flux components, even when their toroidal flow directions are the same.

Consequently, we check equilibrium accuracy component by component. A smooth total pressure and an Ampere-equation-consistent total current do not guarantee that each component satisfies radial force balance. The radial force balance of each of the fluids is shown in Figs. 4-a, b, c, d, e, f of Appendix E, indicating numerical errors within 0.1% of the peak forces.

### B. Quasineutrality and relativistic electron enthalpy

The replacement of Poisson's equation with quasineutrality should be presented as an ordered approximation rather than an assumption with scaled estimates. Because $q_\alpha V_E$ appears in the Bernoulli-like relation for each component, errors in $V_E$ propagate directly into the density, flow, and force balance. Quasineutrality is verified in four example equilibria as shown in Appendix C.

Relativistic corrections are primarily thermal rather than kinematic. The drift Lorentz factor remains close to unity in the reported cases, whereas $h$(T*) reaches approximately 19.1 for the relativistic suprathermal electron population. Thus, relativistic effects arise from random electron energies comparable to or exceeding $m_e c^2$, not from equilibrium drift speeds approaching c. The implemented approximation to $h$(T*) should be compared with the exact Bessel-function ratio across the full temperature range. Appendix A provides the error estimates.

### C. Numerical range, interpretation, and limitations

The four equilibria demonstrate that the formalism and the input functionals can span $\beta_T$=0.058-0.253, $I_p$=2.697-12.59 MA, $T_{eh}$=287 keV - 2.37 MeV, and $h$ values up to 19.1. The high-$\beta_T$ burning-plasma-relevant case is the most demanding, with $R_{major}$ approximately 1.426 m, $I_p$ =12.59 MA, B approximately 3.383 T, $\beta_T$ = 0.253, $\beta_p$ = 1.091, $T_{pl}$ = $T_{bl}$ approximately 132 keV, and $T_{eh}$ approximately 2.37 MeV. The lower-$\beta_T$ case is more conservative, with $R_{major}$ approximately 1.482 m, $I_p$= 6.36 MA, B approximately 2.971 T, $\beta_T$ = 0.095, and $T_{pl}$ = $T_{bl}$ approximately 44.8 keV. The two EHL-2-like cases test whether the same model can represent lower-temperature plasmas at lower-current and suprathermal-temperature regimes.

These cases should be viewed as fusion-relevant equilibrium studies that do not demonstrate p-$^{11}$B burn. The profile functions are prescribed, and the present equilibrium solution does not address energy balance, alpha-particle heating, radiation, ash accumulation, transport, kinetic slowing down, and stability. Suprathermal populations are treated as scalar-temperature fluids, so anisotropy, pitch-angle structure, finite-orbit-width effects, loss cones, slowing-down distributions, and wave-particle resonances remain future kinetic tasks. [16,17]

High $\beta_T$, energetic-electron current, electrostatic-field structure, and flow shear can affect kink, ballooning, interchange, tearing, energetic-particle, and wave-particle dynamics. The present equilibria are best treated as inputs to future stability, kinetic, radiation, alpha-particle, and transport assessments. [18,19]

## VII. CONCLUSION

This paper describes the development of a relativistic six-fluid, flowing-plasma-equilibrium formulation for p-$^{11}$B spherical-torus plasmas. The paper treats thermal and suprathermal proton, boron, and electron fluids as distinct components, each with its own density, flow, temperature, and profile-function structure. The component-modified flux $Y_\alpha$, associated with canonical angular momentum, serves as the organizing label for axisymmetric force balance, while $h$(T*) provides the relativistic thermal enthalpy correction required for the high-temperature electron fluids.

The reported cases span reactor-relevant and EHL-2-like spherical-torus parameters, with $\beta_T$ = 0.058-0.253, $I_p$ = 2.697-12.59 MA, $T_{eh}$= 287 keV-2.37 MeV, and $h$ up to approximately 19.1. The range of these parameters indicates that multi-fluid high-temperature p-$^{11}$B ST equilibria can differ qualitatively from single-fluid static tokamak equilibria and require a plasma-component-resolved model developed here.

The numerical accuracies relative to the average values of the equilibrium solutions, in satisfying Ampere's law, the electrostatic potential, and force balance, are provided in Table II in Section IV. It is seen that, while Ampere's law is well satisfied, the thermal electron solution can have errors of up to 3% in the electrostatic potential, and the thermal boron solution can have errors of up to 1%.

These equilibria do not account for plasma power balance, alpha heating, radiation losses, ash accumulation, the kinetic slowing down of suprathermal ions, plasma transport, or plasma stability. Quasineutrality, the scalar-temperature suprathermal fluid, and the h(T*) approximations introduce simplification errors relative to the plasma kinetic model. With experimental verification and computational reproducibility, the model can still provide an equilibrium foundation for subsequent studies of stability, transport, wave-plasma interactions, radiation losses, and alpha-particle heating and transport in ST p-11B fusion plasmas. Recent p-B plasma data on EXL-50U showed[20,21], in separate discharges, $T_{i0}$ up to 10 keV, $T_{e0}$ up to 4 keV, $I_p$ up to 1 MA with boron density up to 10% of the proton density, and non-inductively sustained $I_p$ up to 270 kA using 350 kW ECRH power. These plasmas serve as new targets for equilibrium reconstruction to refine the multi-fluid equilibrium model and its applications.

**ACKNOWLEDGMENTS**

We acknowledge discussions with Dr. Bing Liu on leveraging suprathermal plus thermal plasma in ST p-$^{11}$B fusion plasmas; with Drs. Yuejiang Shi and Dong Guo on high-performance p-boron plasma data from EXL-50U; with Drs. Yunfeng Liang and Jiaqi Dong on the physics design issues of the upcoming EHL-2 experiment; with Drs. Wenjun Liu, Tiantian Sun, Yuan Ying, and Hairon Huang on the special properties of ST p-boron plasmas associated with substantial suprathermal ions and electrons; and with Dr. Eduard Karutdinov on improving the transport physics basis for multi-fluid p-boron plasma equilibrium.

## APPENDIX A: RELATIVISTIC THERMAL ENTHALPY FUNCTION h

In the first equilibrium, $T_{eh_{max}}$ reaches about 2.4 MeV. So, we adopt the following form of $h$.

$h(T^*) = 1 + 2.5T^* + 1.1495T^{*2}$ for small $T^*$, and

$h(T^*) = constant + 4T^*$ for large $T^*$,

according to Eqs. (A18) and (A19) of [11]. The continuity condition of $h$ and $dh/dT^*$ then gives

$$h(T^*) = 1 + 2.5T^* + 1.1495T^{*2} \quad \text{for } T^* \leq T_c^* = 0.652458, \text{ and}$$

$$h(T^*) = 1 - 1.5T_c^* + 1.1495T_c^{*2} + 4T^* \quad \text{for } T^* \geq T_c^*. \qquad \text{(A.1)}$$

See Fig. 1 below.

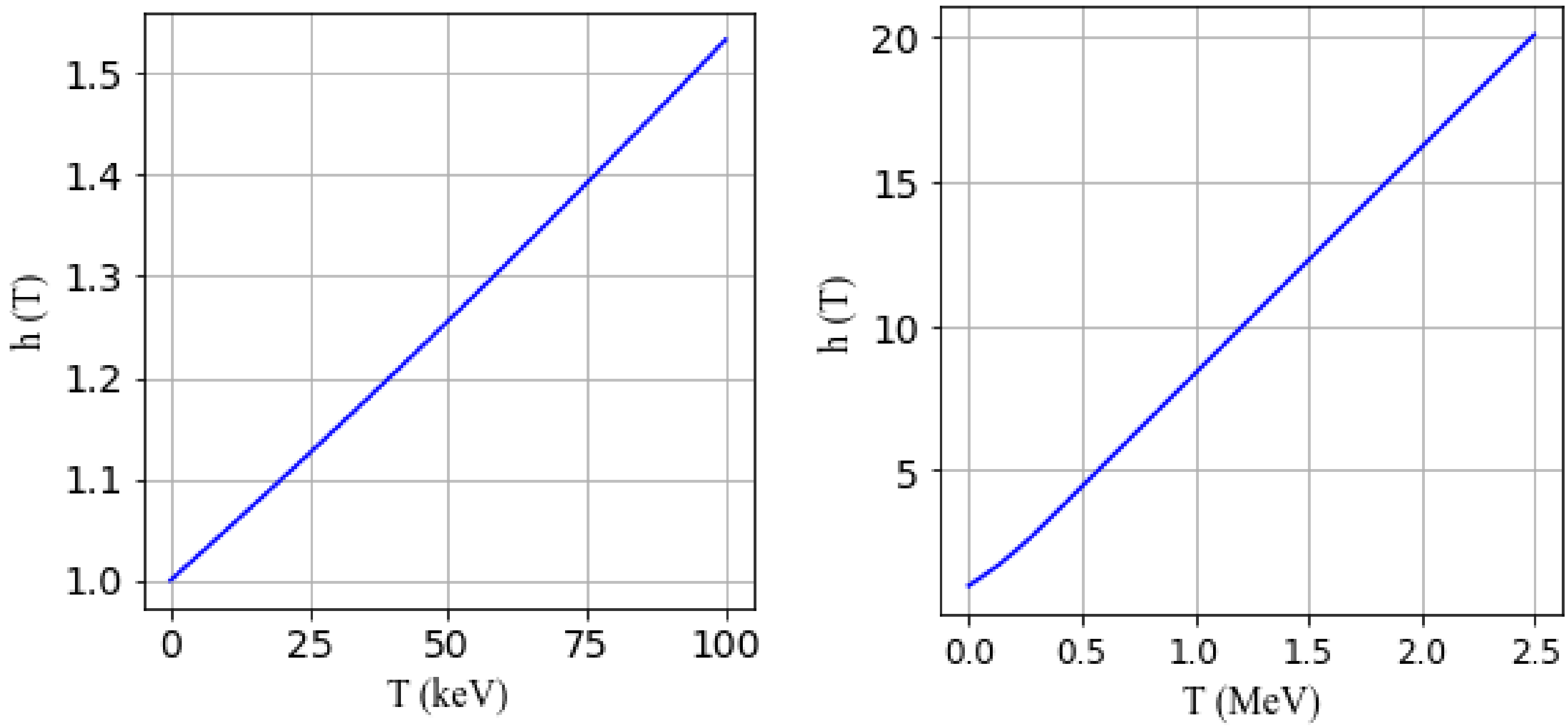

Figure 1. The values of $h(T)$ based on Eq. (A.1).

Previous calculations treated the nonrelativistic thermal electron fluid; the present parameters require a relativistic enthalpy correction even for the thermal electrons. For equilibrium example 1, $h_{el_{max}} = 1.35$ for $T_{el_{max}} = 66.7keV$ and $h_{eh_{max}} = 19.1$ for $T_{eh_{max}} = 2.37MeV$.

The above piecewise $h$ function can be compared with the integral representation of $h$ given by

$$h = \frac{z}{5}\frac{I_3(z)}{I_2(z)}, \text{ where } z = \frac{1}{T^*} \text{ and } I_n(z) \equiv \int_0^\infty dx e^{-zx}\left(x(x+2)\right)^{n-\frac{1}{2}}, \qquad \text{(A.2)}$$

according to Eq. (A16) of [11]. Its comparison with the values of Eqn. A.1 and relative errors are shown in Fig. 2.

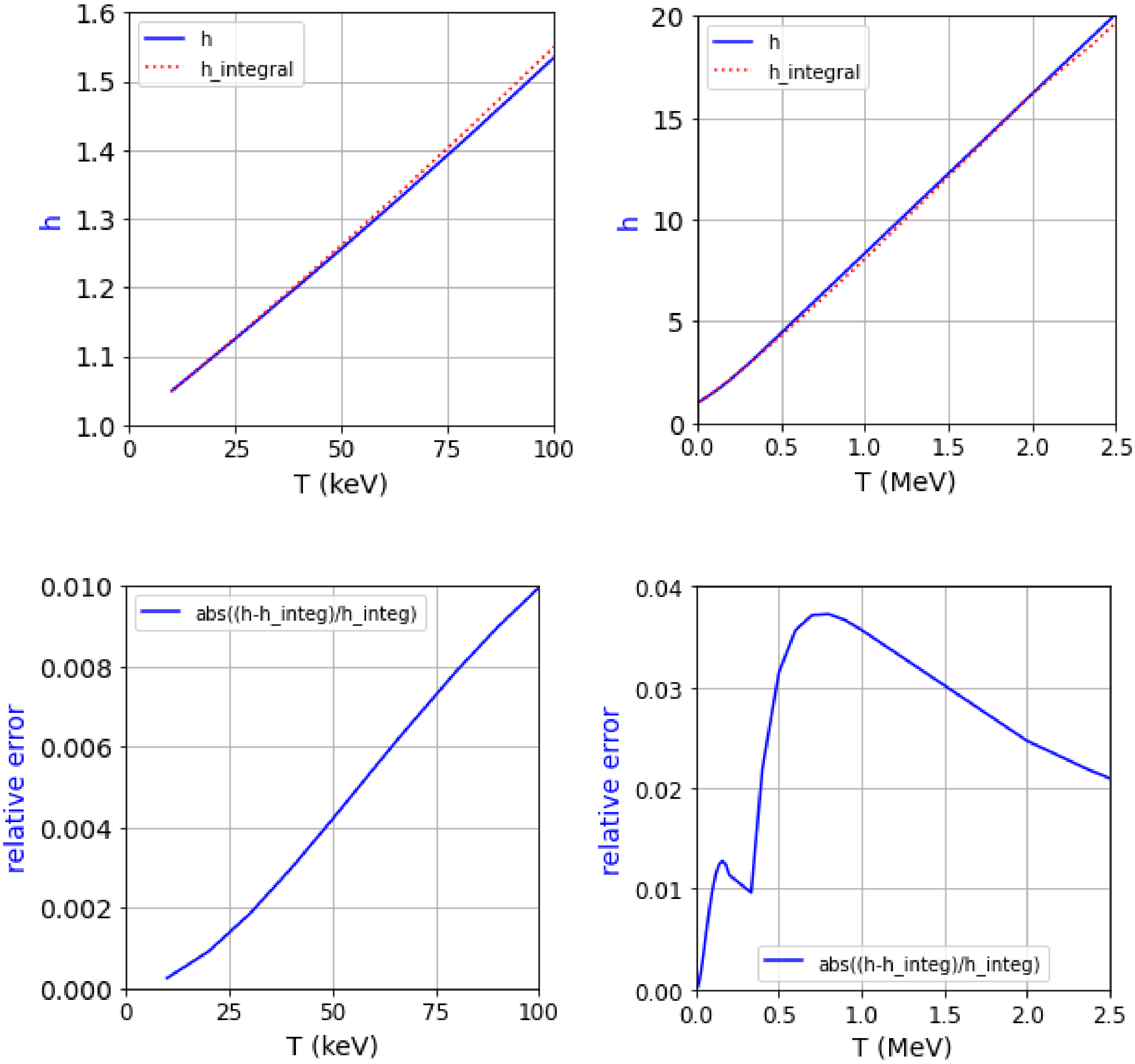

Figure 2. The $h$ function defined by Eq. (A.1) compared with the $h$ function defined by Eq. (A.2), showing a maximum relative error of 0.037 at T = 0.8 MeV. Here the relative error is defined by $\left|\frac{h-h_{integral}}{h_{integral}}\right|$.

## APPENDIX B: THERMAL VELOCITY, PLASMA FREQUENCY, AND SOUND VELOCITY

The effect of low-density fluid species is small on the dispersion relation of the electrostatic mode; we neglect the ph- and bh- fluids here. However, we keep the relativistic eh-fluid to account for thermal velocity and the plasma frequency of the relativistic fluid species.

The continuity equation is given by

$$\frac{\partial \gamma n}{\partial t} + \nabla \cdot (\gamma n \boldsymbol{u}) = 0. \quad \text{(B.1)}$$

The equation of fluid motion is given by

$$\frac{\partial}{\partial t}(mn\gamma^2 h(T^*)\boldsymbol{u}) + \frac{\partial}{\partial x_j}\left(mn\gamma^2 h(T^*)\boldsymbol{u}u_j\right) + \boldsymbol{\nabla} p = q\gamma n\boldsymbol{E} + q\gamma n\boldsymbol{u} \times \boldsymbol{B}. \quad \text{(B.2)}$$

As for heat balance, we adopt the isothermal relation,

$$\frac{dT}{dt} = \frac{\partial T}{\partial t} + \boldsymbol{u} \cdot \boldsymbol{\nabla} T. \quad \text{(B.3)}$$

The equilibrium state requires that $\boldsymbol{u_0} \cdot \boldsymbol{\nabla} T_0 = 0$, which is consistent with Eqs. (20) and (19) for axisymmetric equilibrium. In the equilibrium example 1 reported in Appendix E, $\gamma - 1 < 2 \times 10^{-4}$ for the eh-fluid, allowing $\gamma = 1$ in this appendix.

We assume in this appendix that the equilibrium plasma flow and its perturbation have only a z-component; the equilibrium quantity is constant along the z-direction, allowing the perturbation form of $\tilde{f} = \hat{f} e^{ikz - i\omega t}$.

From Eq. (B.3), one obtains

$$\frac{\partial T_1}{\partial t} + u_0 \frac{\partial T_1}{\partial z} = 0, \text{ leading to } T_1 = 0. \quad \text{(B.4)}$$

From Eq. (B.1) and $p=nT$, one obtains

$$u_1 = \frac{\omega - k u_0}{k n_0} n_1 \text{ and } p_1 = T_0 n_1. \quad \text{(B.5)}$$

Since $T_1 = 0$, no perturbation of $g_{ep}(T^*)$ exists. Eq. (B.2) becomes

$$-imh(T_0^*)(\omega - k u_0) n_0 u_1 + ikT_0 n_1 = -ikqn_0 V_{E1}, \quad \text{(B.6)}$$

Substituting Eq. (B.5) in Eq. (B.6) gives Eq. (B.7),

$$n_1 = \frac{\frac{qn_0}{mh}}{(\omega - ku_0)^2 - k^2\frac{T_0}{mh}} k^2 V_{E1}. \quad \text{(B.7)}$$

Include these results for the four fluid species, pl, bl, el, and eh. The charge neutrality condition becomes

$$en_{pl1} + eZ_b n_{bl1} - en_{el1} - en_{eh1} = 0, \quad \text{(B.8)}$$

which gives the following dispersion relation,

$$\frac{\frac{e^2 n_{pl}}{\varepsilon_0 m_p}}{\left(\omega - ku_{pl}\right)^2 - k^2\frac{T_{pl}}{m_p}} + \frac{\frac{(eZ_b)^2 n_{bl}}{\varepsilon_0 m_b}}{(\omega - ku_{bl})^2 - k^2\frac{T_{bl}}{m_b}} + \frac{\frac{e^2 n_{el}}{\varepsilon_0 m_e h\left(T_{el}^*\right)}}{(\omega - ku_{el})^2 - k^2\frac{T_{el}}{m_e h\left(T_{el}^*\right)}} + \frac{\frac{e^2 n_{eh}}{\varepsilon_0 m_e h\left(T_{eh}^*\right)}}{(\omega - ku_{eh})^2 - k^2\frac{T_{eh}}{m_e h\left(T_{eh}^*\right)}} = 0. \quad \text{(B.9)}$$

The plasma frequency and thermal velocity of the pl-and bl-fluids are given by

$$\omega_{pl} = \sqrt{\frac{e^2 n_{pl}}{\varepsilon_0 m_p}} \text{ and } v_{th_{pl}} = \sqrt{\frac{T_{pl}}{m_p}}, \ \omega_{bl} = \sqrt{\frac{[\![(eZ_b)]\!]^2 n_{bl}}{\varepsilon_0 m_b}} \text{ and } v_{th_{bl}} = \sqrt{\frac{T_{bl}}{m_b}}, \quad \text{(B.10)}$$

where the plasma frequencies and thermal velocities of the el-fluid and eh-fluid are given by,

$$\omega_{el} = \sqrt{\frac{e^2 n_{el}}{\varepsilon_0 m_e h\left(T_{el}^*\right)}} \text{ and } v_{th_{el}} = \sqrt{\frac{T_{el}}{m_e h\left(T_{el}^*\right)}}, \quad \text{(B.11)}$$

$$\omega_{eh} = \sqrt{\frac{e^2 n_{eh}}{\varepsilon_0 m_e h\left(T_{eh}^*\right)}} \text{ and } v_{th_{eh}} = \sqrt{\frac{T_{eh}}{m_e h\left(T_{eh}^*\right)}}. \quad \text{(B.12)}$$

Eq. (B.12) keeps $\frac{v_{th_{eh}}}{c} < 0.5$, i.e., thermal velocity is less than the speed of light.

Since $n_{eh} \ll n_{el}$ and $T_{eh} \gg T_{el}$ in the plasma considered here, we neglect the fourth term of Eq. (B.9) and consider the sound mode, i.e., $\omega = kC_s$. Since the drift velocity is much smaller than the thermal velocity, we neglect it. Then the dispersion relation for the sound velocity $C_s$ is given by

$$\frac{\omega_{pl}^2}{C_s^2 - v_{th_{pl}}^2} + \frac{\omega_{bl}^2}{C_s^2 - v_{th_{bl}}^2} = k_{el}^2, \text{ where } k_{el}^2 = \frac{\omega_{el}^2}{v_{th_{el}}^2}, \quad \text{(B.13)}$$

The contribution appears in the form of $k_{ph}^2 = \frac{\omega_{ph}^2}{v_{th_{ph}}^2}$ because its thermal velocity is large. This is much less than $k_{el}^2$. We neglect this. As for the suprathermal boron fluid, it makes $\frac{\omega_{bh}^2}{C_s^2 - v_{th_{bh}}^2}$. Since $\omega_{bh}^2 \ll \omega_{bl}^2$, its contribution is significant only if $C_s$ is close to $v_{th_{bh}}$, Here we neglect this effect.

For numerical computation, the following expression is useful.

$$C_s^4 - \left(v_{th_{pl}}^2 + v_{th_{bl}}^2 + \left(\frac{\omega_{pl}}{k_{el}}\right)^2 + \left(\frac{\omega_{bl}}{k_{el}}\right)^2\right) C_s^2 + v_{th_{pl}}^2 v_{th_{bl}}^2 + \left(\frac{\omega_{pl}}{k_{el}}\right)^2 v_{th_{bl}}^2 + \left(\frac{\omega_{bl}}{k_{el}}\right)^2 v_{th_{pl}}^2 = 0 \quad \text{(B.14)}$$

A dimensionless form of the dispersion relation becomes,

$$C_s^2 = \frac{A_1}{2} \pm \sqrt{\left(\frac{A_1}{2}\right)^2 - A_2} \qquad \text{or} \qquad C_s = \sqrt{\frac{A_1}{2} \pm \sqrt{\left(\frac{A_1}{2}\right)^2 - A_2}}\ , \qquad \text{(B.15)}$$

$A_1 \equiv \left(v_{th_{pl}}^2 + v_{th_{bl}}^2 + \left(\frac{\omega_{pl}}{k_{el}}\right)^2 + \left(\frac{\omega_{bl}}{k_{el}}\right)^2\right)$, and

$A_2 \equiv v_{th_{pl}}^2 v_{th_{bl}}^2 + \left(\frac{\omega_{pl}}{k_{el}}\right)^2 v_{th_{bl}}^2 + \left(\frac{\omega_{bl}}{k_{el}}\right)^2 v_{th_{pl}}^2$, with

$$v_{th_{pl}}^2 = T_{pl}\ ,\ v_{th_{bl}}^2 = \frac{T_{bl}}{\frac{m_b}{m_p}}\ ,\ \left(\frac{\omega_{pl}}{k_{el}}\right)^2 = T_{el}\frac{n_{pl}}{n_{el}}\ ,\ \left(\frac{\omega_{bl}}{k_{el}}\right)^2 = Z_b^2 \frac{T_{el}}{\frac{m_b}{m_p}} \frac{n_{bl}}{n_{el}}\ .$$

(B.15) shows that two types of Cs can exist, i.e., fast Cs and slow Cs.

An effective sound velocity emerges with

$$Z_{eff} \equiv \frac{\sum_j Z_j Z_j n_j}{\sum_j Z_j n_j},\quad T_{eff} \equiv \frac{\sum_j \llbracket T_j Z)_j n_j}{\sum_j \llbracket Z_j n_j)},\quad m_{eff} \equiv \frac{\sum_j \llbracket m_j Z)_j n_j}{\sum_j \llbracket Z_j n_j)}, \text{ and } C_s^{eff} \equiv \sqrt{\frac{Z_{eff} T_e + T_{eff}}{m_{eff}}},\quad \text{(B.16)}$$

where j-sum is taken for ions, in contrast to Eq. (B.16).

$$C_s^{eff} = \sqrt{\frac{\left(1 + z_b(z_b - 1)\frac{n_{bl}}{n_{el}}\right) T_{el} + T_{pl}\frac{n_{pl}}{n_{el}} + T_{bl}\frac{z_b n_{bl}}{n_{el}}}{m_p \frac{n_{pl}}{n_{el}} + m_b \frac{z_b n_{bl}}{n_{el}}}}\ . \qquad \text{(B.17)}$$

**APPENDIX C. VALIDITY OF THE QUASINEUTRALITY APPROXIMATION**

Poisson equation in dimensional form is given by Eq. (C.1).

$$-\varepsilon_0 \Delta V_E = e(n_{pl} + Z_b n_{bl} + n_{ph} + Z_b n_{bh} - n_{el} - n_{eh}) \quad \text{(C.1)}$$

The effective density of the Laplacian term is defined by $n_{Lap} \equiv \frac{\varepsilon_0}{e} \Delta V_E$ (C.2)

If $n_{Lap}$ is evaluated by the potential $V_E$, which is derived using the charge-neutrality condition, is much less than the lowest density, then the charge-neutrality is justified.

In the numerical computation, we use dimensionless variables. The dimensionless form of $n_{Lap}$ is given by Eq. (C.3).

$$n_{Lap} = \varepsilon^2 \left(\frac{u_{ref}}{c}\right)^2 \left(\frac{\partial^2 V_E}{\partial R^2} + \frac{1}{R}\frac{\partial V_E}{\partial R} + \frac{\partial^2 V_E}{\partial Z^2}\right) \quad \text{(C.3)}$$

Here c is speed of light in vacuum and $\varepsilon \equiv \frac{l_{pR}}{L_{ref}}$, where $l_{pR}$ is the proton inertial length for the reference density $n_{ref}$

For Example 1 equilibrium, $\varepsilon$ =0.014401, $u_{ref}$ =13868.37 km/s. $\varepsilon^2 \left(\frac{u_{ref}}{c}\right)^2 = 4.44 \times 10^{-7}$.

For Example 2 equilibrium, $\varepsilon$ =0.014401 and $u_{ref} =$ 8099.127 km/s. $\varepsilon^2 \left(\frac{u_{ref}}{c}\right)^2 = 1.51 \times 10^{-7}$.

For Example 3 equilibrium, $\varepsilon = 0.024175$ and $u_{ref}$ = 6172.3 km/s. $\varepsilon^2 \left(\frac{u_{ref}}{c}\right)^2 = 2.48 \times 10^{-7}$.

For Example 4 equilibrium, $\varepsilon = 0.03198$ and $u_{ref}$ = 10328.7 km/s. $\varepsilon^2 \left(\frac{u_{ref}}{c}\right)^2 = 1.21 \times 10^{-6}$.

We compare $n_{Lap}$ with the density of the suprathermal boron. The following four results show that the density of the Laplacian term $n_{Lap}$ is much less than the density of the suprathermal boron. This shows that charge neutrality is satisfied, as shown in Figs. 3 – 6.

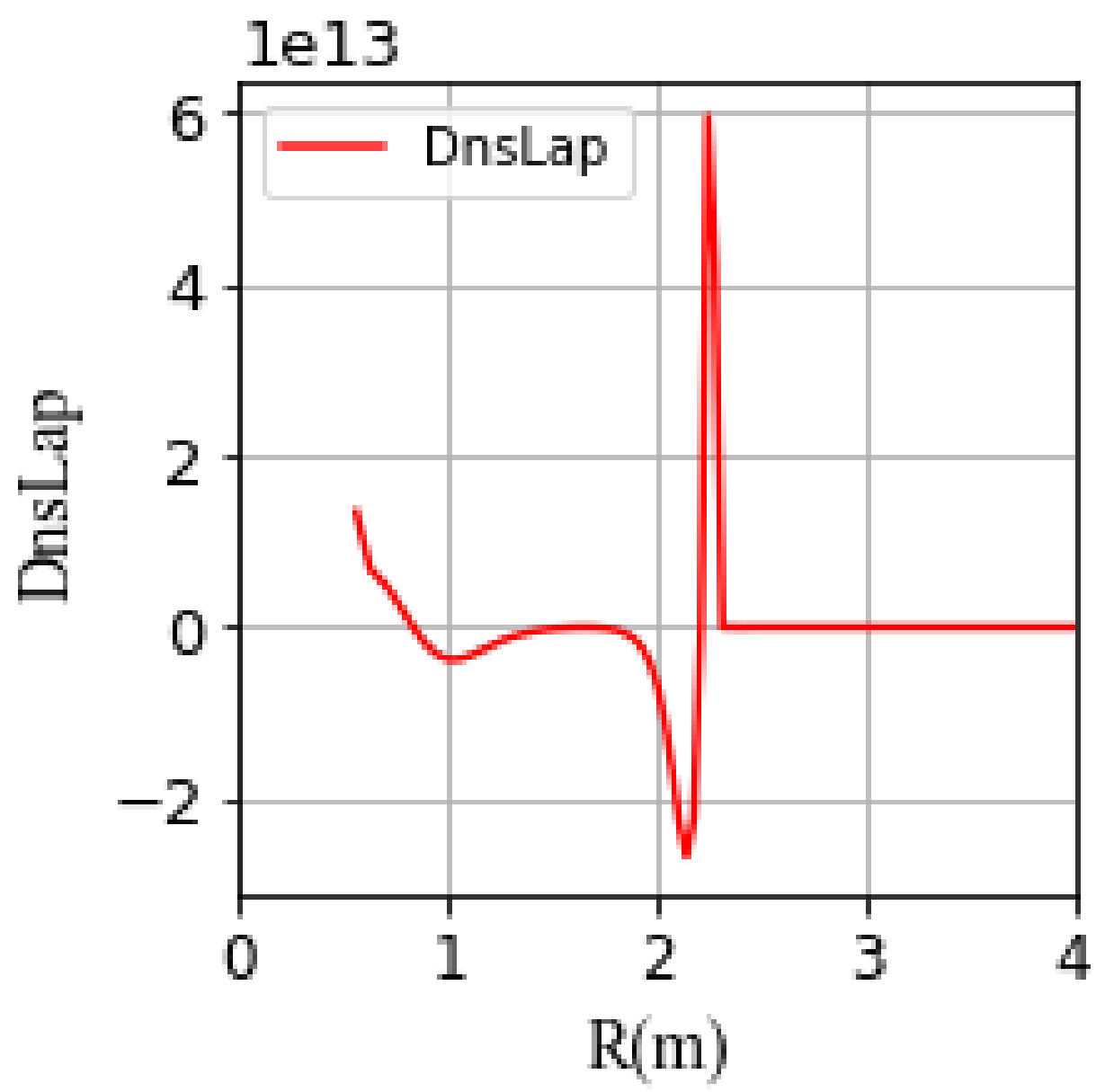

Figure 3. Relative error in charge neutrality of the Example 1 equilibrium, showing the average of $|DnsLap|$ / the average of $|Z_b Dnsbh| = 1.2 \times 10^{-5}$ .

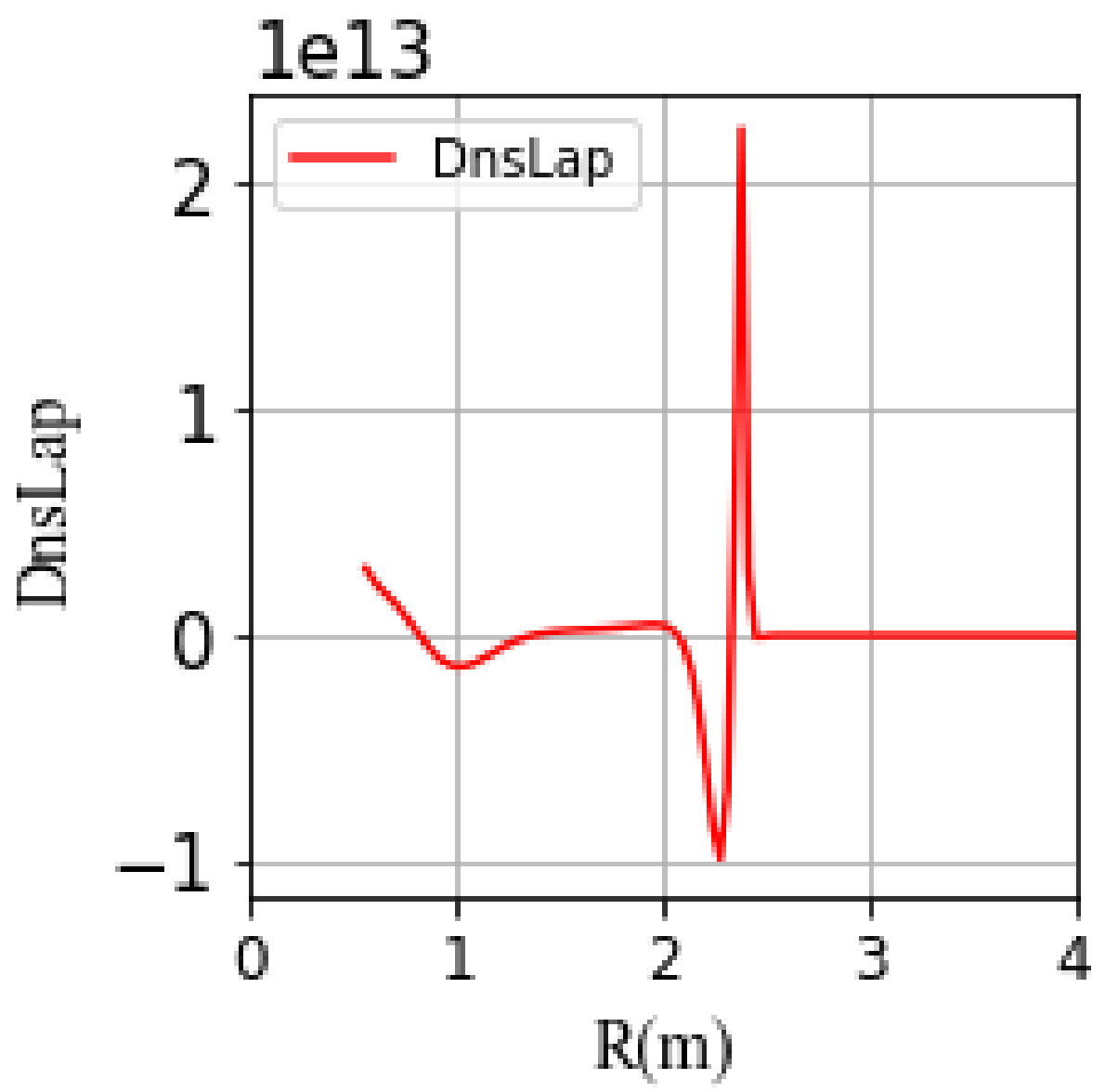

Figure 4. Relative error in charge neutrality of the Example 2 equilibrium, showing the average of $|DnsLap|$ / the average of $|Z_b Dnsbh| = 1.5 \times 10^{-5}$ .

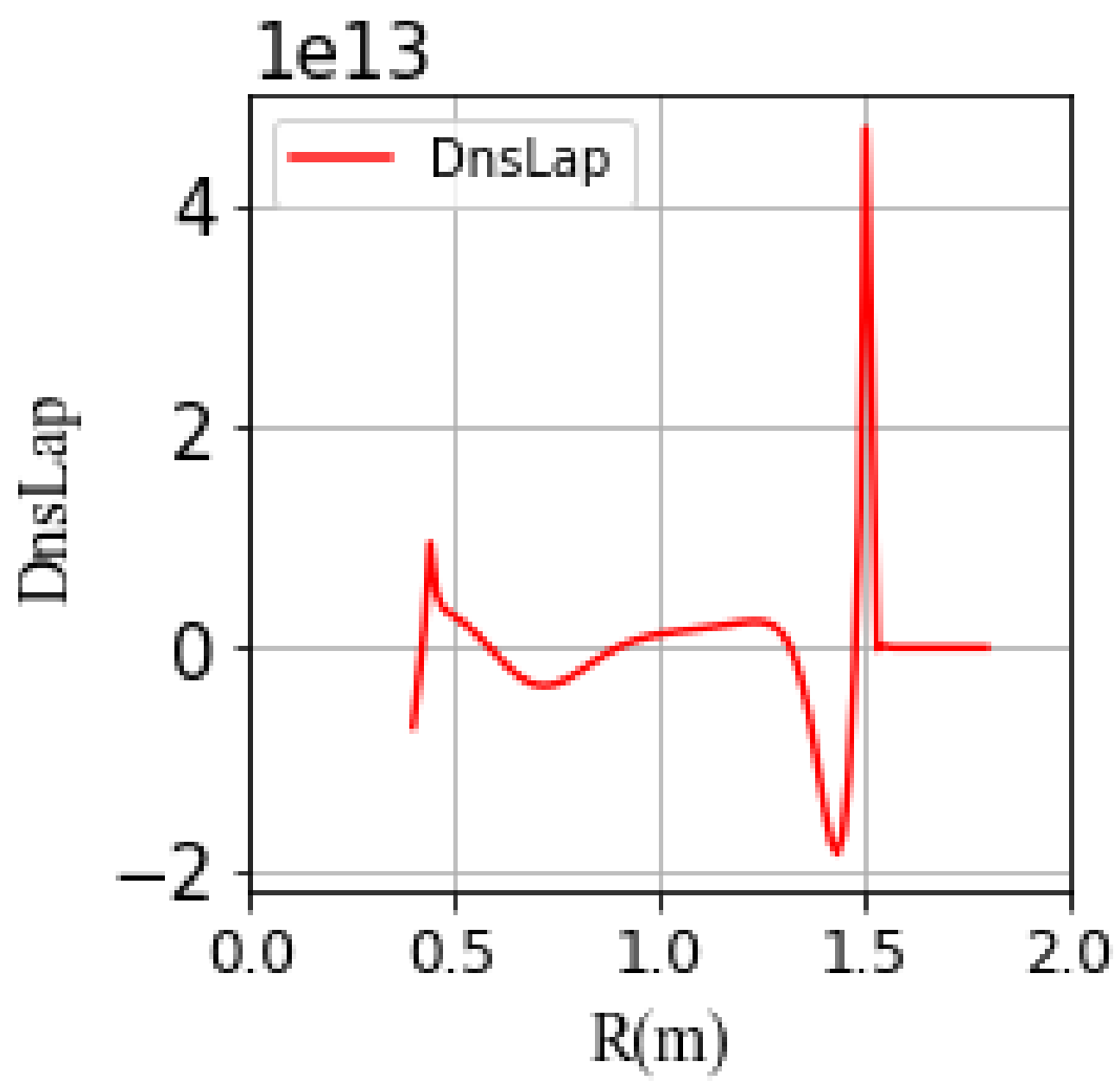

Figure 5. Relative error in charge neutrality of the Example 3 equilibrium, showing the average of $|DnsLap|$ / the average of $|Z_b Dnsbh| = 6.5 \times 10^{-5}$.

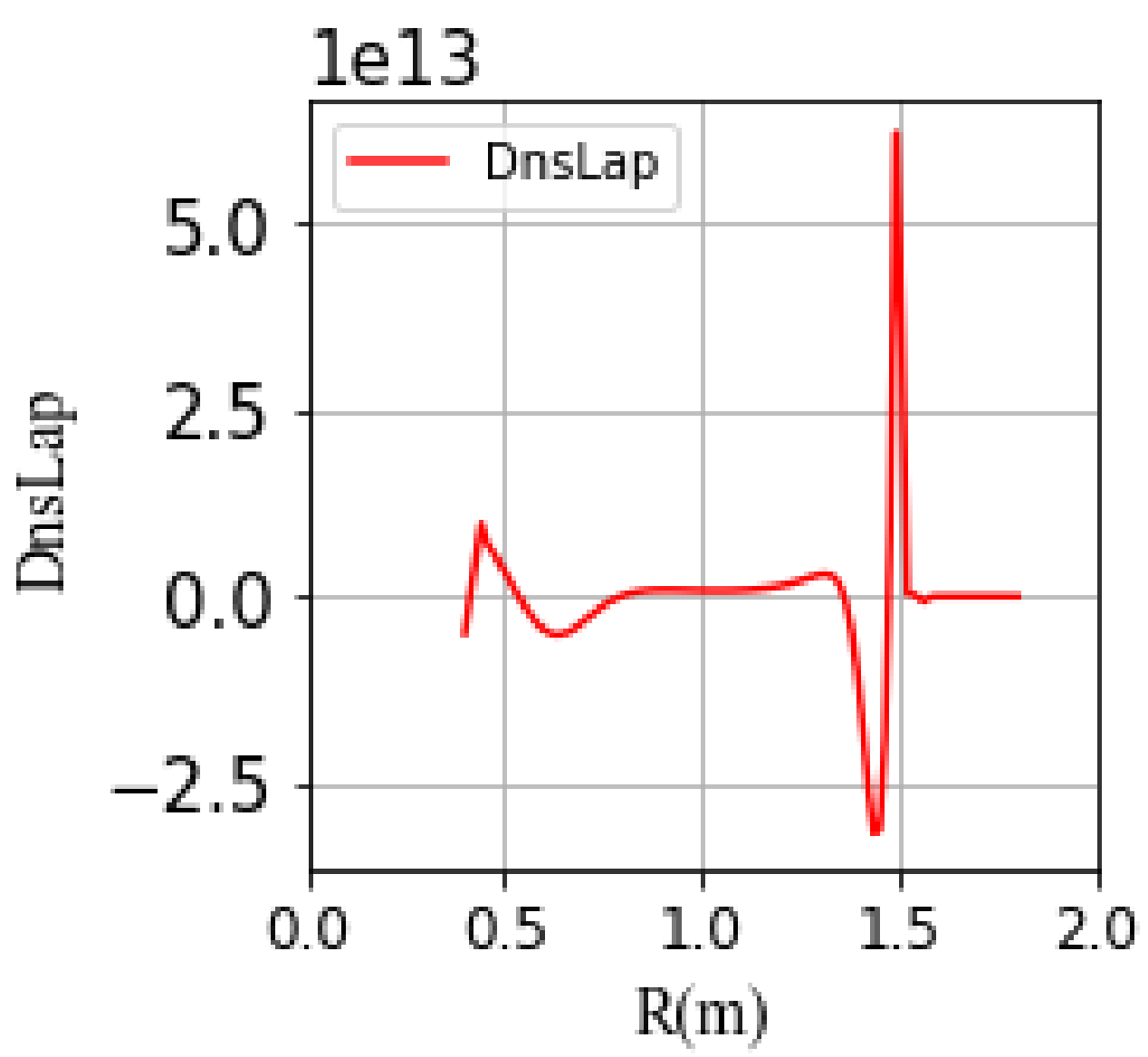

Figure 6. Relative error in charge neutrality of the Example 4 equilibrium, showing the average of $|DnsLap|$ / the average of $|Z_b Dnsbh| = 7.5 \times 10^{-6}$.

## APPENDIX D: Debye length $\boldsymbol{\lambda_D}$

Debye length $\lambda_D$ is given by $\frac{1}{\lambda_D^2} = \sum_\alpha \frac{q_\alpha^2 n_\alpha}{\varepsilon_0 T_\alpha}$ in dimensional quantities, where

$$\frac{1}{\lambda_D^2(m)} = 1.8095 \times 10^{-9} \left\{ \frac{n_{el}(m^{-3})}{T_{el}(keV)} + \frac{n_{eh}}{T_{eh}} + \frac{n_{pl}}{T_{pl}} + \frac{n_{ph}}{T_{ph}} + \frac{Z_b^2 n_{bl}}{T_{bl}} + \frac{Z_b^2 n_{bh}}{T_{bh}} \right\} .$$

The Debye lengths on the mid-plane of the four equilibria are plotted below and are quite small compared with typical lengths that characterize the plasma parameters, as shown in Fig. 7.

Debye length of Example 1 equilibrium

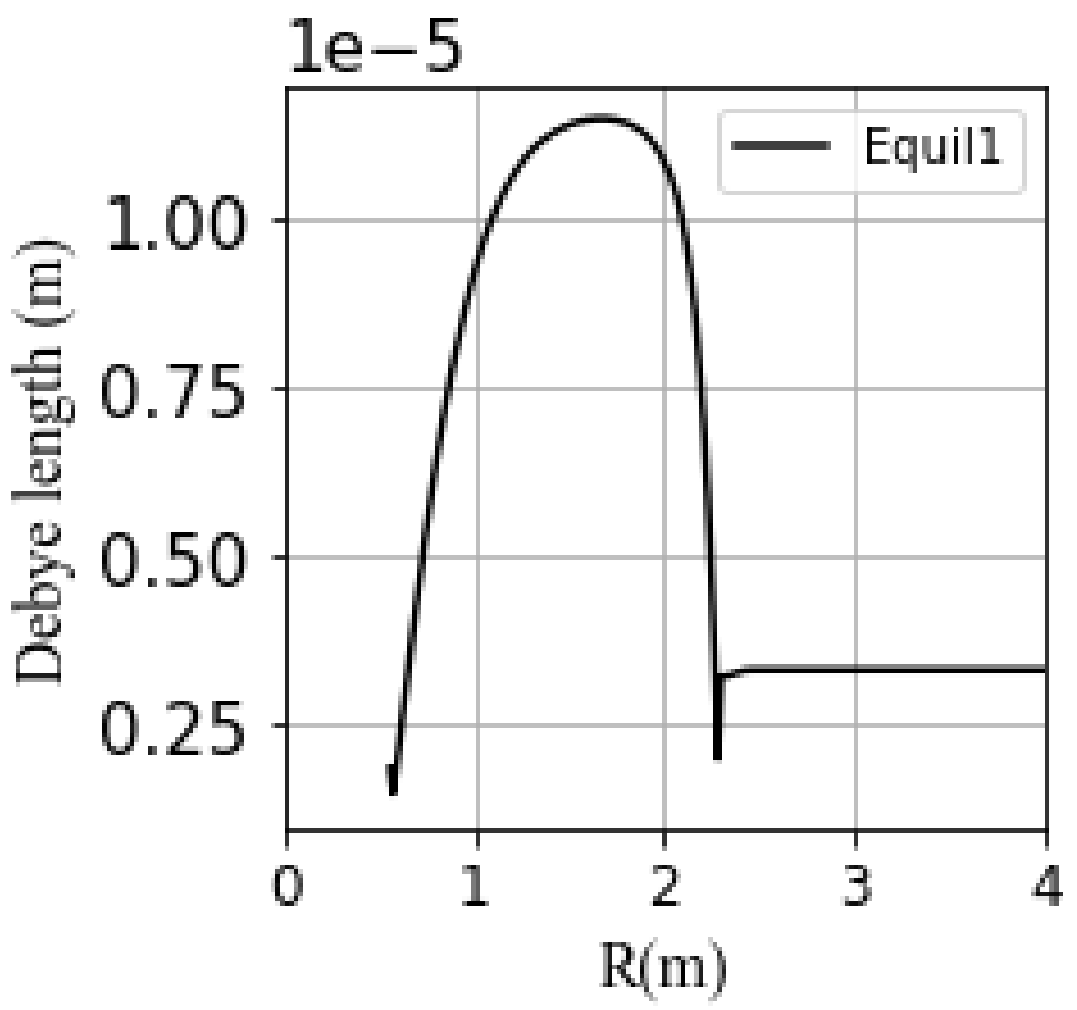

Debye length of Example 2 equilibrium

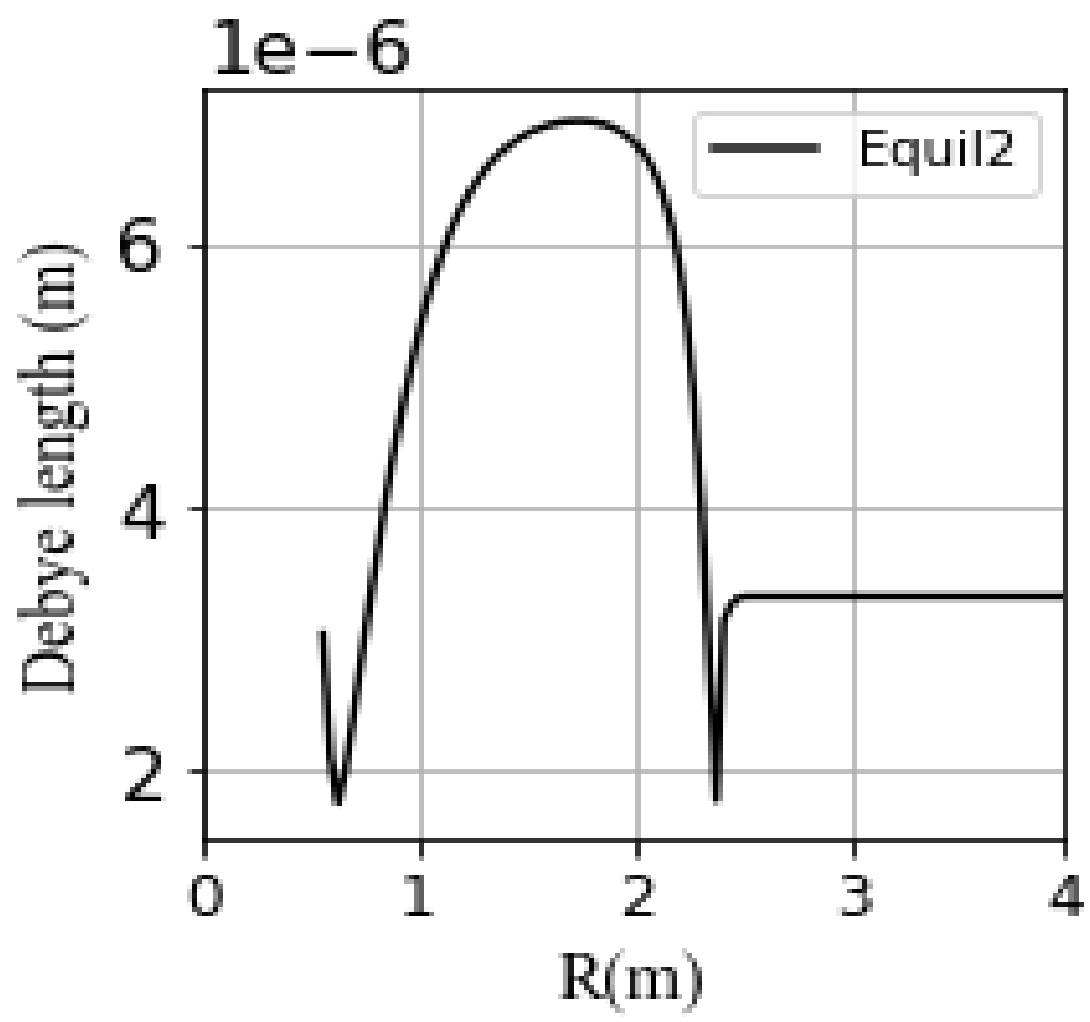

Debye length of Example 3 equilibrium

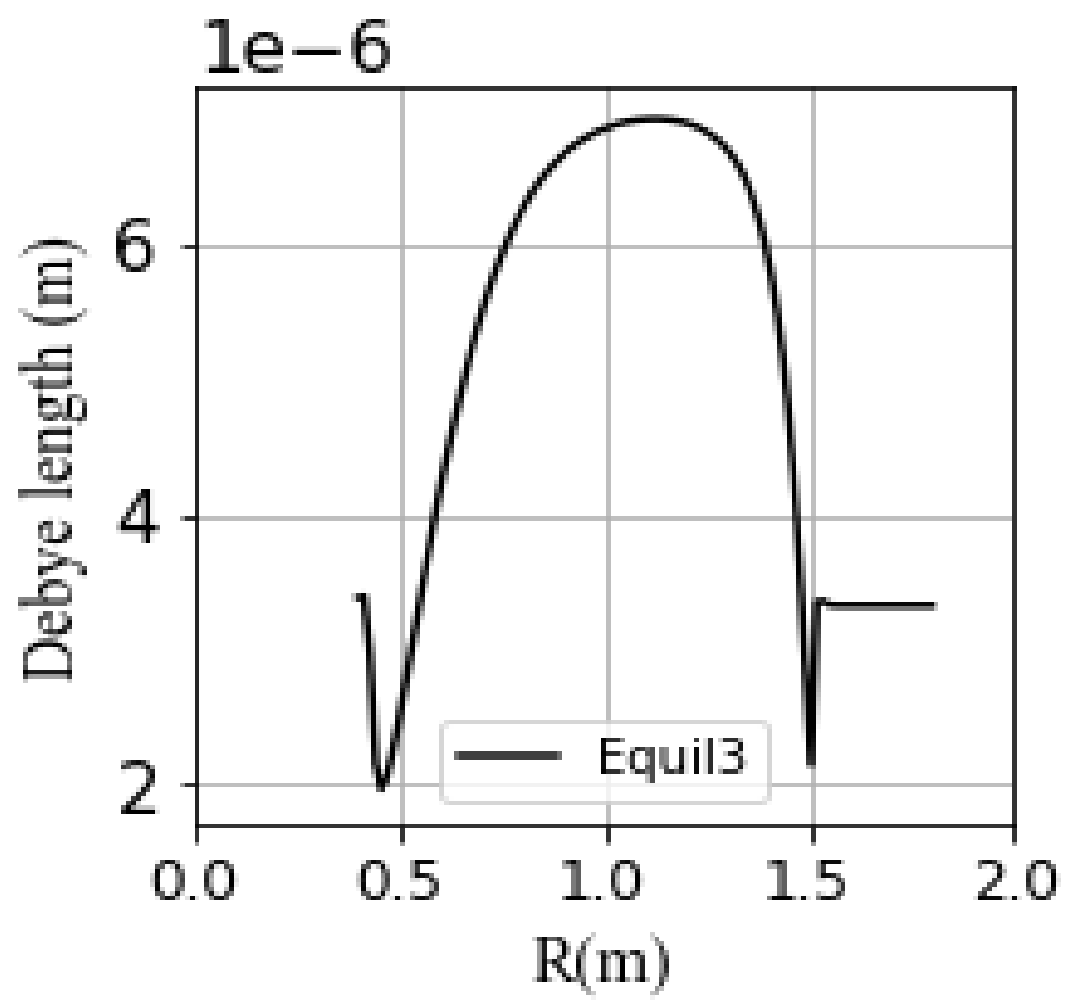

Debye length of Example 4 equilibrium

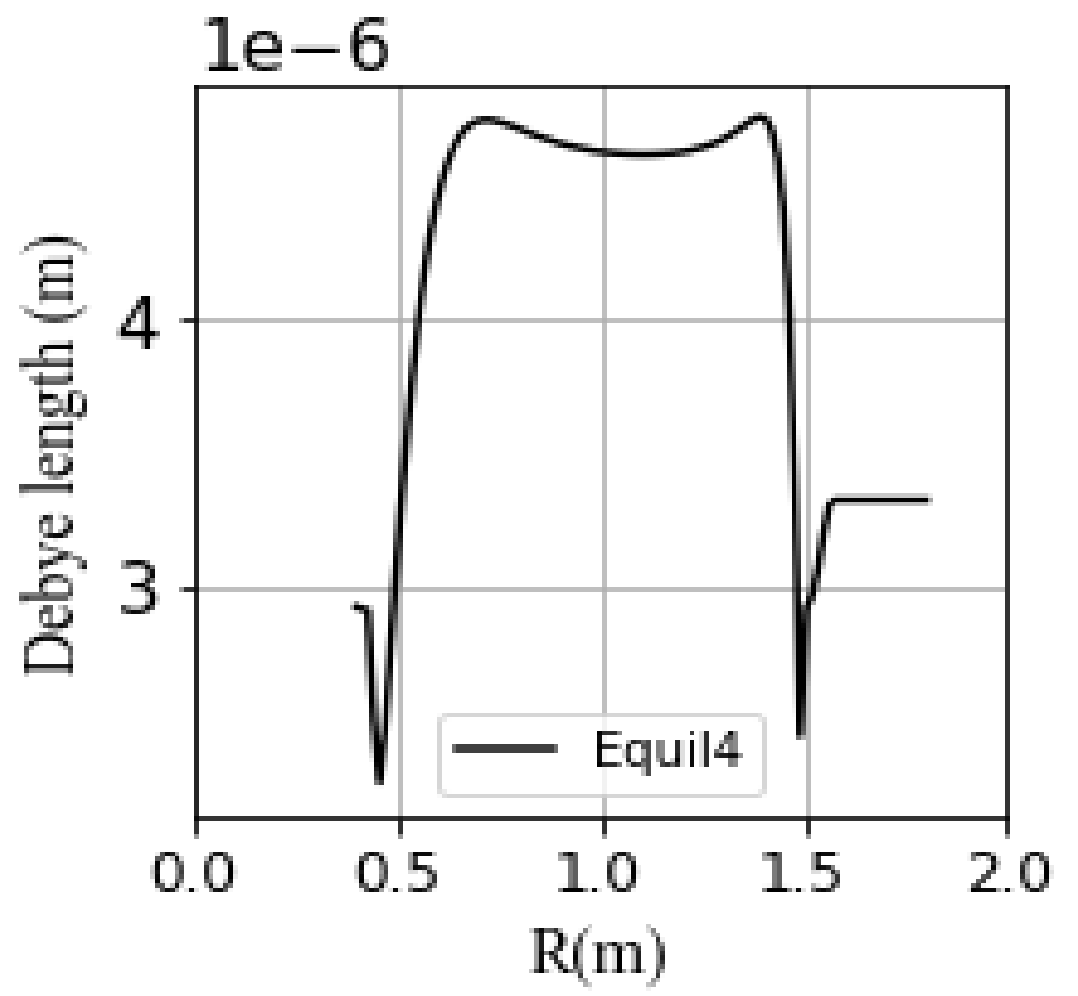

Figure 7. The Debye length of the four Example equilibria along the mid-plane.

## APPENDIX E: EXAMPLE EQUILIBRIA - COMPLETE INPUTS, OUTPUTS, AND DIAGNOSTICS

Four 6-fluid equilibria are presented here:

Example 1 equilibrium: High $\beta_T$ equilibrium near fusion burn condition,

Example 2 equilibrium: Low $\beta_T$ equilibrium near fusion burn condition,

Example 3 equilibrium: Low $\beta_T$ EHL-2 equilibrium, and

Example 4 equilibrium: High $\beta_T$ EHL-2 equilibrium.

### A. Example 1: High-$\beta_T$ burn test EHL-3 equilibrium

The computational boundary is rectangular with $0.56 \leq R(m) \leq 4.0$ and $-3 \leq Z(m) \leq 3$. In the Example 1 computation, we adopt $L_{ref} = 2.5m$, $I_{ref} = 8000kA$ and $n_{ref} = 4 \times 10^{19}\ m^{-3}$ as primary scales. Therefore, $B_{ref} = 4.0212T$, $\psi_{ref} = 25.1327Wb/rad$, $u_{ref} = 13868.37km/s$, $T_{ref} = 2007.365keV$ and $\varepsilon$ =0.014401.

The equilibrium result of Example 1 shown below is computed using the $200$ in $R \times 100$ in $Z$ mesh numbers. The poloidal field coil locations and currents are provided in Table III.

Table III. Poloidal field coil locations and currents used for Example equilibrium 1.

| Poloidal field coil location and current (Equilibrium 1) | | | | | |
|---|---|---|---|---|---|
| | R (m) | Z (m) | current(kA) | factor | net current (dimensionless) |
| PF1 | 1.5 | 3.5 | 700 | -0.98 | -0.085750 |
| PF2 | 1.5 | -3.5 | 700 | -0.98 | -0.085750 |
| PF3 | 2.75 | 3.5 | 700 | 0.98 | 0.085750 |
| PF4 | 2.75 | -3.5 | 700 | 0.98 | 0.085750 |
| PF5 | 4.5 | 2.6 | 2000 | 0 | 0.000000 |
| PF6 | 4.5 | -2.6 | 2000 | 0 | 0.000000 |
| PF7 | 4.5 | 1.0 | 2500 | 2.156 | 0.673750 |
| PF8 | 4.5 | -1.0 | 2500 | 2.156 | 0.673750 |
| PF9 | 0.55 | 3.0 | 700 | -8.82 | -0.771750 |
| PF10 | 0.55 | -3.0 | 700 | -8.82 | -0.771750 |

Ampere's equation, the electrostatic potential $V_E$, the poloidal magnetic flux $\psi$, and the radial force balance of each fluid of this equilibrium are solved within the relative errors shown below:

Ampere's equation is solved within an error of $\frac{average|GSop+RJt|}{average|RJt|}$ = 0.000614 (Fig. 8).

The potential $V_E$ of each species fluid is derived from Eq. (41) and solved within an error of $\frac{average|VE_{eh}-VE_{el}|}{average|VE_{el}|} = 0.029$ (Fig. 9).

The quantity $(\psi - \psi_{plasma})$ is solved within an error of $\frac{average|\psi-\psi_{plasma}-\psi_{pfcoil}|}{average|\psi_{pfcoil}|} = 0.0032$ (Fig. 10).

Radial force balance of the pl-fluid is solved within an error of $\frac{average|balance|}{average|FplP|} = 0.00189$ (Fig. 11).

Radial force balance of the bl-fluid is solved within an error of $\frac{average|balance|}{average|FblP|} = 0.00213$ (Fig. 12).

Radial force balance of the el-fluid is solved within an error of $\frac{average|balance|}{average|FelP|} = 0.00223$ (Fig. 13).

Radial force balance of the ph-fluid is solved within an error of $\frac{average|balance|}{average|FphP|} = 0.00157$ (Fig. 14).

Radial force balance of the bh-fluid is solved within an error of $\frac{average|balance|}{average|FbhP|} = 0.00164$ (Fig. 15).

Radial force balance of the eh-fluid is solved within an error of $\frac{average|balance|}{average|FehP|} = 0.00163$ (Fig. 16).

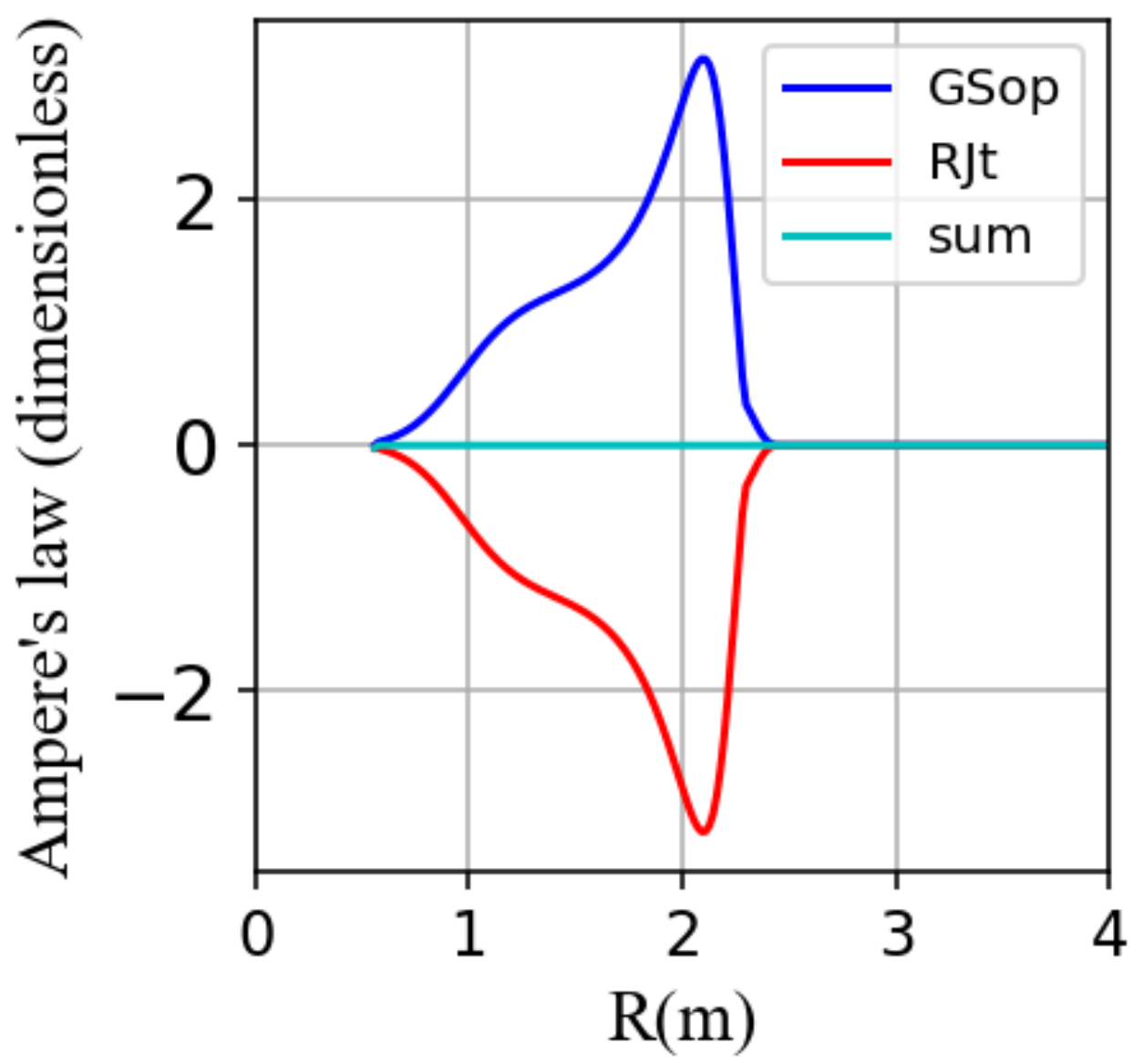

Figure 8. Ampere's equation of Eq. (38) on the mid-plane. GSop (blue) shows the L.H.S. of Eq. (38), and RJt (red) shows the negative R.H.S. of Eq. (38). The sum (green) shows the balance.

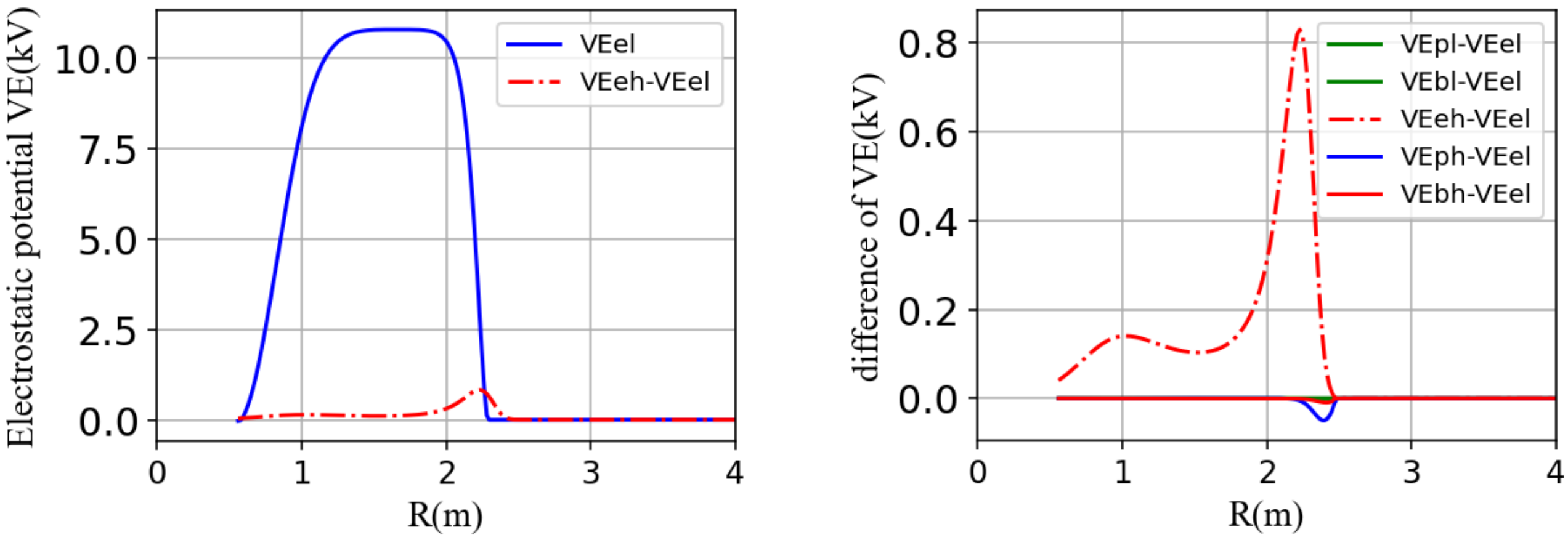

Figure 9. Electrostatic potential $V_E$ calculated separately from the charge-neutrality condition of each equilibrium fluid component.

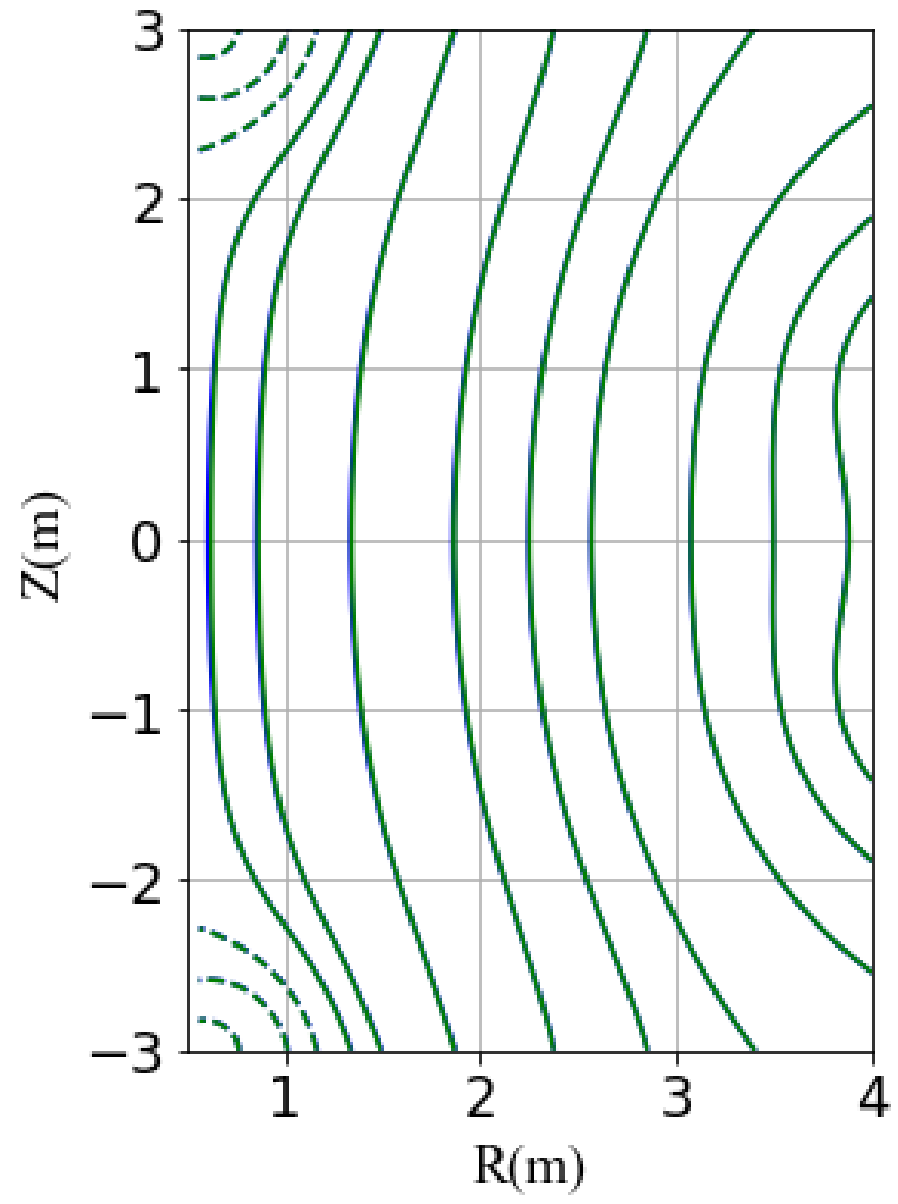

Figure 10. Contours of $\psi_{pfcoil}$ (blue) and contours of $(\psi - \psi_{plasma})$ (green) overlap to within an error of 0.32%.

The radial force balance of each species fluid on the mid-plane is given by:

$$-\frac{\partial p}{\partial R} - \frac{q}{e}\gamma n \frac{\partial V_E}{\partial R} + j_\phi B_z - j_z B_\phi + \frac{m}{m_p}\gamma^2 h n \frac{u_\phi^2}{R} = 0.$$

Here the 1st, 2nd, 3rd, 4th, and 5th terms are plotted as FαP, FαE, $\text{FαLor}_{tz}$, $\text{FαLor}_{zt}$, and Fαcent, which are the sums of the forces on the mid-plane from the pl, ph, bl, bh, el, and eh fluids.

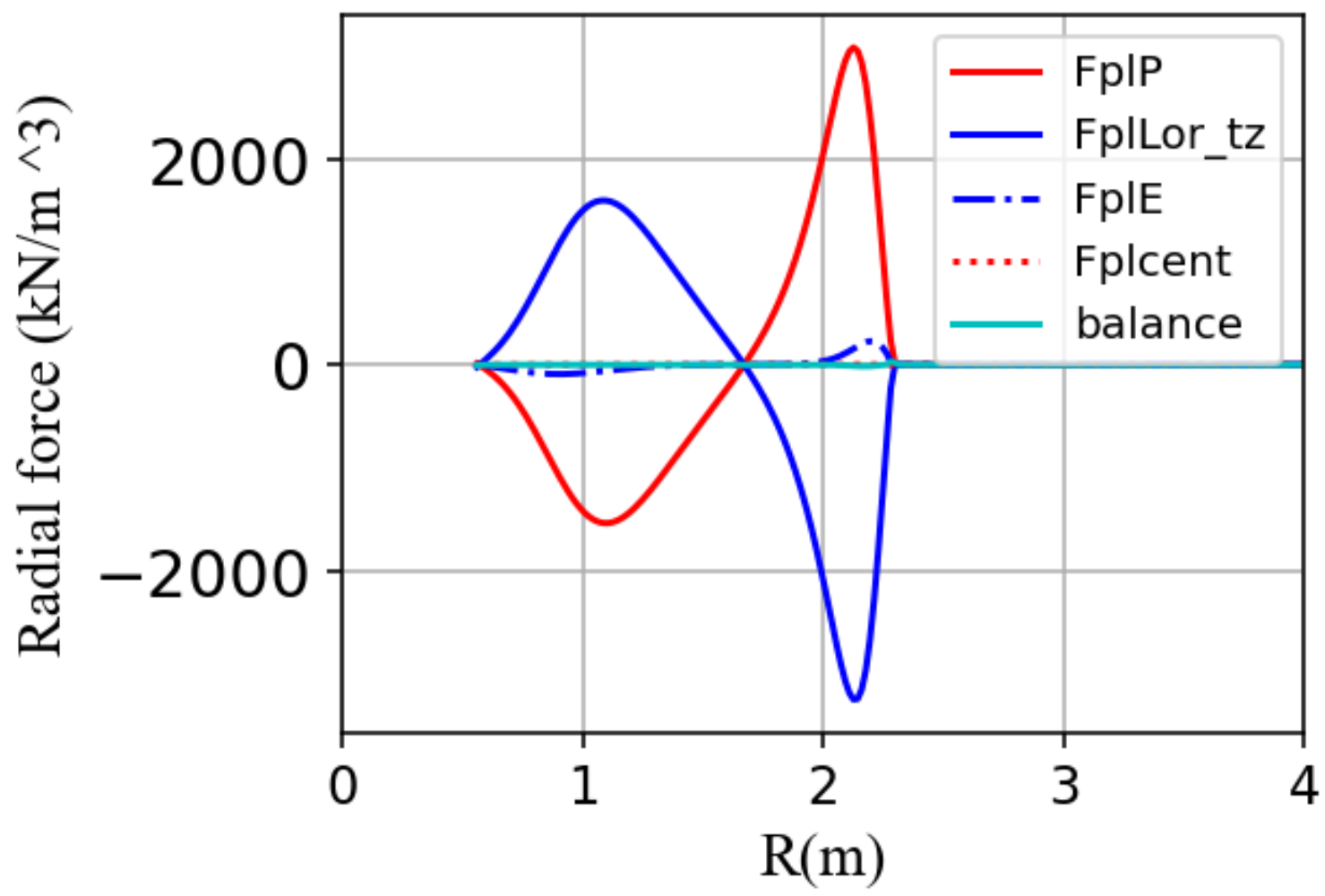

Figure 11. Radial force balance of the pl-fluid.

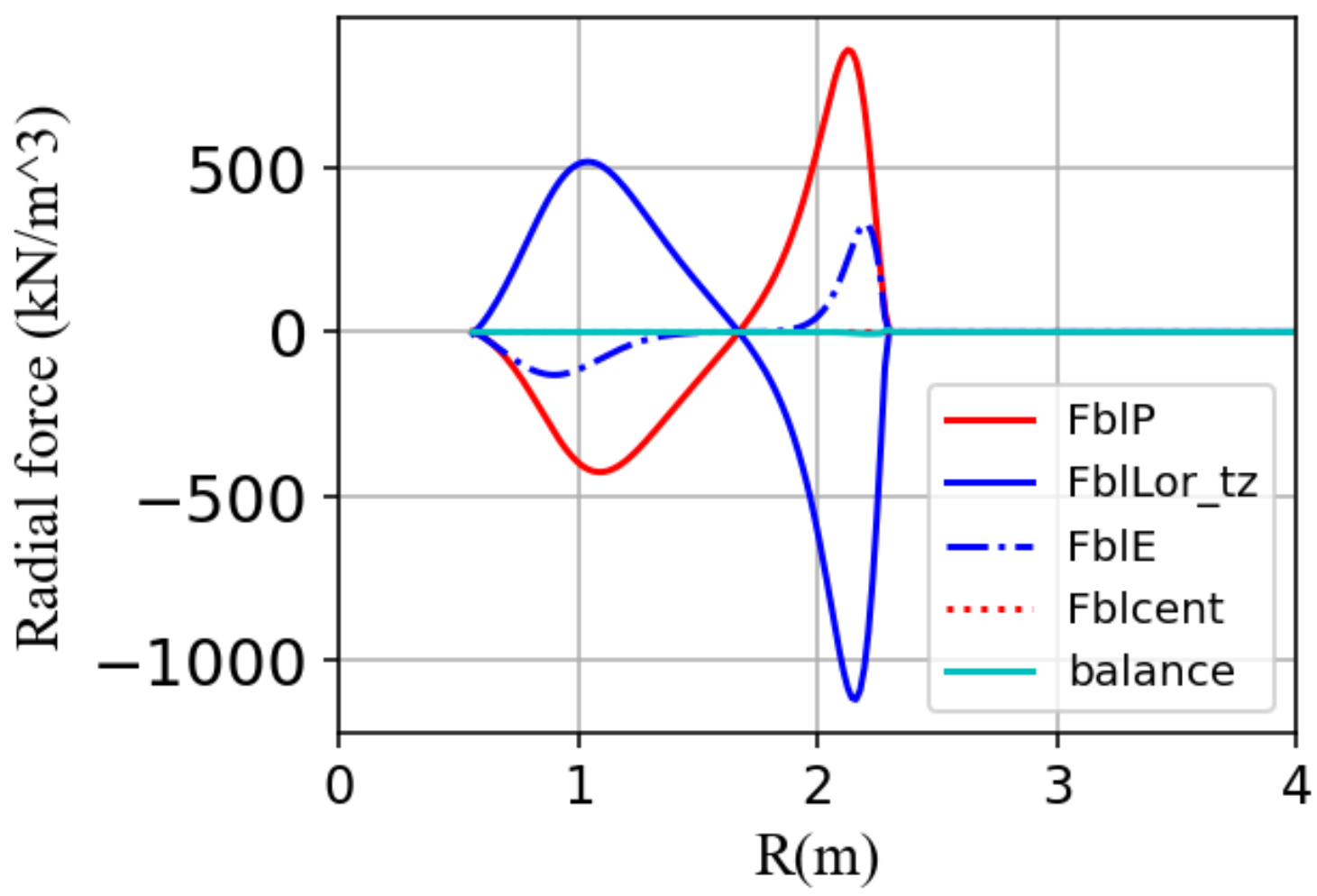

Figure 12. Radial force balance of the bl-fluid;

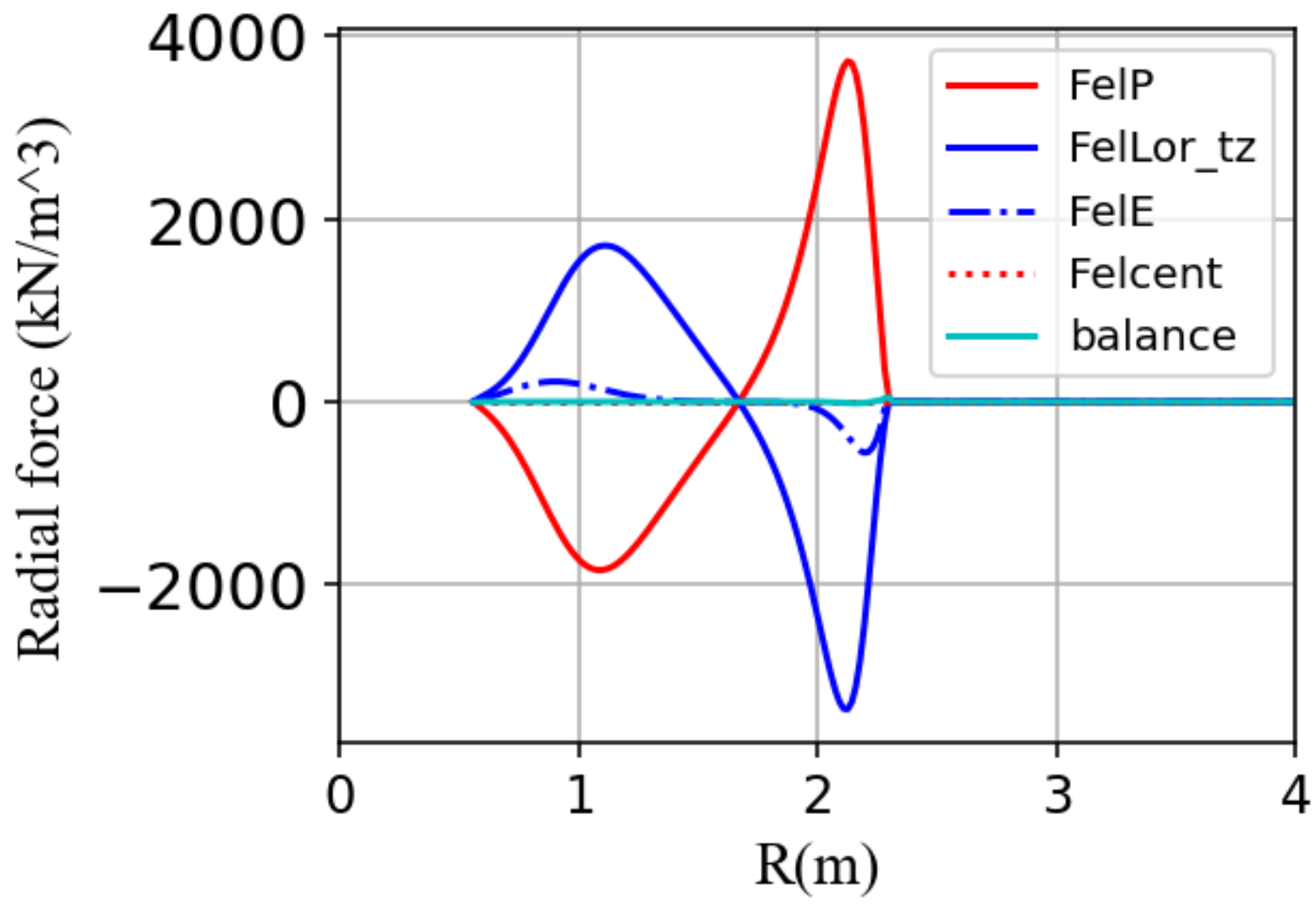

Figure 13. Radial force balance of the el-fluid.

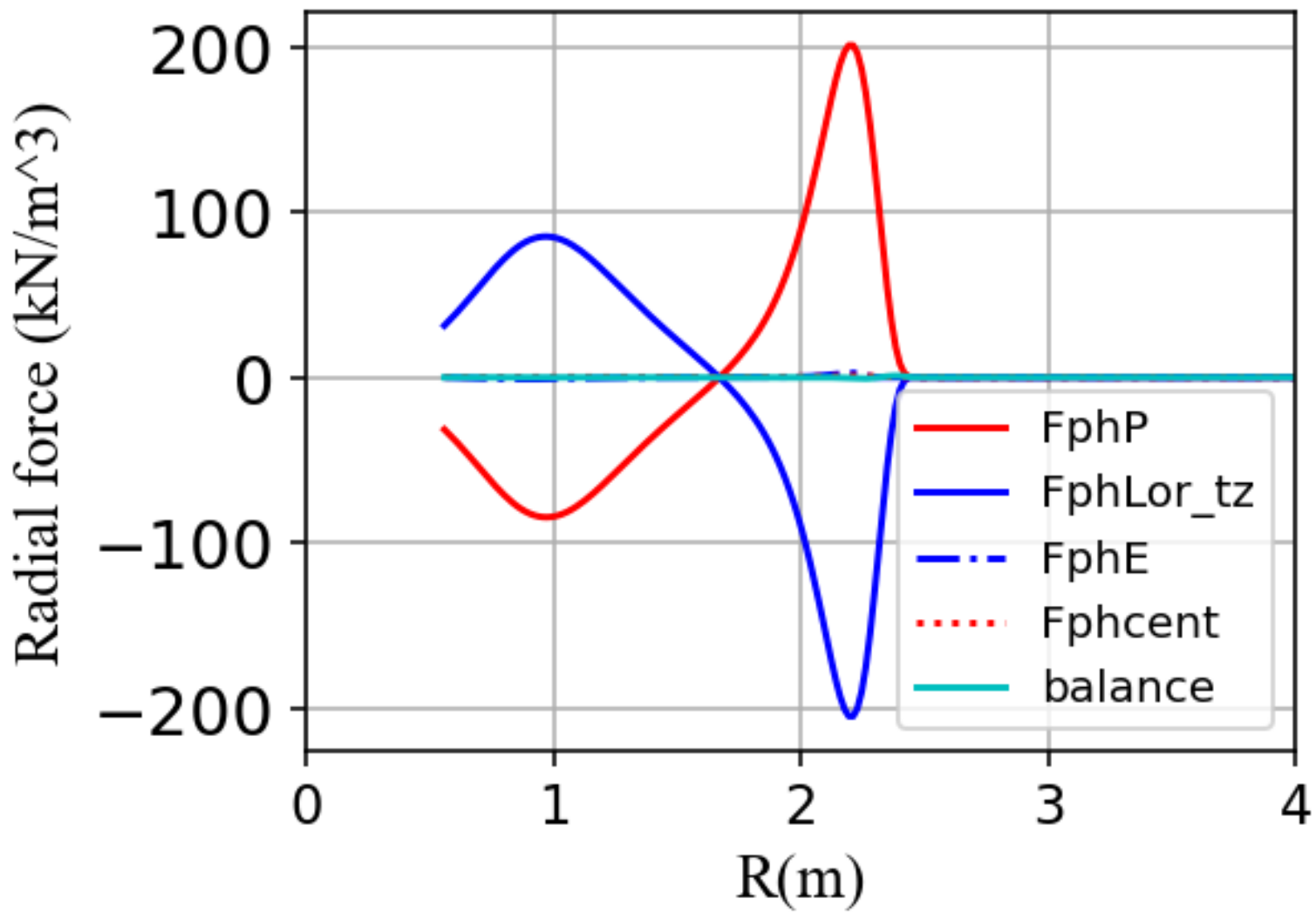

Figure 14. Radial force balance of the ph-fluid.

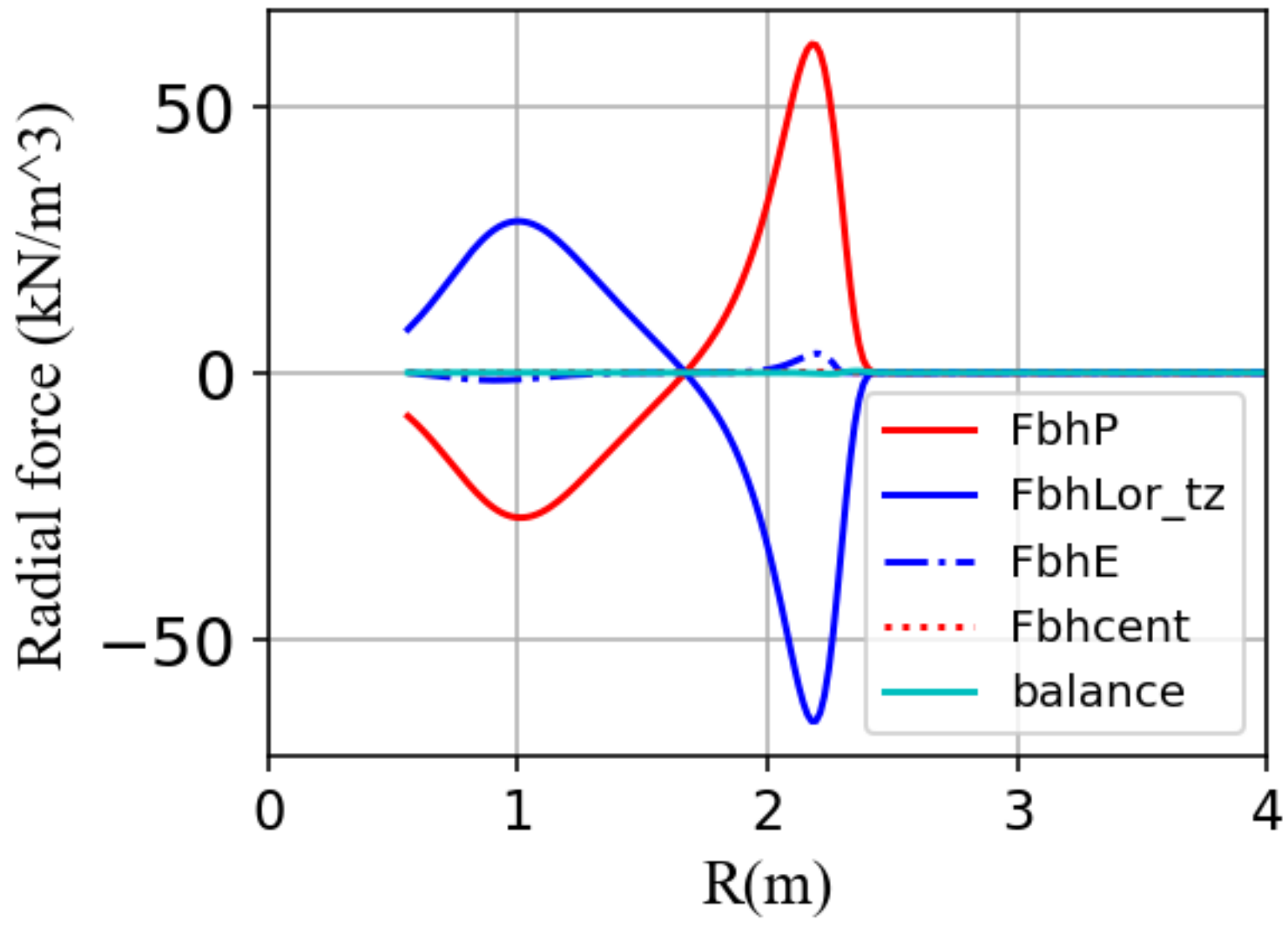

Figure 15. Radial force balance of the bh-fluid.

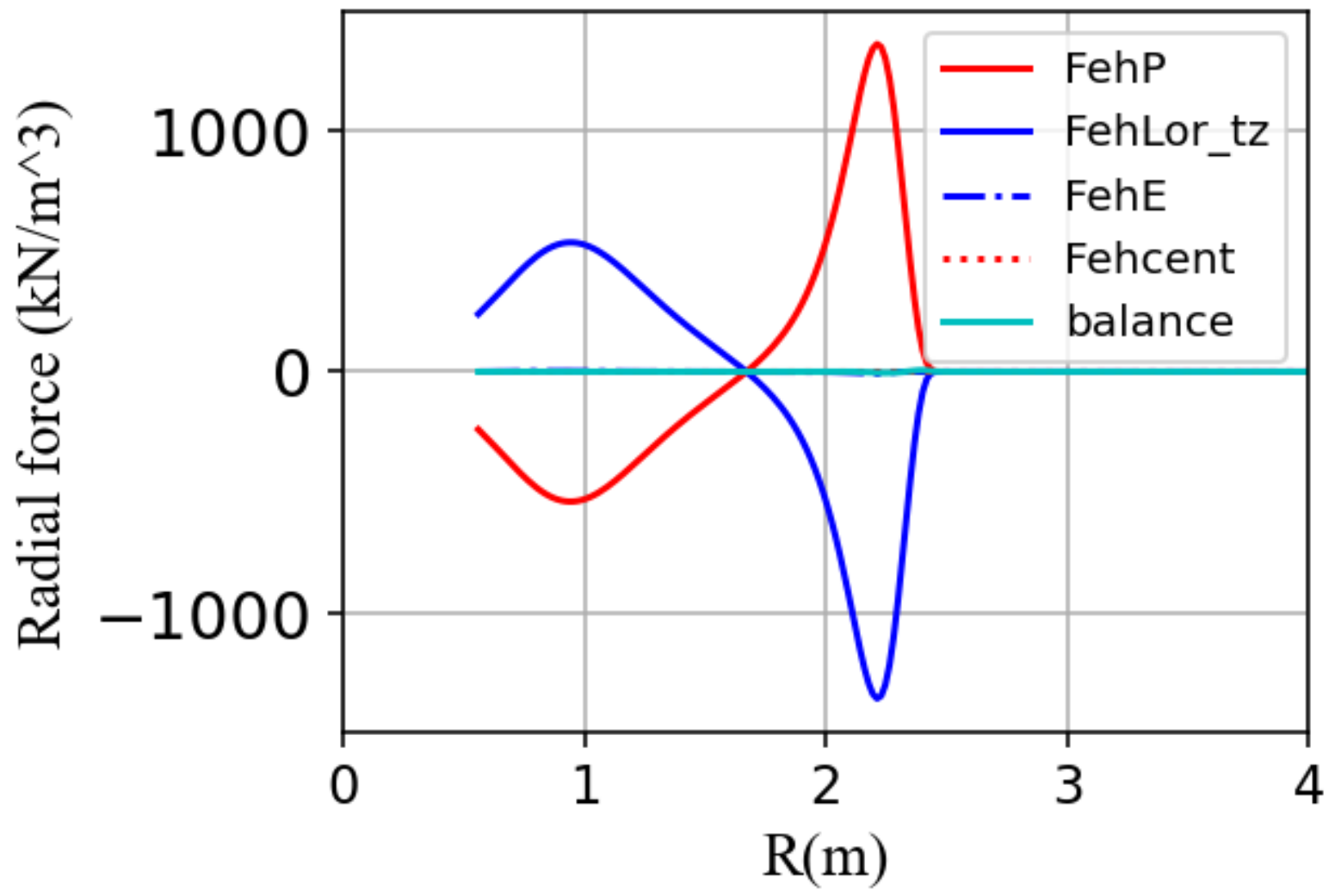

Figure 16. Radial force balance of the eh-fluid.

Figs. 17 – 22 show the poloidal magnetic flux, the toroidal plasma current density (substantially shifted outward), the centrally peaked total plasma pressure, the toroidal current density of the suprathermal electrons (substantially shifted outward), the centrally peaked electrostatic potential, and the magnetic field strength (with a strong outboard magnetic well) of a high-$\beta_T$ burn test EHL-3 equilibrium. The solid red line marks the flux-surface boundary of the suprathermal components. This boundary is determined by the parameter Psi_crit2. The LCFS is determined by a mid-plane limiter or by the x-point of the divertor configuration. The $\beta_T$ and $\beta_p$ are computed within the LCFS.

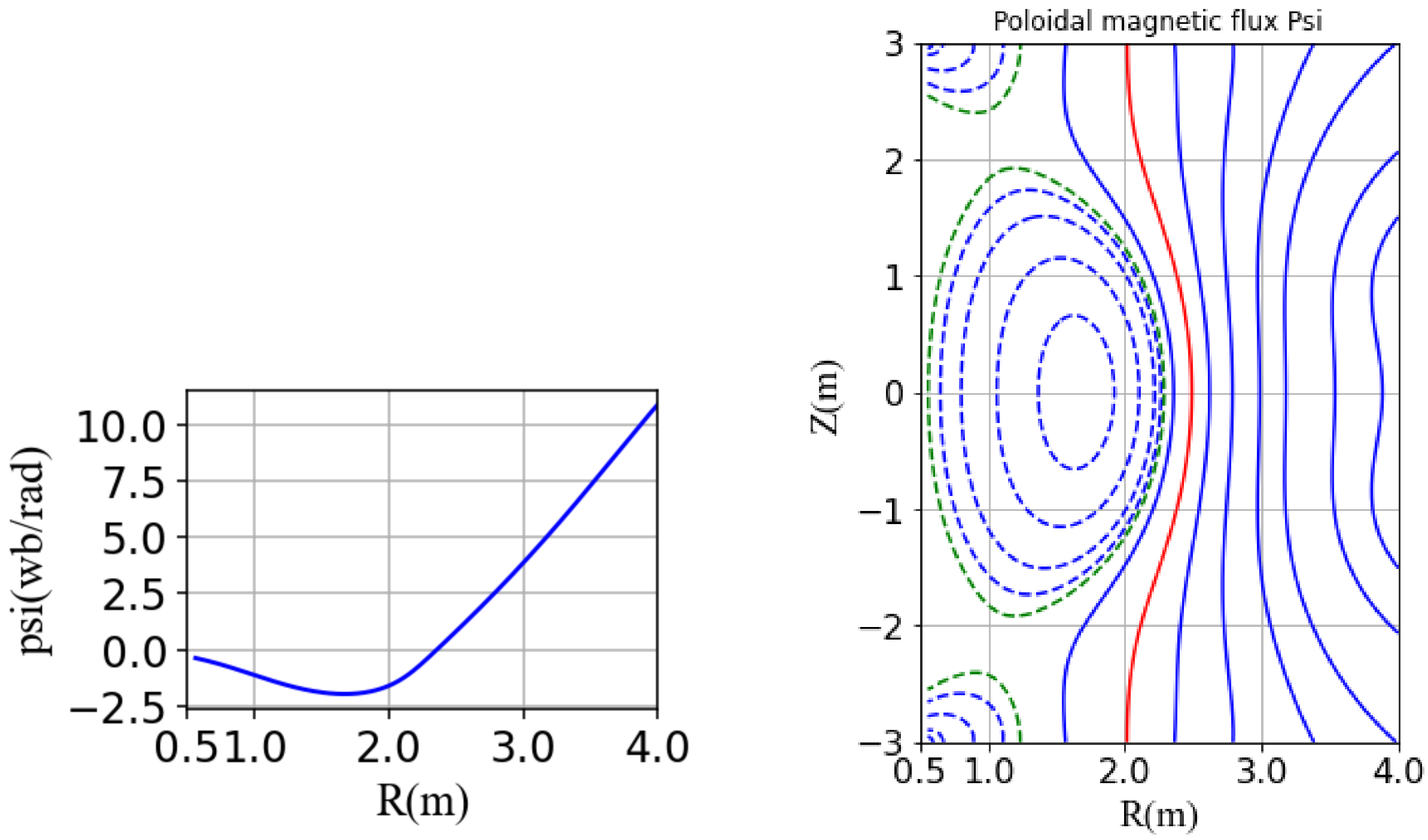

Figure 17. Poloidal magnetic flux on the mid-plane and over the R-Z plane.

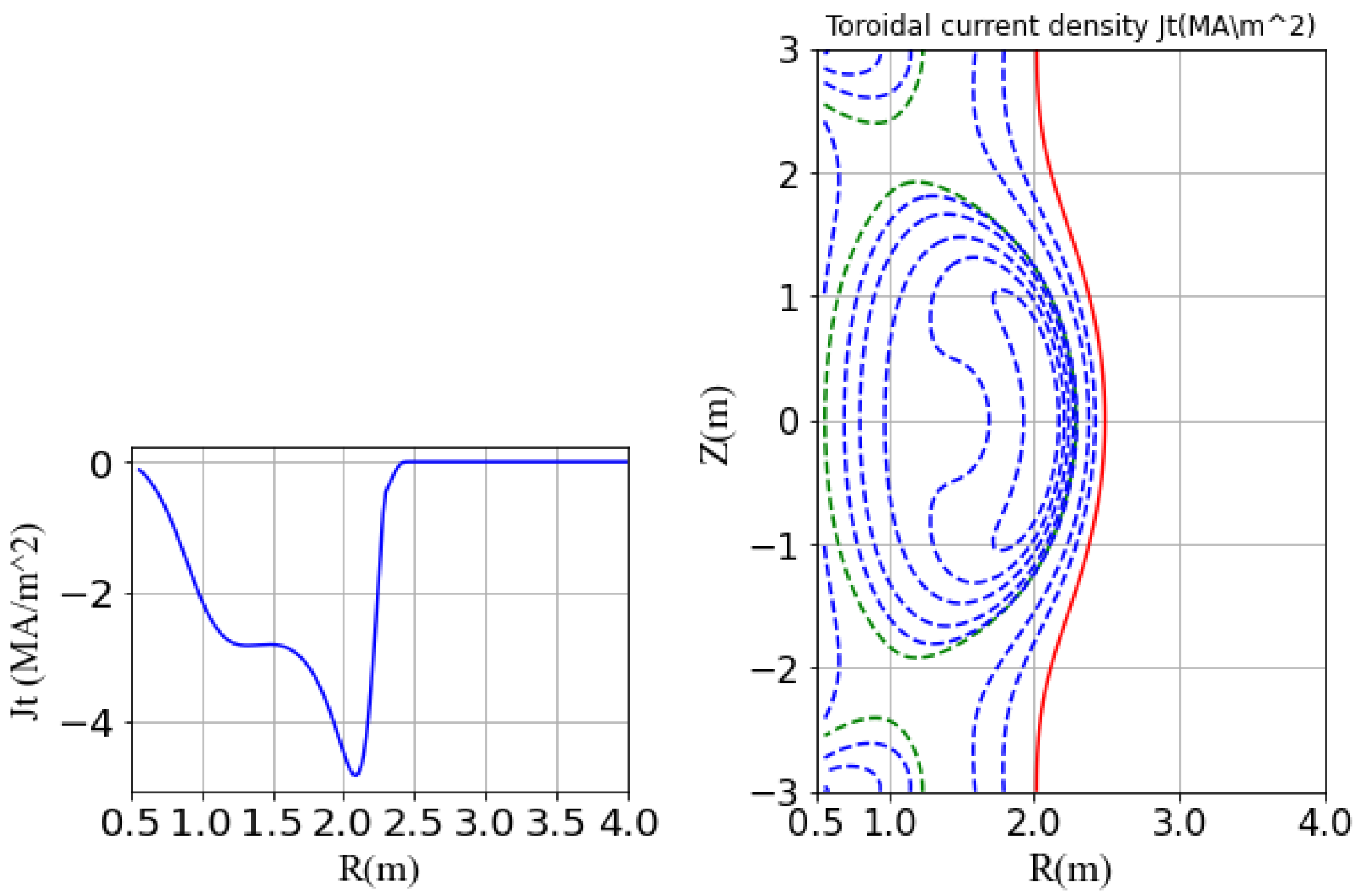

Figure 18. Total toroidal plasma current density on the mid-plane and over the R-Z plane. Note that the current shows a pronounced outward shift, spills beyond the LCFS (green dashes), and vanishes at the wall material boundary (red).

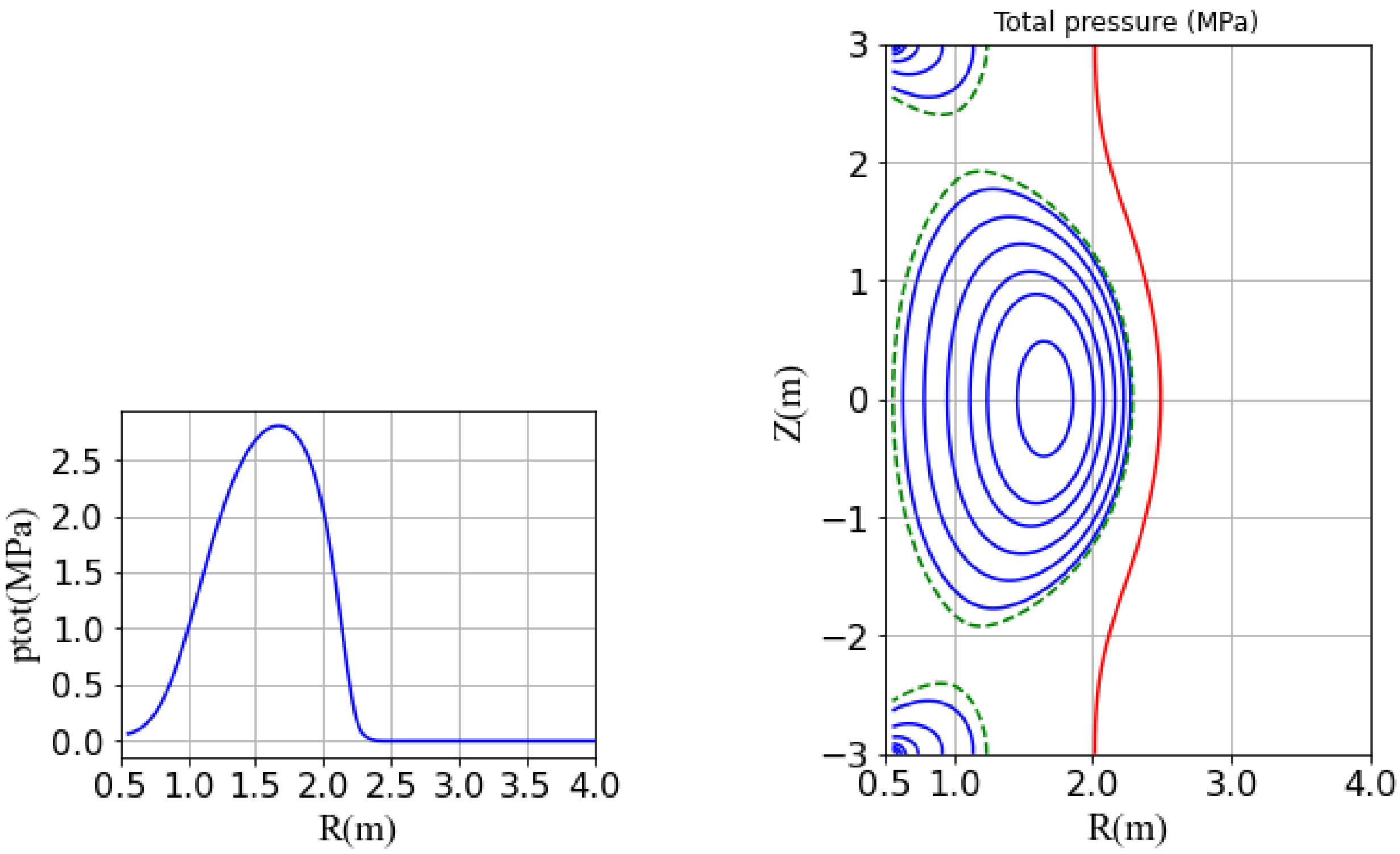

Figure 19. Centrally peaked total plasma pressure on the mid-plane and over the R-Z plane.

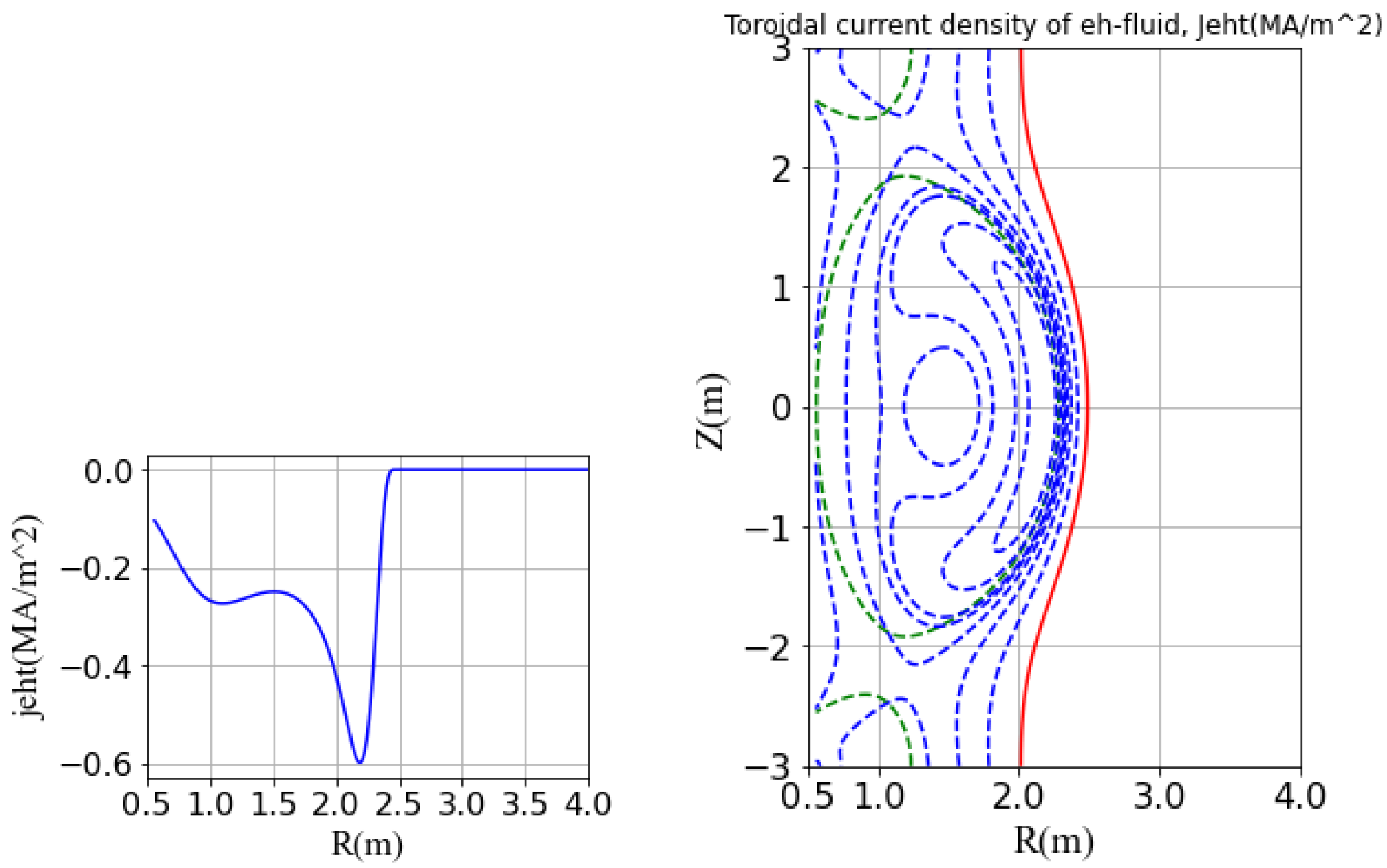

Figure 20. Toroidal current density of the suprathermal electron fluid (eh) on the mid-plane and over the R-Z plane, showing a pronounced radial and vertical outward shift.

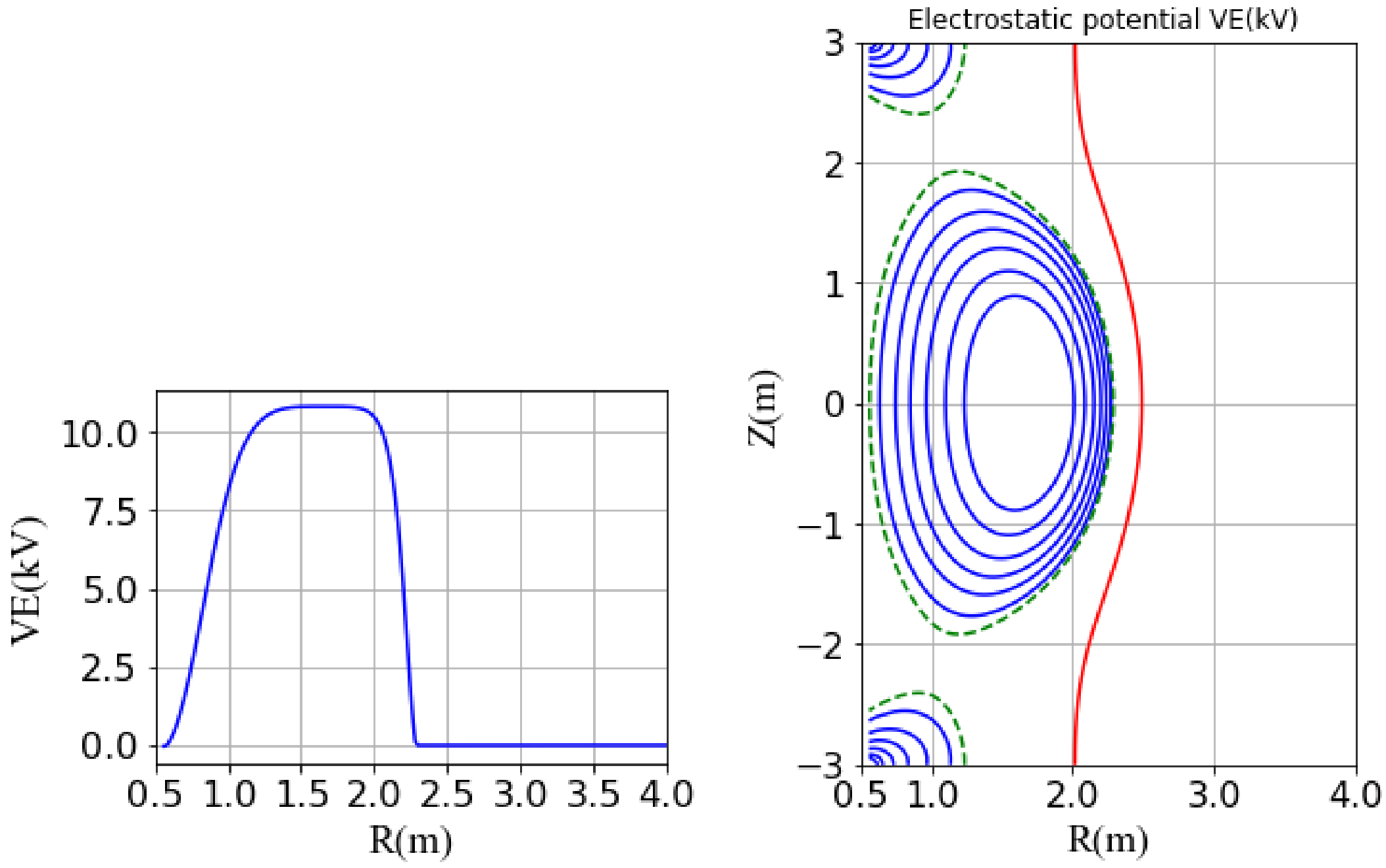

Figure 21. Electrostatic potential $V_E$ on the mid-plane and over the R-Z plane. Note that $V_E$ is constant on the poloidal magnetic flux surface.

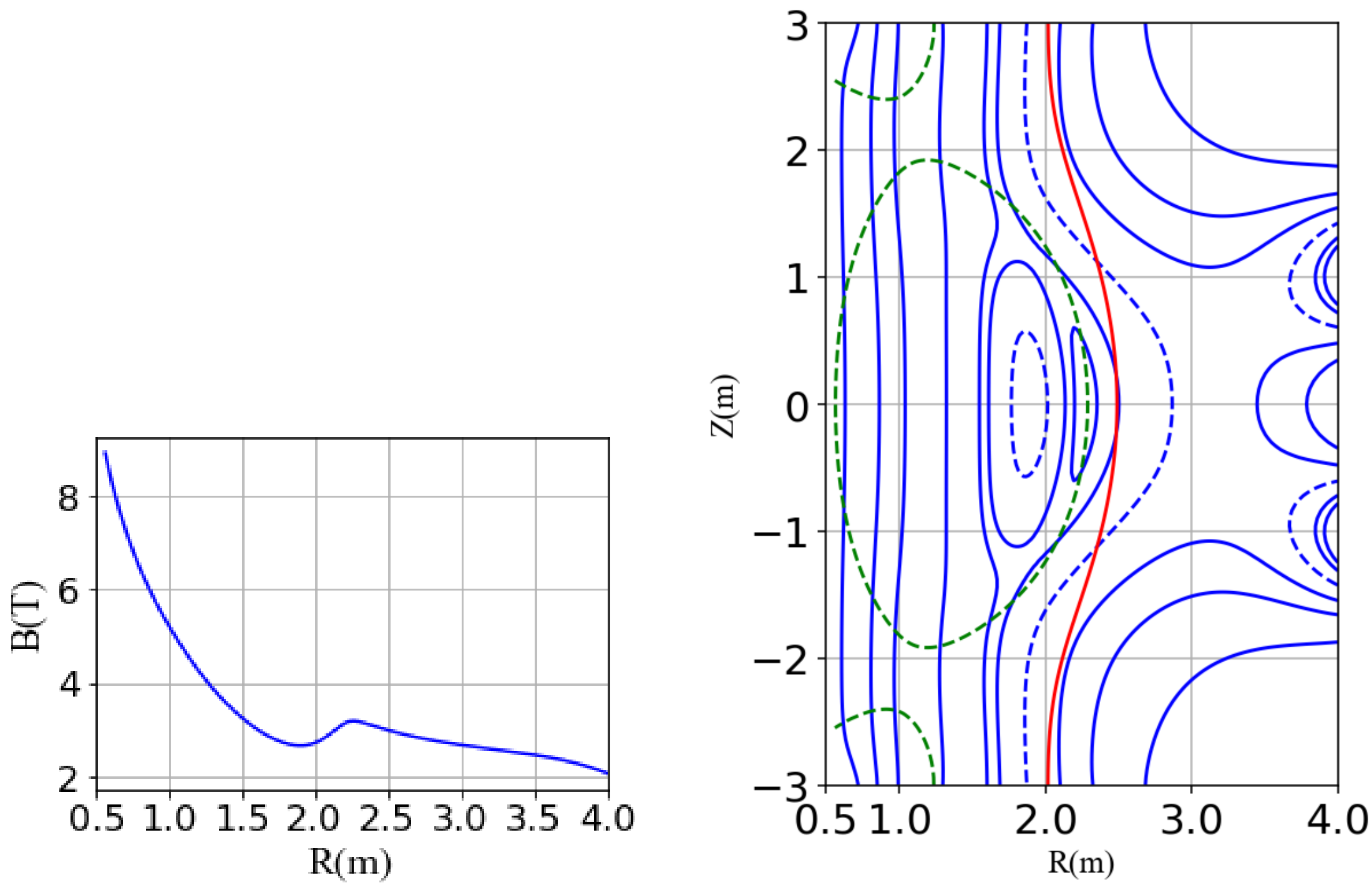

Figure 22. Magnetic field strength on the mid-plane and over the R-Z plane. Note that an absolute magnetic well spans nearly the entire outboard region of the plasma.

Figs. 23 and 24 show the mid-plane density and temperature profiles of the six fluid components.

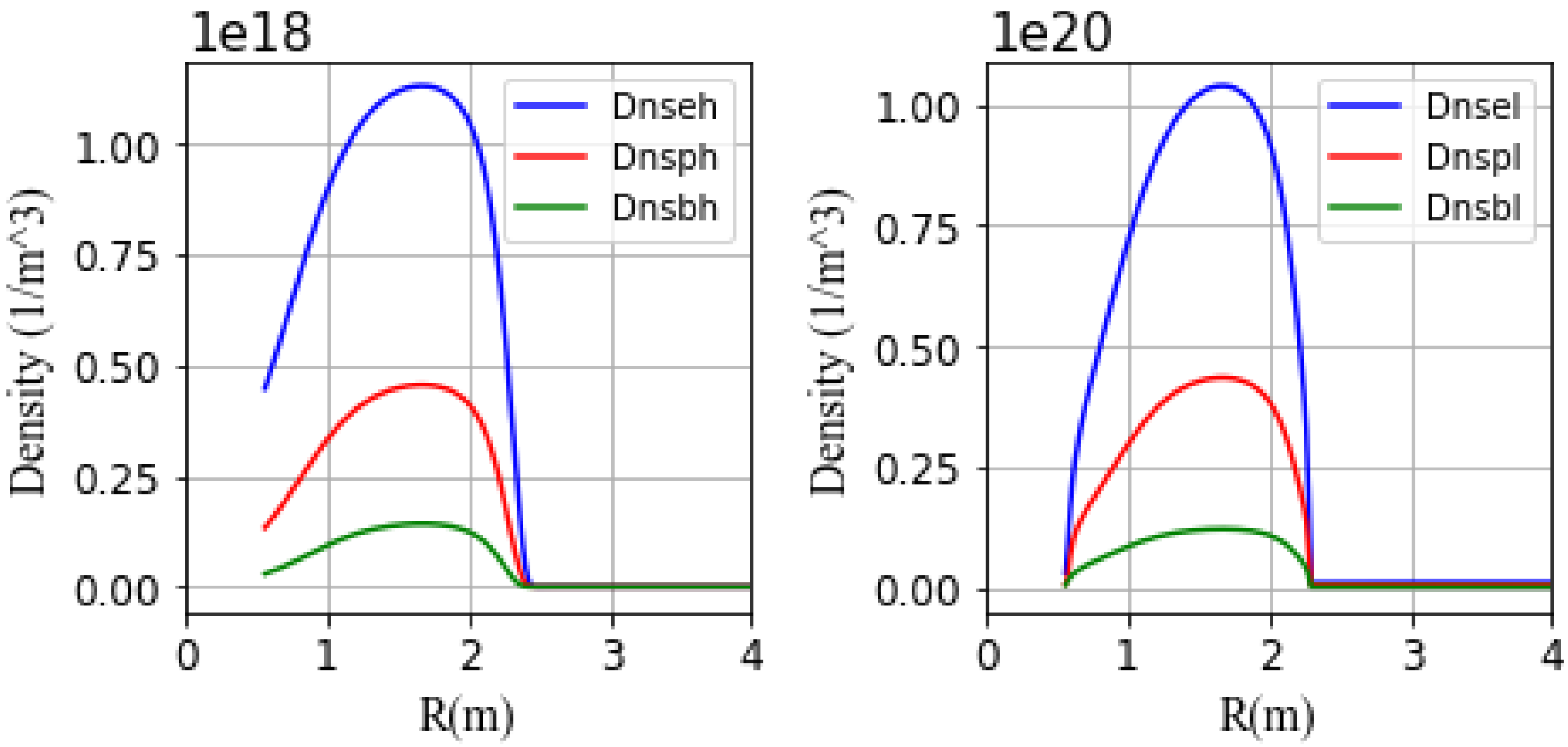

Figure 23. Mid-plane **d**ensity profiles of the six fluid components.

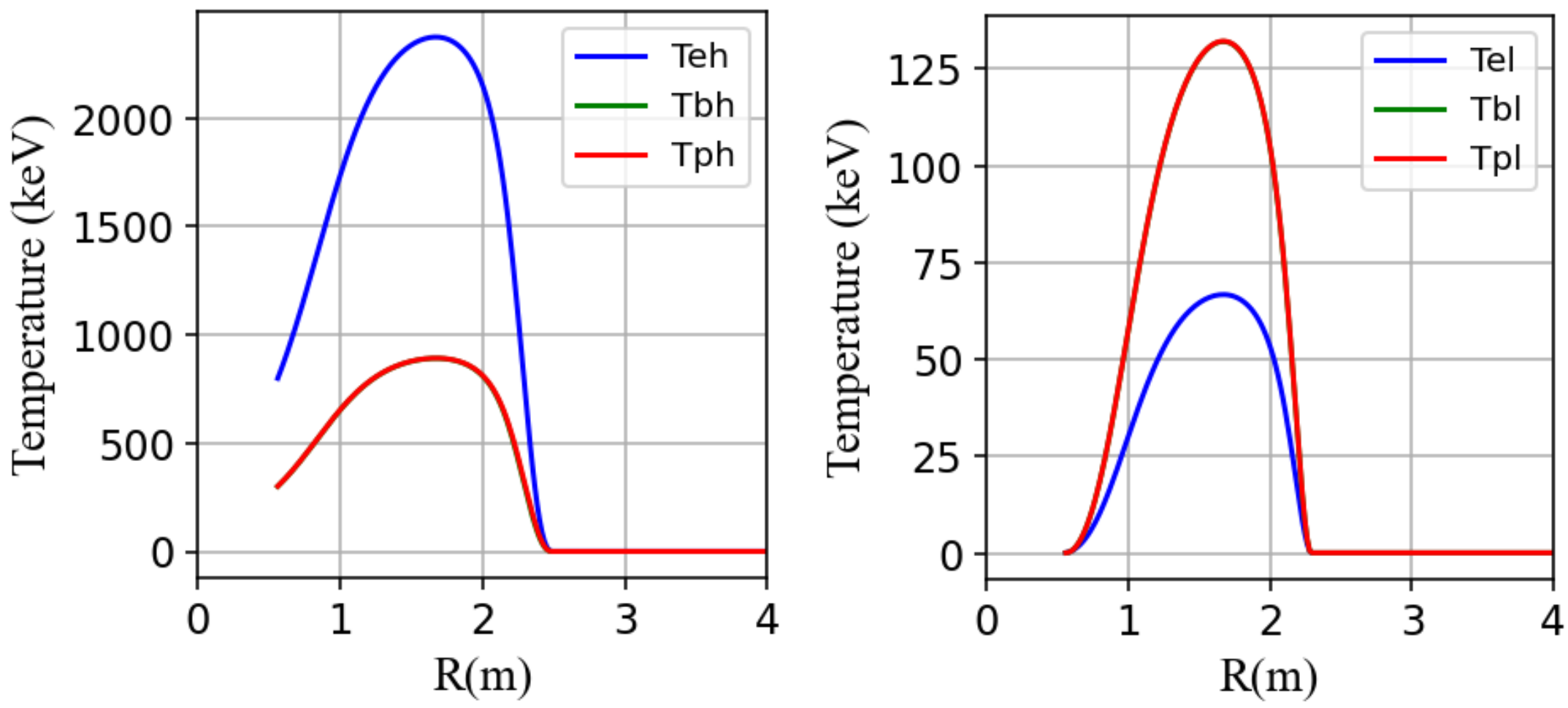

Figure 24. Mid-plane temperature profiles of the six fluid components with $T_{pl} = T_{bl}$.

Fig. 25 shows the h functions, which represent the relativistic effect of thermal motion, for the suprathermal and thermal electrons.

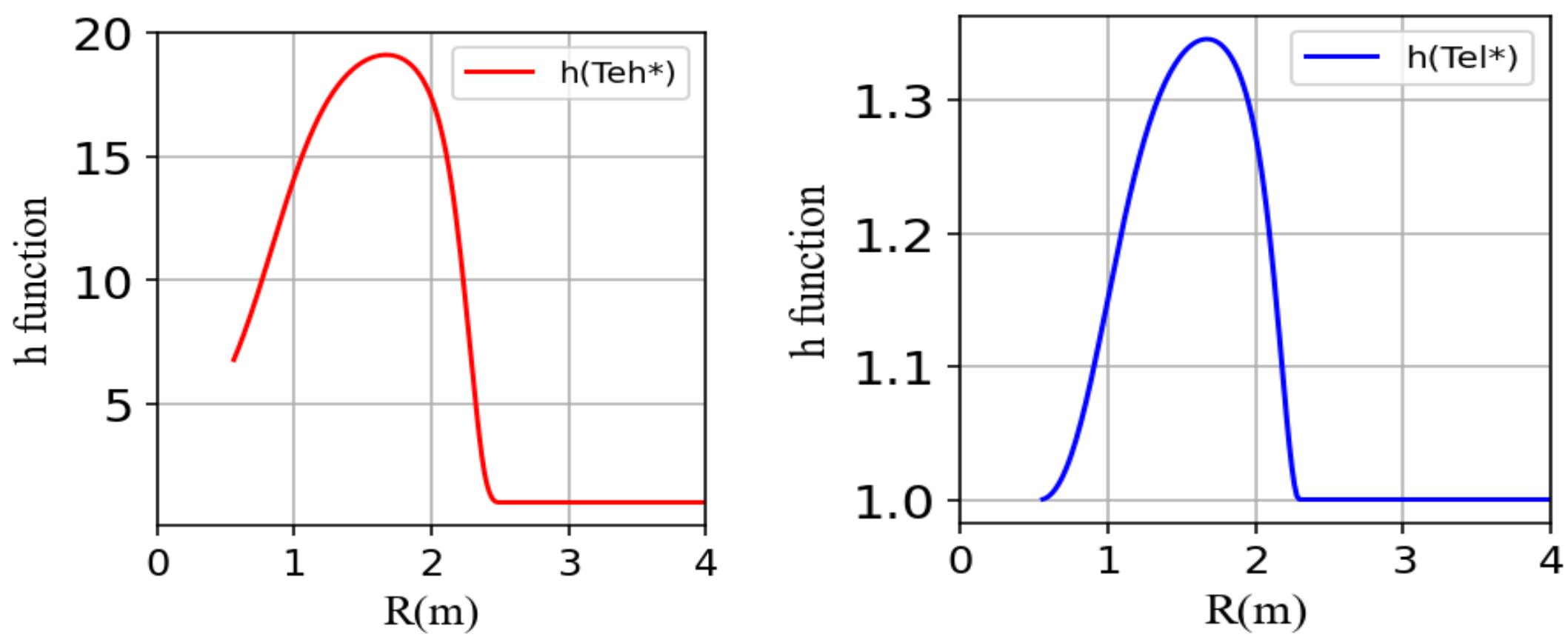

Figure 25. The $h$ functions for the suprathermal electrons with temperature $T_{eh}$ and the thermal electron fluid with temperature $T_{el}$.

Fig. 26 shows the mid-plane cyclotron frequencies of the thermal electrons, $eB/m_e h(T_{el}^*)$, and the suprathermal electrons, $eB/m_e h(T_{eh}^*)$, which is an order of magnitude lower due to the much larger electron mass.

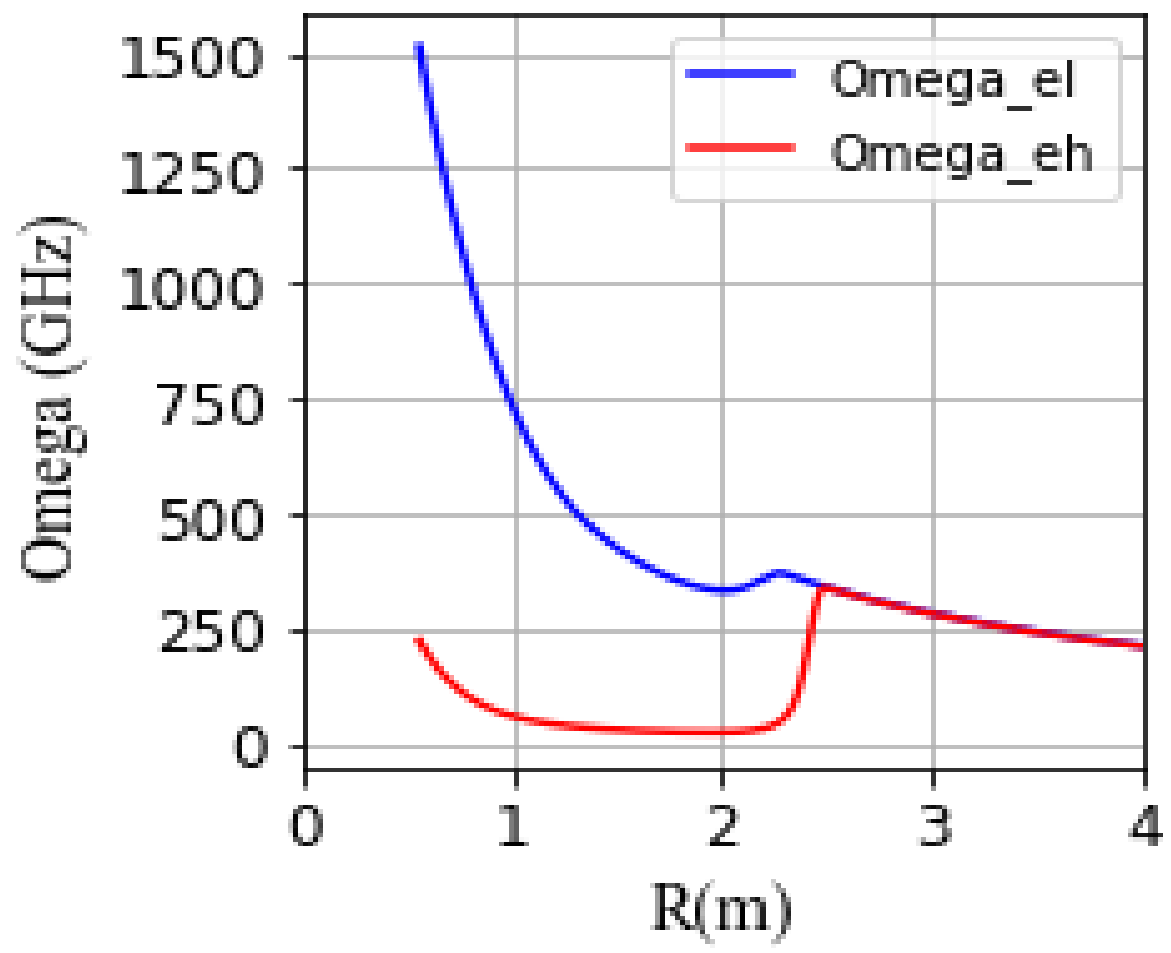

Figure 26. Cyclotron frequency of the el-fluid, $eB/m_e h(T_{el}^*)$, and the eh-fluid, $eB/m_e h(T_{eh}^*)$.

Fig. 27 shows the toroidal flow velocities of the six fluid components. Note that the peak relative velocity between the suprathermal protons and the thermal boron ions can be greater than 2700 km/s, which can exceed the slow and effective sound velocities shown in Fig. 28.

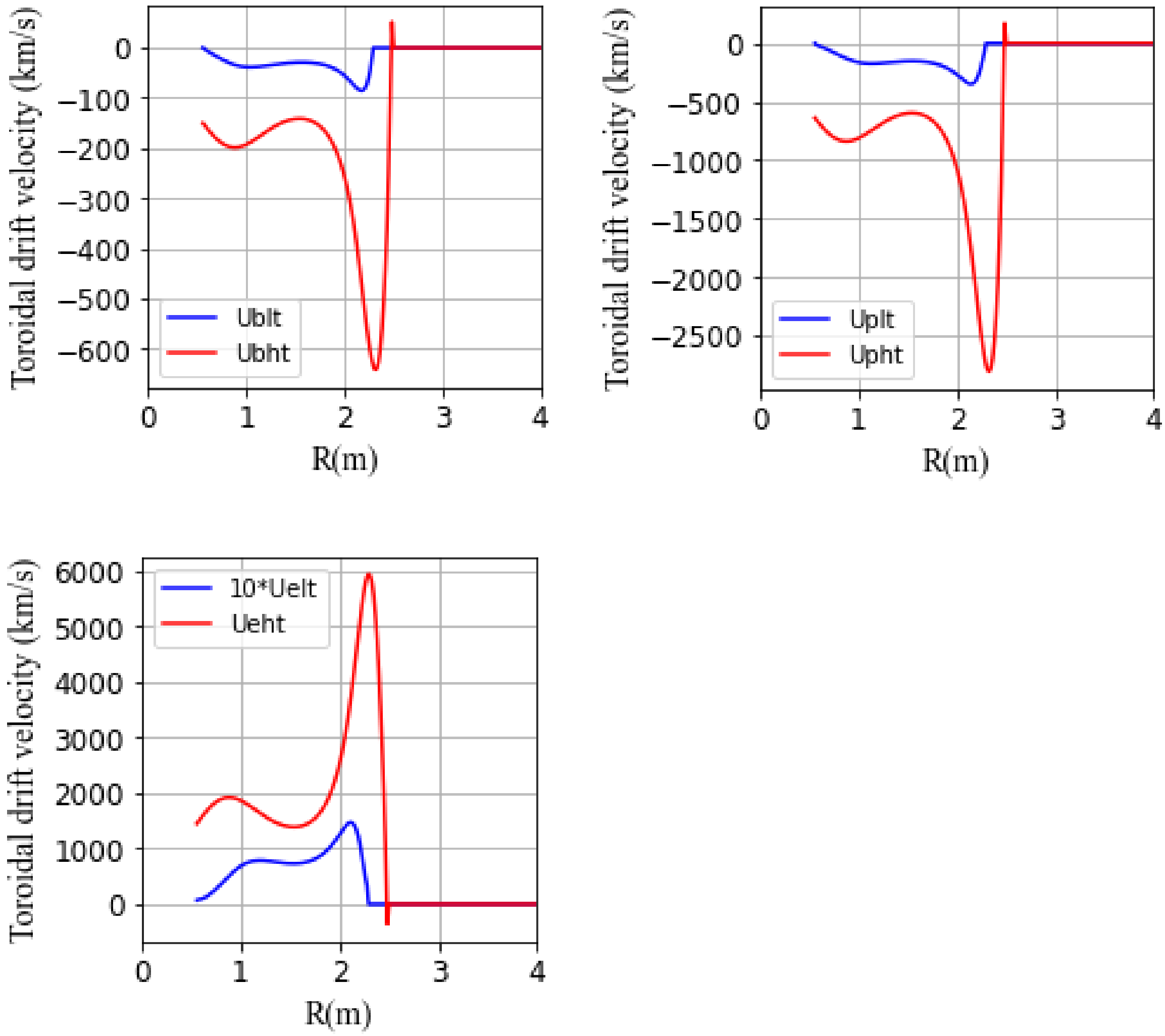

Figure 27. Toroidal flow velocities of the six fluid components.

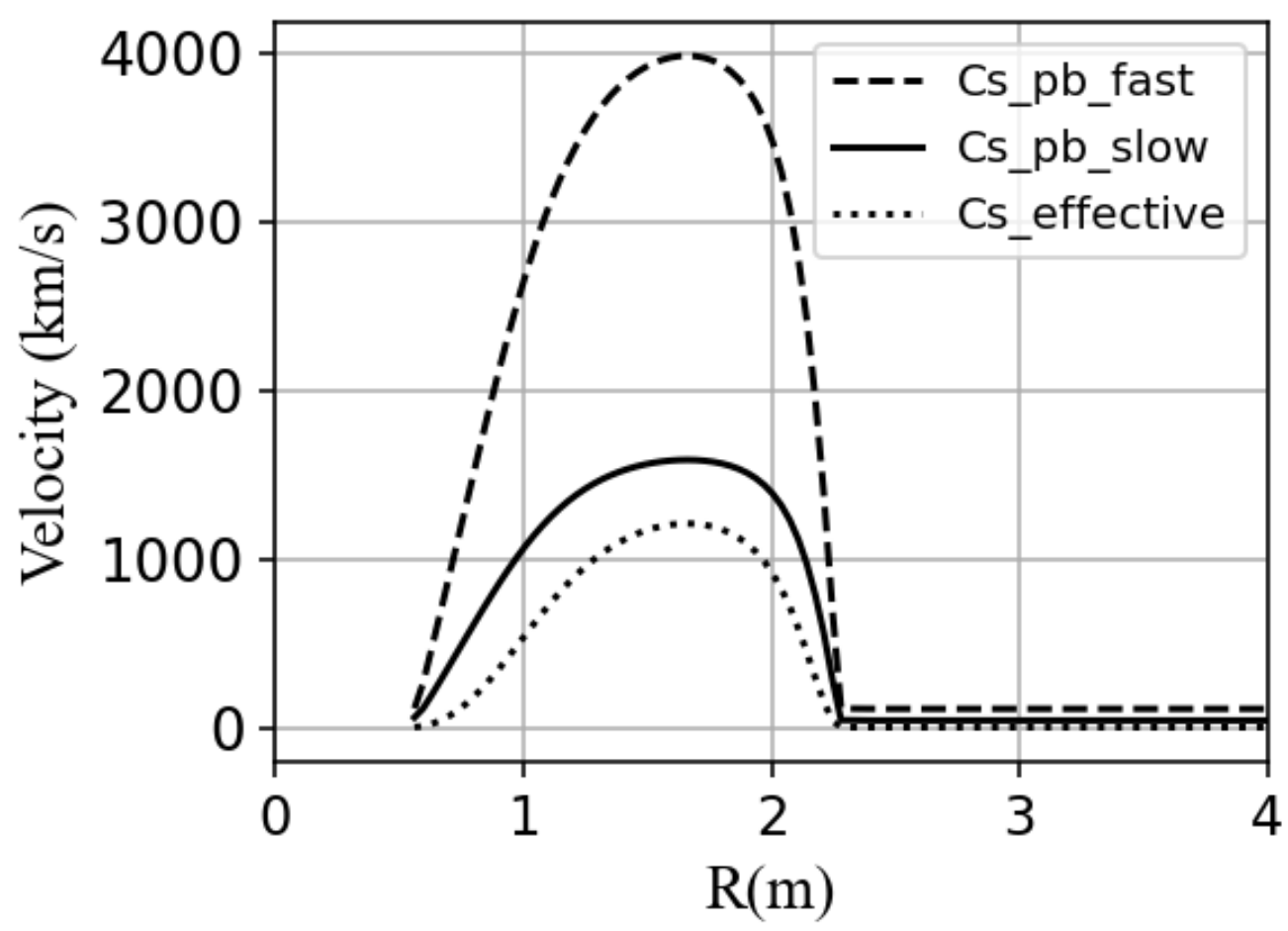

Figure 28. The mid-plane values of the fast, slow, and effective sound velocities.

Fig. 29 compares the initial current density and poloidal magnetic flux model profiles with their final profiles for Example 1 Equilibrium, which took 80 s to compute on a personal computer with an Intel Core Ultra 9 285 (2.5 GHz, 24-core).

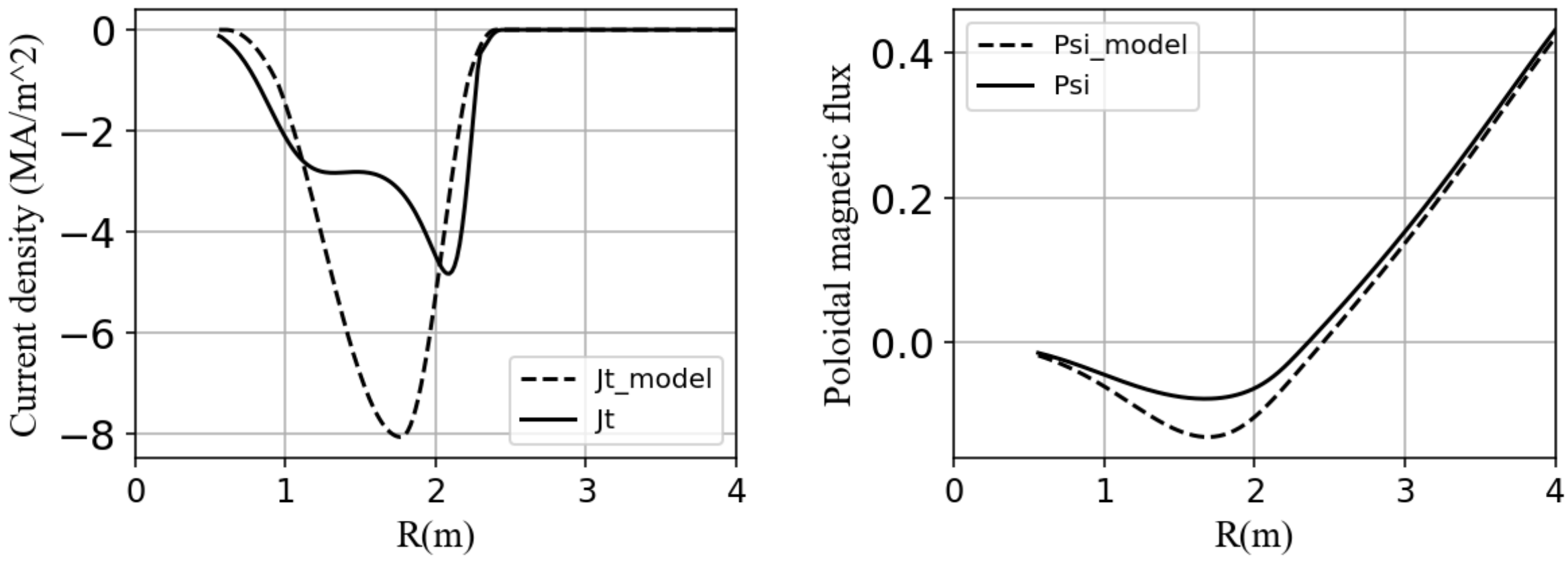

Figure 29. The initial (dashed) and final (solid) mid-plane current density profiles of Example equilibrium 1, corresponding to Ip_model = 13.3 MA and Ip = 12.6 MA, and the initial (dashed) and final (solid) mid-plane poloidal magnetic flux.

Table IV. Input data of Example equilibrium 1.

| | | | |
|---|---|---|---|
| gfun_C1 | 0.0014591 | CFpl1 | 78.4112 |
| gfun_nf | 2.5 | CTpl1 | 60 |
| gfun_nt | 1.8 | CFbl1 | 14.524084 |
| gfun_C2 | 0.064 | CTbl1 | 60 |
| gfun_C2l | 0.05 | CFel1 | 60 |
| CFpl0 | -0.0001685 | CTel1 | 30 |
| CTpl0 | 4.982E-5 | CFph1 | -1187.394252 |
| CFbl0 | -0.0002487 | CTph1 | 360 |
| CTbl0 | 4.982E-5 | CFbh1 | -1570.71676 |
| CFel0 | -6.698E-5 | CTbh1 | 360 |
| CTel0 | 2.491E-5 | CFeh1 | -2350.905576 |
| CFph0 | -0.0029083 | CTeh1 | 960 |
| CTph0 | 0.0002491 | | |
| CFbh0 | -0.0033092 | Psi_crit1 | -0.015 |
| CTbh0 | 0.0002491 | Psi_crit2 | 0.03 |
| CFeh0 | -0.0054713 | Psi_crit3 | -0.014 |
| CTeh0 | 0.0004982 | | |

Table V. Output data of Example equilibrium 1.

| | | | |
|---|---|---|---|
| Rmajor/m | 1.426 | Tpl-axis/keV | 132.1 |
| Rminor,a/m | 0.862 | Tph-axis/keV | 892.28 |
| Iplasma/MA | 12.59 | Tbl-axis/keV | 132.0 |
| B at Rmajor/T | 3.383 | Tbh-axis/keV | 891.3 |
| kappa | 2.16 | Upl-peak/km/s | -347 |
| A | 1.65 | Uph-peak/km/s | -2810 |
| delta | 0.269 | Ubl-peak/km/s | -86 |
| q0 | 1.35 | Ubh-peak/km/s | -642 |
| q-LCFS | 5.33 | Uel-peak/km/s | 147 |
| q95 | 4.28 | Ueh-peak/km/s | 5938 |
| li (int.inductance) | 0.284 | h(Teh)-peak | 19.1 |
| beta-T-LCFS | 0.253 | h(Tel)-peak | 1.35 |
| beta-pol-LCFS | 1.091 | Ipl/kA | 4160.0 |
| beta-axis | 0.83 | Ibl/kA | 1375.9 |
| nel-axis/e20/m3 | 1.039363 | Iel/kA | 4353.8 |
| neh-axis/e18/m3 | 1.1296 | Iph/kA | 336.8 |
| npl-axis/e20/m3 | 0.434668 | Ibh/kA | 103.8 |
| nph-axis/e18/m3 | 0.4553 | Ieh/kA | 2259.4 |
| nbl-axis/e20/m3 | 0.120868 | Ipl+Ibl+Iel/kA | 9889.7 |
| nbh-axis/e18/m3 | 0.1420 | Iph+Ibh+Ieh/kA | 2699.9 |
| Tel-axis/keV | 66.7 | | |
| Teh-axis/keV | 2373.80 | | |

**B. Example 2: Low-$\beta_T$ burn test EHL-3 equilibrium**

The computational boundary is rectangular with $0.56 \leq R(m) \leq 4.0$ and $-3 \leq Z(m) \leq 3$. In the Example 2 computation, we adopt $L_{ref} = 2.5m$, $I_{ref} = 4672kA$ and $n_{ref} = 4 \times 10^{19}\frac{1}{m^3}$ as the primary scales. Therefore, $B_{ref} = 2.3484T$, $\psi_{ref} = 14.6775\frac{Wb}{rad}$, $u_{ref} = \frac{8099.127km}{s}$, $T_{ref} = 684.624keV$, and $\varepsilon$ =0.014401. The poloidal field coil locations and currents used are provided in Table VI.

Table VI. Poloidal field coil locations and currents used for Example equilibrium 2.

| Poloidal field coil location and current (Equilibrium 2) | | | | | |
|---|---|---|---|---|---|
| | R (m) | Z (m) | current(kA) | factor | net current (dimensionless) |
| PF1 | 1.5 | 3.5 | 700 | -0.882 | -0.077180 |
| PF2 | 1.5 | -3.5 | 700 | -0.882 | -0.077180 |
| PF3 | 2.75 | 3.5 | 700 | 0.882 | 0.077175 |
| PF4 | 2.75 | -3.5 | 700 | 0.882 | 0.077175 |
| PF5 | 4.5 | 2.6 | 2000 | 0 | 0.000000 |
| PF6 | 4.5 | -2.6 | 2000 | 0 | 0.000000 |
| PF7 | 4.5 | 1 | 2500 | 1.9404 | 0.606375 |
| PF8 | 4.5 | -1 | 2500 | 1.9404 | 0.606375 |
| PF9 | 0.55 | 3 | 700 | -7.938 | -0.694580 |
| PF10 | 0.55 | -3 | 700 | -7.938 | -0.694580 |

The coil locations are the same as those in Example 1 equilibrium.

Ampere's equation, the electrostatic potential $V_E$, the poloidal magnetic flux $\psi$, and the radial force balance of each fluid of this equilibrium are solved within the relative errors shown below:

Ampere's equation is solved within an error of $\frac{average|GSop+RJt|}{average|RJt|} =$ 0.00098.

The electrostatic potential VE is solved within an error of $\frac{averag|VEeh-VEel|}{average|VEel|}$ = 0.00168.

The quantity $(\psi - \psi_{plasma})$ is solved within an error of $\frac{average|\psi-\psi_{plasma}-\psi_{pfcoil}|}{average|\psi_{pfcoil}|} =$ 0.0104.

Radial force balance of the pl-fluid is solved within an error of $\frac{average|balance|}{average|FplP|} = 0.0083.$

Radial force balance of the bl-fluid is solved within an error of $\frac{average|balance|}{average|FblP|} = 0.0102.$

Radial force balance of the el-fluid is solved within an error of $\frac{average|balance|}{average|FelP|} = 0.0096.$

Radial force balance of the ph-fluid is solved within an error of $\frac{average|balance|}{average|FphP|} = 0.0058.$

Radial force balance of the bh-fluid is solved within an error of $\frac{average|balance|}{average|FbhP|} = 0.0057.$

Radial force balance of the eh-fluid is solved within an error of $\frac{average|balance|}{average|FehP|} = 0.0046.$

Figs. 30 – 35 show the poloidal magnetic flux, the toroidal current density, the total pressure, the toroidal current density of the suprathermal electron fluid (eh), the electrostatic potential $V_E$, and the magnetic field strength, which shows a weak magnetic well, of a low-$\beta_T$ burn test EHL-3 equilibrium. The green dotted lines in (a), (b), (c), (d), (e), and (f) represent the LCFS. The solid red line indicates the flux-surface boundary of the suprathermal components. This boundary is determined by the parameter Psi_crit2. The LCFS, indicated by green dashes, is determined via the convergent equilibrium. The $\beta_T$ and $\beta_p$ are computed within the LCFS.

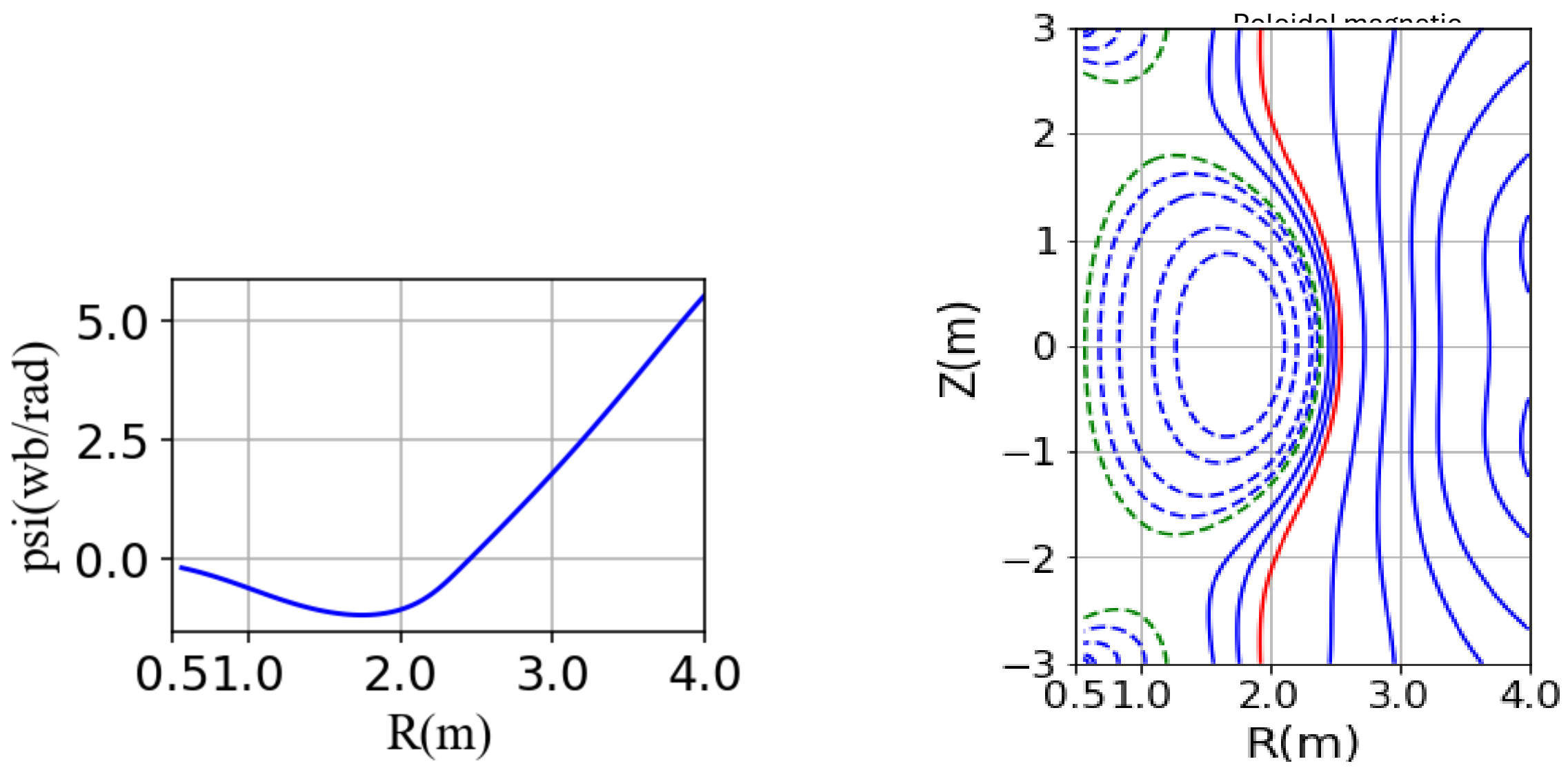

Figure 30. The poloidal magnetic flux across $R$ on the mid-plane, and over the $R$ and $Z$ coordinates.

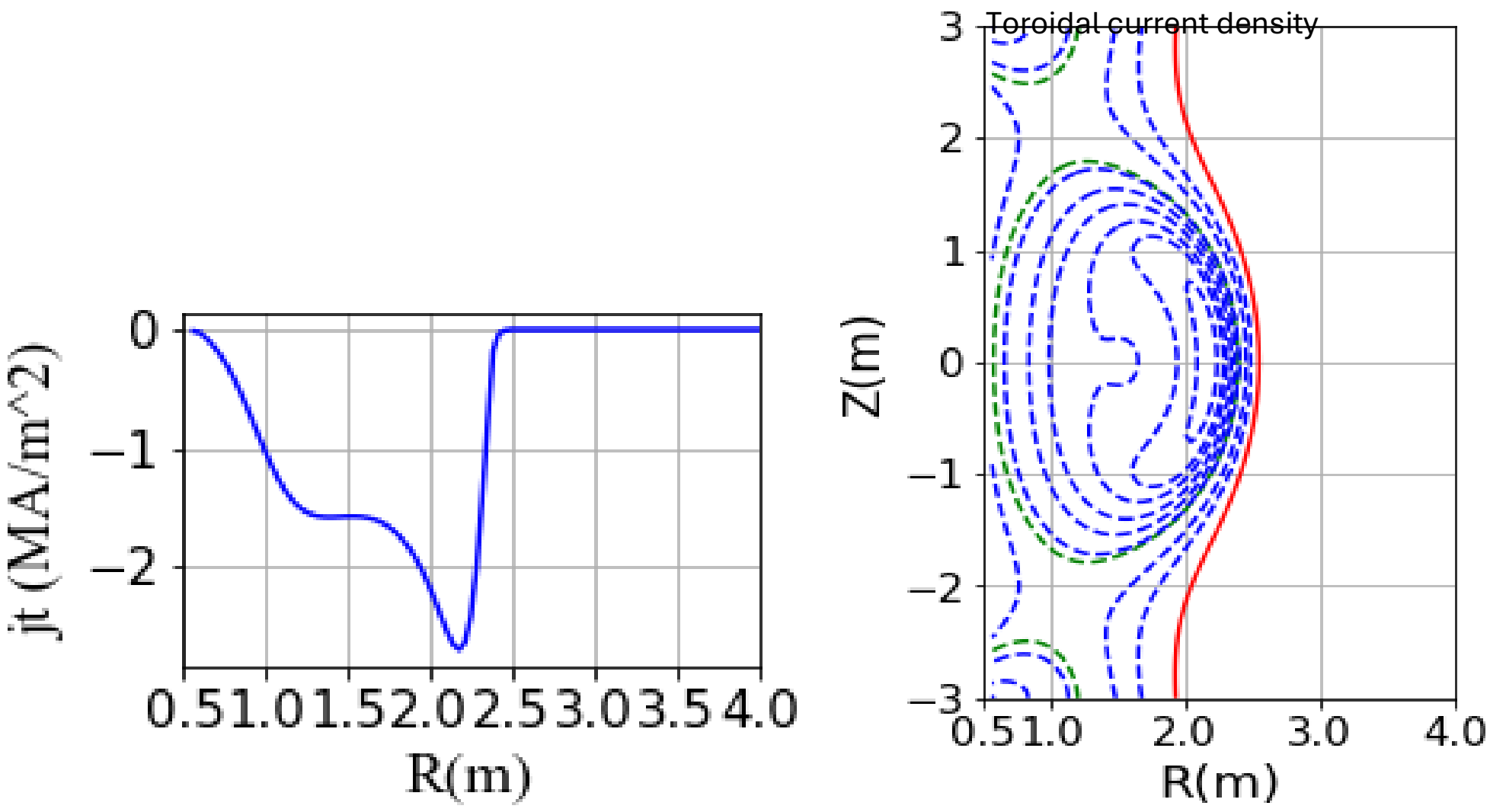

Figure 31. The toroidal plasma current density across *R* on the mid-plane, and over the *R* and *Z* coordinates.

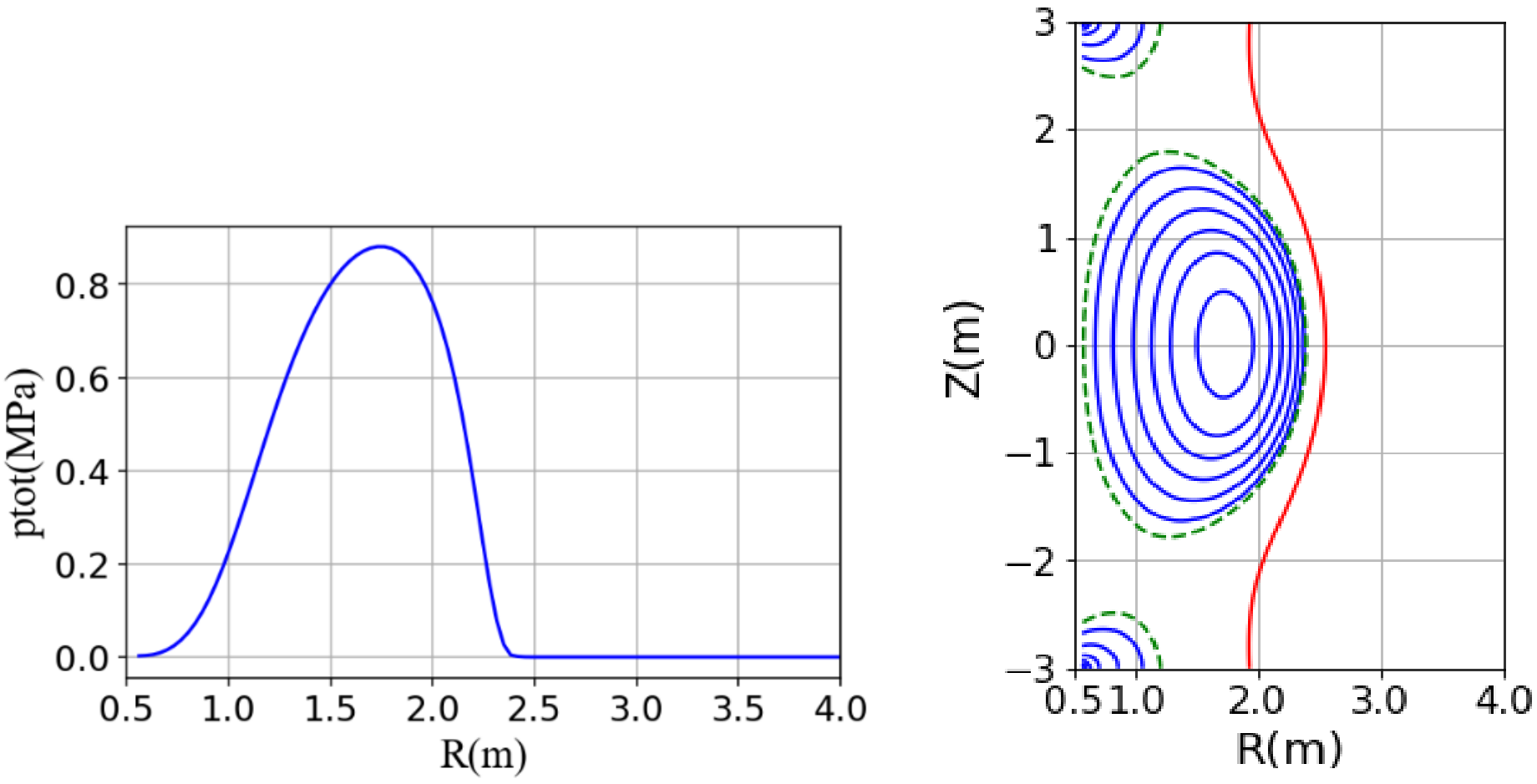

Figure 32. Total pressure across *R* on the mid-plane, and over the *R* and *Z* coordinates.

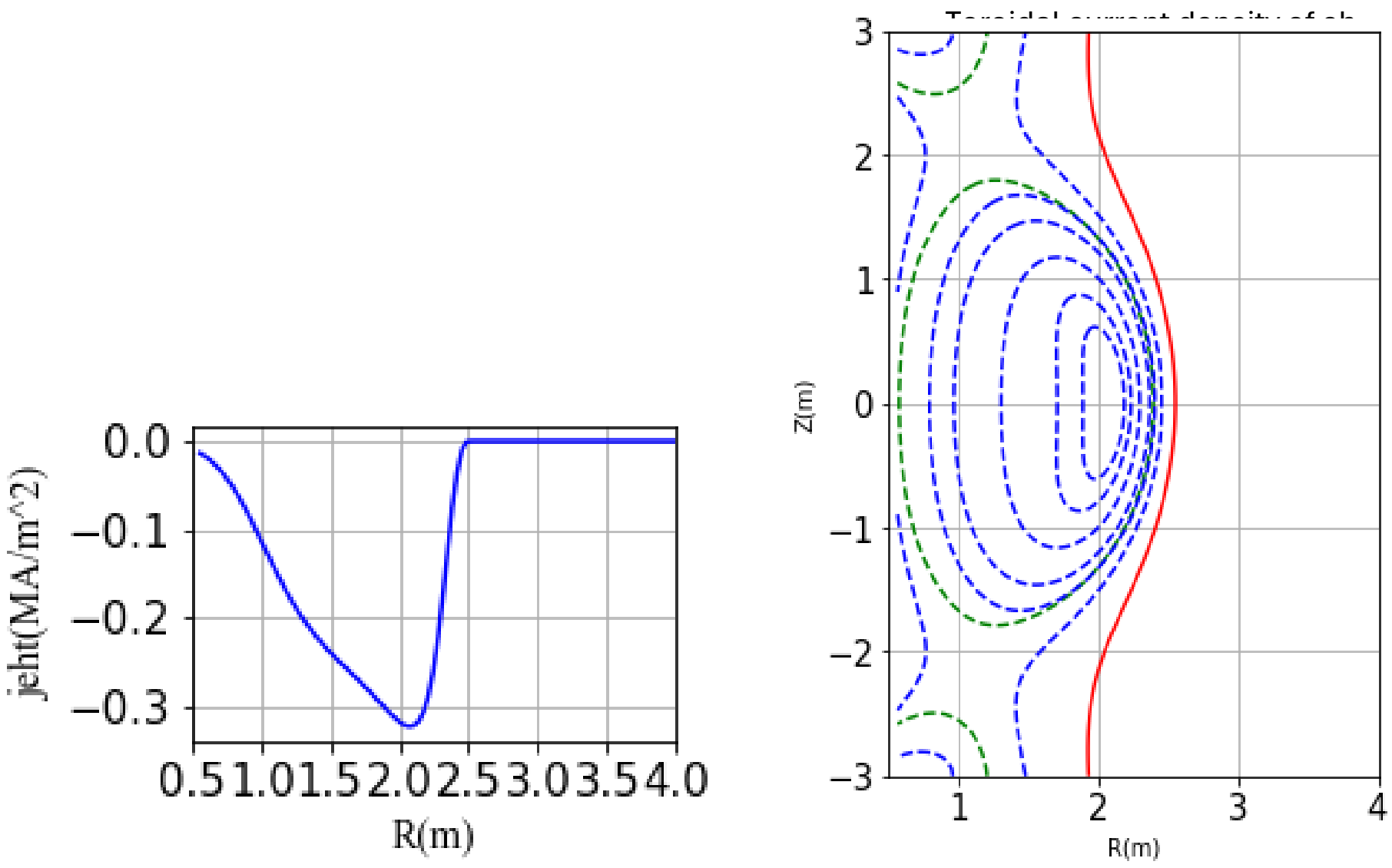

Figure 33. Toroidal current density of the suprathermal electron fluid, i.e., eh-fluid across *R* on the mid-plane, and over the *R* and *Z* coordinates.

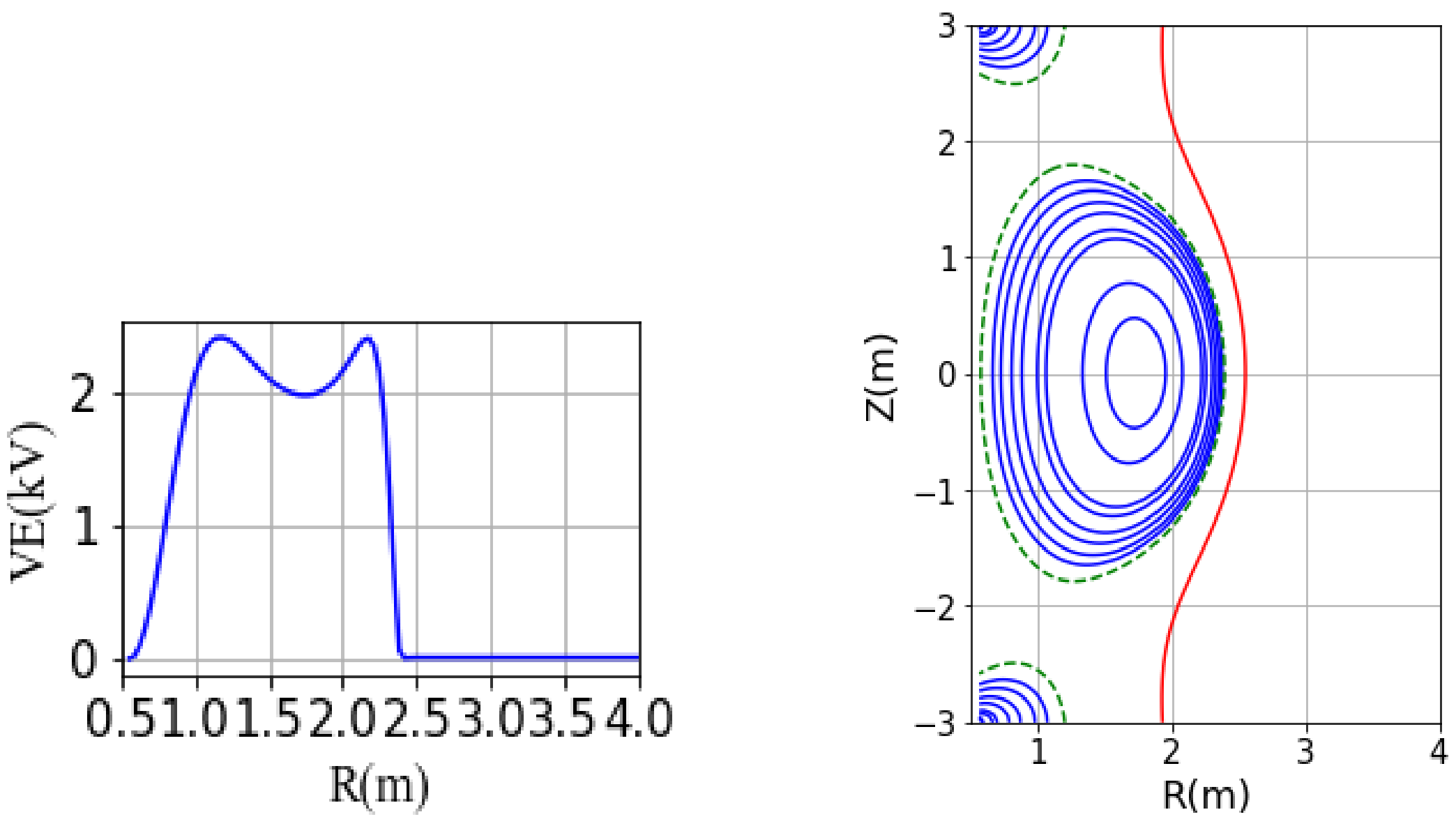

Figure 34. Electrostatic potential $V_E$ across *R* on the mid-plane, and over the *R* and *Z* coordinates.

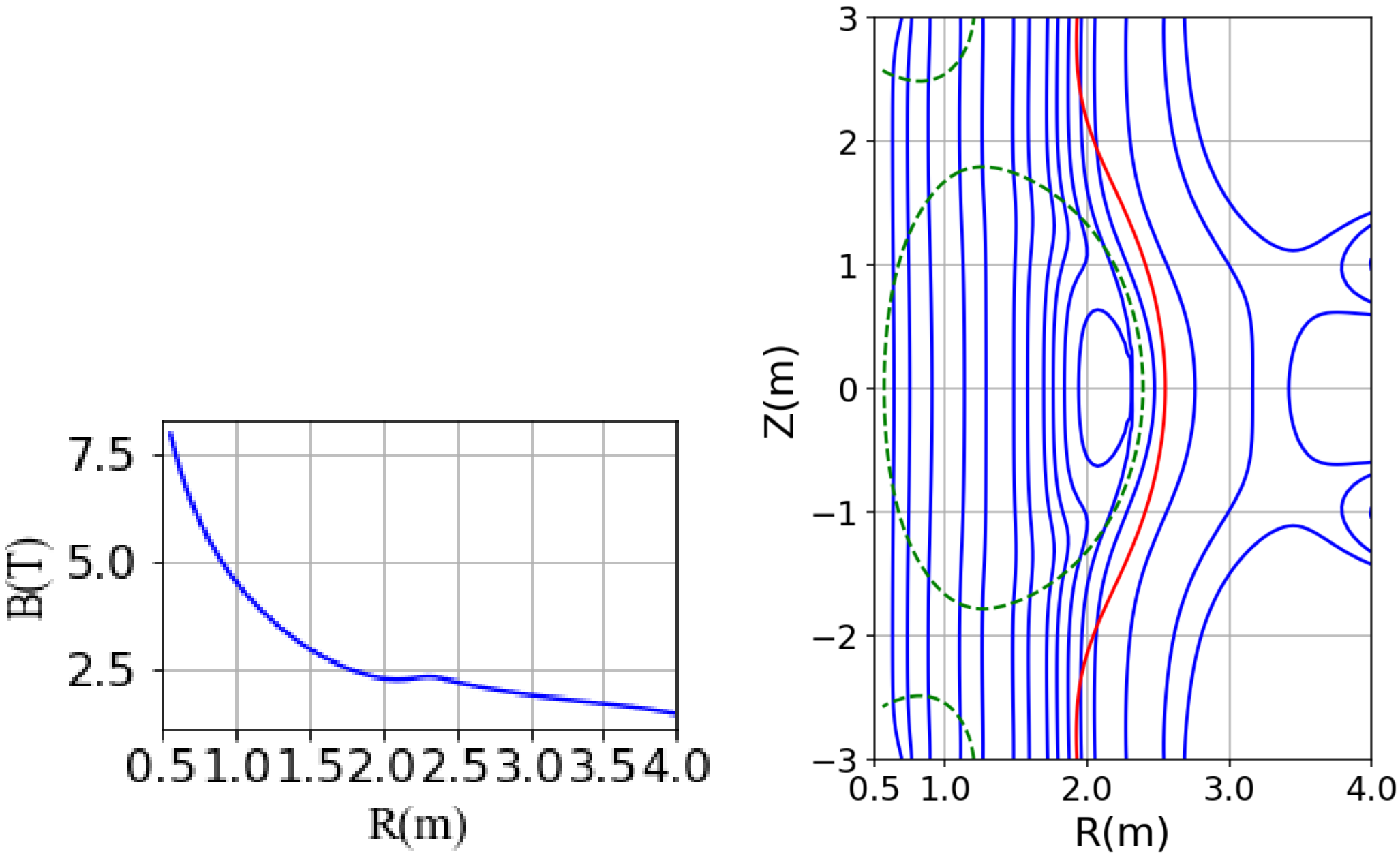

Figure 35. Magnetic field across $R$ on the mid-plane, and over the $R$ and $Z$ coordinates.

Figs. 36 and 37 show the mid-plane density and temperature profiles of the six fluid components.

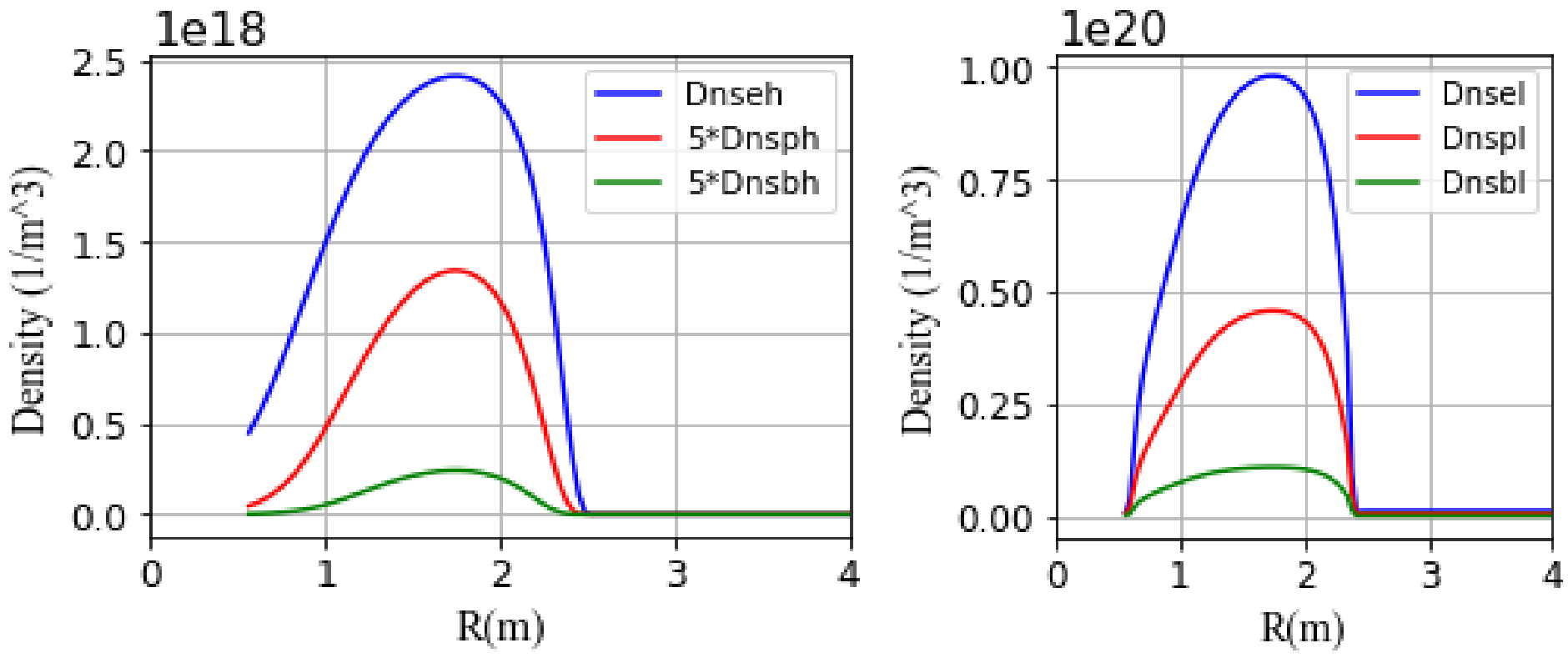

Figure 36. Densities of the six fluid components.

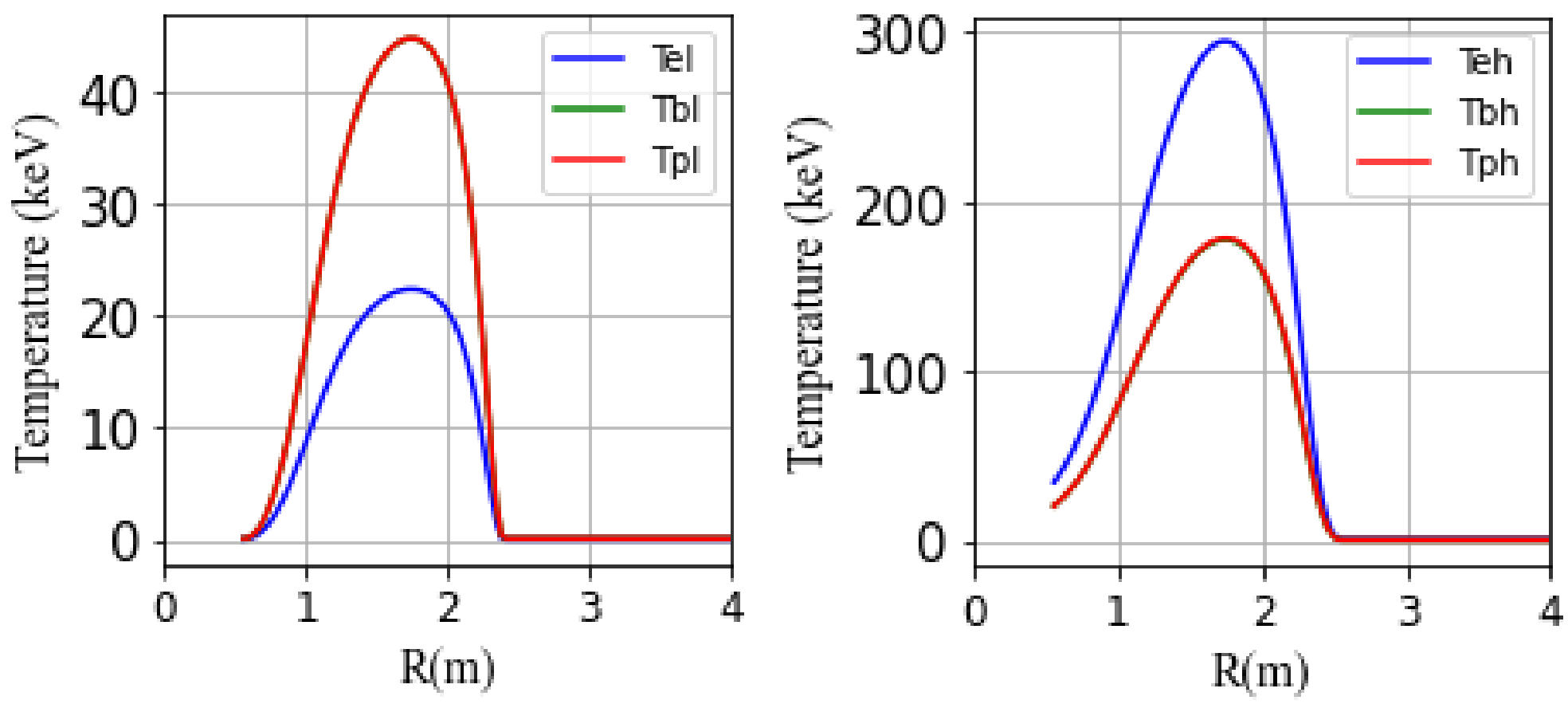

Figure 37. Temperatures of six fluids with Tpl = Tbl.

Fig. 38 shows the *h* functions, which represent the relativistic effects due to thermal motion, for the suprathermal electron fluid with temperature $T_{eh}$, and the thermal electron fluid with temperature $T_{el}$.

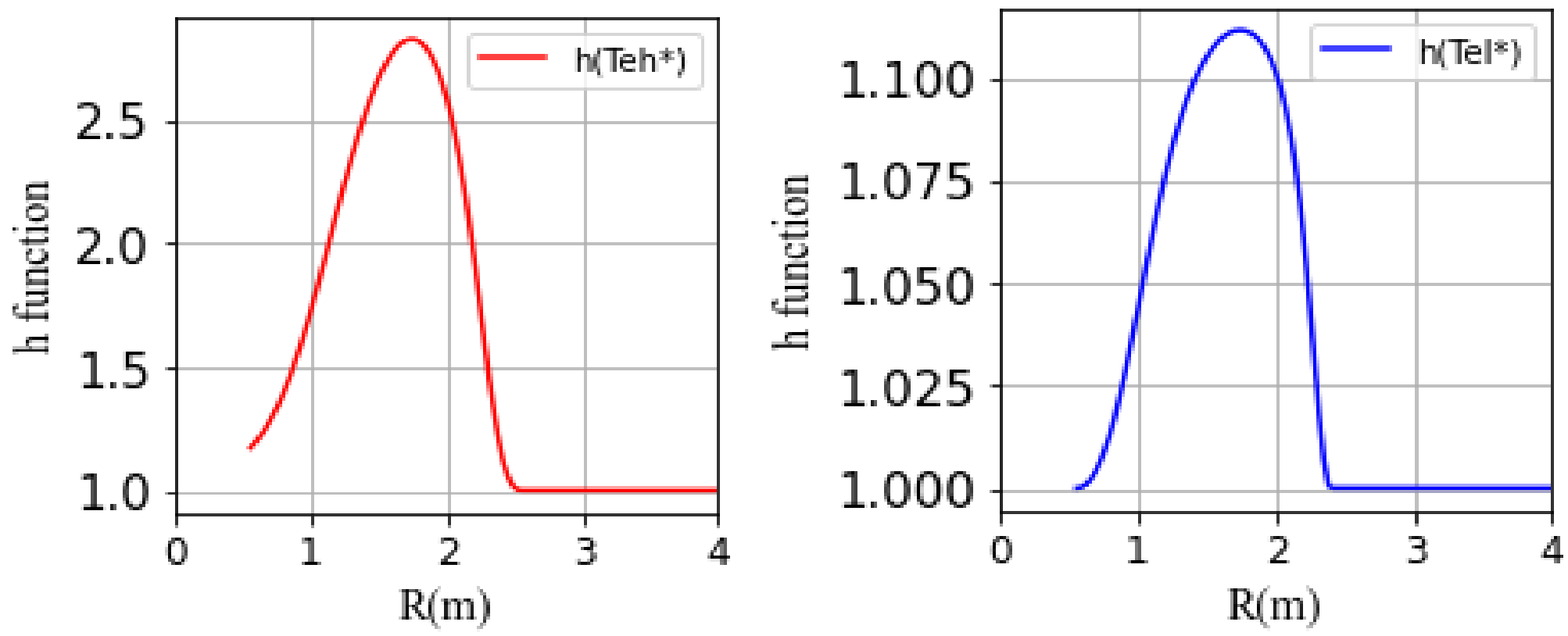

Figure 38. The *h* functions for the suprathermal electron and the thermal electron fluids.

Fig. 39 shows the cyclotron frequency of the el-fluid and the eh fluid.

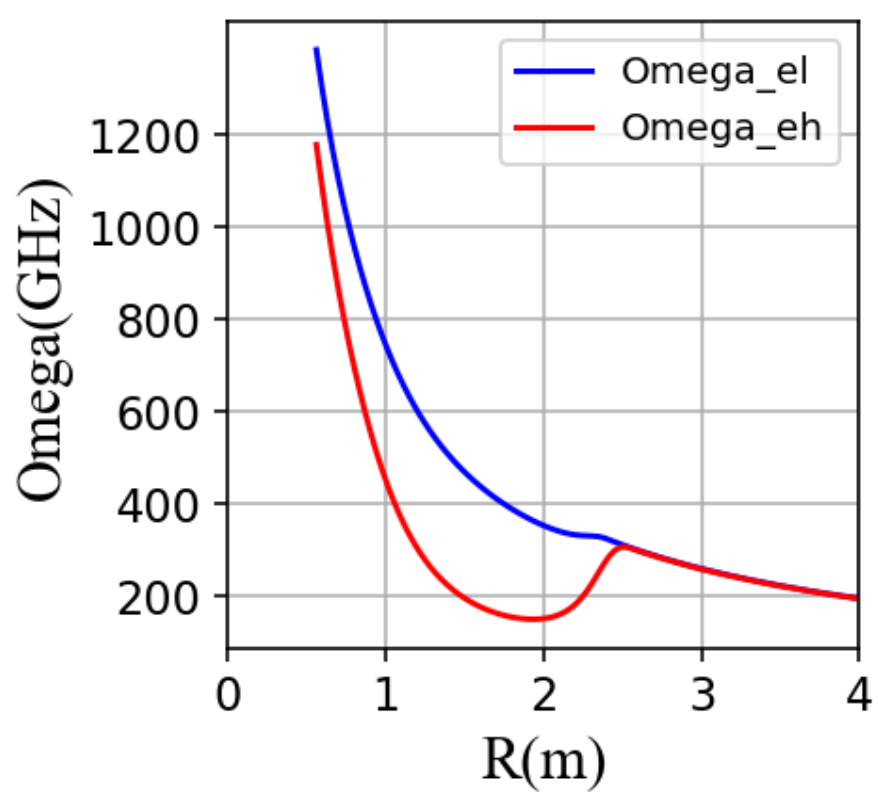

Figure 39. Cyclotron frequency of the el-fluid, $eB/m_e h(T_{el}^*)$, and the eh-fluid, $eB/m_e h(T_{eh}^*)$.

Fig. 40 shows toroidal rotation velocities of six fluid components.

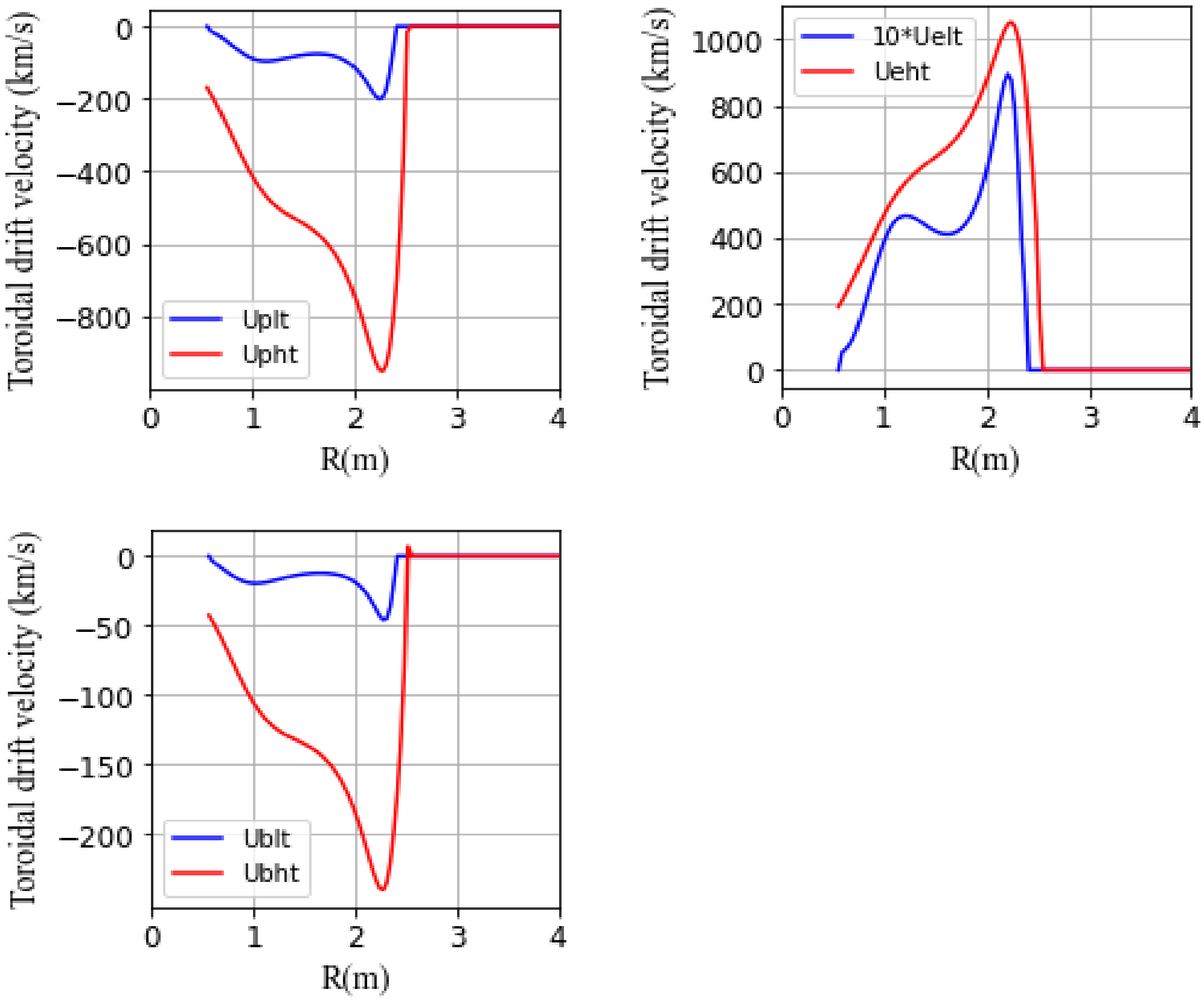

Figure 40. Toroidal rotation velocities of six fluid components.

Table VII. Input coefficients of Example equilibrium 2

| | | | |
|---|---|---|---|
| gfun_C1 | 0.0013892 | CFpl1 | 78.4112 |
| gfun_nf | 2.5 | CTpl1 | 60 |
| gfun_nt | 1.8 | CFbl1 | -4.766493 |
| gfun_C2 | 0.1 | CTbl1 | 60 |
| gfun_C2l | 0.05 | CFel1 | 60 |
| CFpl0 | -0.000494 | CTel1 | 30 |
| CTpl0 | 0.000146 | CFph1 | -1187.39425 |
| CFbl0 | -0.000729 | CTph1 | 360 |
| CTbl0 | 0.000146 | CFbh1 | -1686.46022 |
| CFel0 | -0.000196 | CTbh1 | 360 |
| CTel0 | 7.3E-05 | CFeh1 | -888.907794 |
| CFph0 | -0.008527 | CTeh1 | 600 |
| CTph0 | 0.00073 | | |
| CFbh0 | -0.009703 | Psi_crit1 | -0.014 |
| CTbh0 | 0.00073 | Psi_crit2 | 0.02 |
| CFeh0 | -0.016042 | Psi_crit3 | -0.014 |
| CTeh0 | 0.001461 | | |

Table XIII. Plasma parameters of Example equilibrium 2.

| | | | |
|---|---|---|---|
| Rmajor/m | 1.482 | Tpl-axis/keV | 44.8 |
| Rminor,a/m | 0.911 | Tph-axis/keV | 178.5 |
| Iplasma/MA | 6.36 | Tbl-axis/keV | 44.8 |
| B at Rmajor/T | 2.971 | Tbh-axis/keV | 177.7 |
| kappa | 2 | Upl-peak/km/s | -201 |
| A | 1.63 | Uph-peak/km/s | -952 |
| delta | 0.24 | Ubl-peak/km/s | -46 |
| q0 | 1.89 | Ubh-peak/km/s | -241 |
| q-LCFS | 7.64 | Uel-peak/km/s | 90 |
| q95 | 6.31 | Ueh-peak/km/s | 1052 |
| li (int.inductance) | 0.331 | h(Teh)-peak | 2.824 |
| beta-T-LCFS | 0.095 | h(Tel)-peak | 1.11 |
| beta-pol-LCFS | 1.042 | Ipl/kA | 2388.2 |
| beta-axis | 0.34 | Ibl/kA | 616.8 |
| nel-axis/e20/m3 | 0.98054 | Iel/kA | 2442 |
| neh-axis/e18/m3 | 2.411 | Iph/kA | 55.5 |
| npl-axis/e20/m3 | 0.45705 | Ibh/kA | 9.84 |
| nph-axis/e18/m3 | 0.2682 | Ieh/kA | 865.8 |
| nbl-axis/e20/m3 | 0.1085 | Ipl+Ibl+Iel/kA | 5427 |
| nbh-axis/e18/m3 | 0.0483 | Iph+Ibh+Ieh/kA | 931.2 |
| Tel-axis/keV | 22.4 | | |
| Teh-axis/keV | 294.66 | | |

### C. Example 3: Low-$\beta_T$ fusion test EHL-2 equilibrium

The computational boundary is rectangular with $0.4 \leq R(m) \leq 1.8$ and $-1.5 \leq Z(m) \leq 1.5$. In Example 3 computation, we adopt $L_{ref} = 1.78m$, $I_{ref} = 2121kA$ and $n_{ref} = 2.8 \times 10^{19} \frac{1}{m^3}$ as the primary scales. Therefore, $B_{ref} = 1.4974T$, $\psi_{ref} = 4.7443 \frac{Wb}{rad}$, $u_{ref} = \frac{6172.30km}{s}$, $T_{ref} = 397.6213keV$ and $\varepsilon$ =0.024175. The poloidal field coil locations and currents used for Example equilibrium 3 are provided in Table IX.

Table IX. Poloidal field coil locations and currents used for Example equilibrium 3.

| Poloidal field coil location and current (Equilibrium 3) | | | | | |
|---|---|---|---|---|---|
| | R(m) | Z(m) | Current(kA) | factor | net current (dimensionless) |
| PF1 | 0.45 | 1.756 | 720 | -1 | -0.339463 |
| PF2 | 0.45 | -1.756 | 720 | -1 | -0.339463 |
| PF3 | 0.7 | 2.05 | 782 | -1 | -0.368694 |
| PF4 | 0.7 | -2.05 | 782 | -1 | -0.368694 |
| PF5 | 0.835 | 2.35 | 761.6 | 1 | 0.359076 |
| PF6 | 0.835 | -2.35 | 761.6 | 1 | 0.359076 |
| PF7 | 1.89 | 2.31 | 1428 | -1 | -0.673267 |
| PF8 | 1.89 | -2.31 | 1428 | -1 | -0.673267 |
| PF9 | 1.89 | 1.64 | 952 | 1 | 0.448845 |
| PF10 | 1.89 | -1.64 | 952 | 1 | 0.448845 |
| PF11 | 2.5 | 0.53 | 989 | 1 | 0.466289 |
| PF12 | 2.5 | -0.53 | 989 | 1 | 0.466289 |

Ampere's equation, the electrostatic potential VE, the poloidal magnetic flux $\psi$, and the radial force balance of each fluid of this equilibrium are solved within the relative errors shown below:

Ampere's equation is solved within an error of $\frac{average|GSop+RJt|}{average|RJt|} = 0.00106$.

The electrostatic potential VE is solved within an error of $\frac{averag|VEeh-VEel|}{average|VEel|}$ = 0.02036.

The quantity $(\psi - \psi_{plasma})$ is solved within an error of $\frac{average|\psi-\psi_{plasma}-\psi_{pfcoil}|}{average|\psi_{pfcoil}|} = 0.00939$.

The radial force balance of the pl-fluid is solved within an error of $\frac{average|balance|}{average|FplP|} = 0.0022.$

The radial force balance of the bl-fluid is solved within an error of $\frac{average|balance|}{average|FblP|} = 0.0025.$

The radial force balance of the el-fluid is solved within an error of $\frac{average|balance|}{average|FelP|} = 0.0030.$

The radial force balance of the ph-fluid is solved within an error of $\frac{average|balance|}{average|FphP|} = 0.0021.$

The radial force balance of the bh-fluid is solved within an error of $\frac{average|balance|}{average|FbhP|} = 0.0030.$

The radial force balance of the eh-fluid is solved within an error of $\frac{average|balance|}{average|FehP|} = 0.0025.$

Figs. 41 – 46 show the poloidal magnetic flux, the toroidal current density, the total pressure, the toroidal current density of the suprathermal electron fluid (eh), the electrostatic potential $V_E$, and the magnetic field strength, which shows a weak magnetic well, of a low-$\beta_T$ fusion test EHL-2 equilibrium. The green dotted lines in (a), (b), (c), (d), (e), and (f) represent the LCFS. The solid red line indicates the flux-surface boundary of the suprathermal components. This boundary is determined by the parameter Psi_crit2. The LCFS, indicated by green dashes, is determined via the convergent equilibrium. The $\beta_T$ and $\beta_p$ are computed within the LCFS.

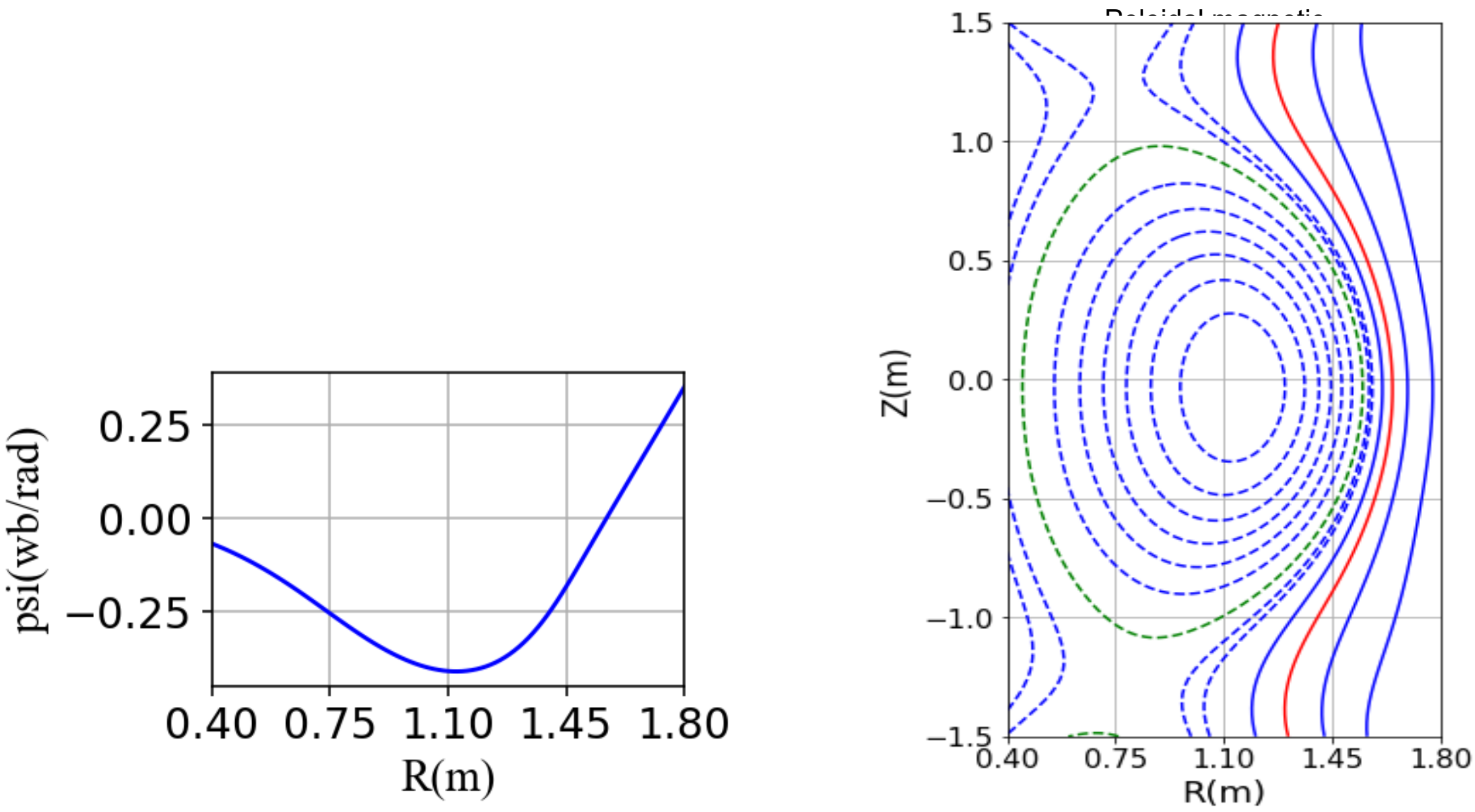

Figure 41. (a) Poloidal magnetic flux across *R* on the mid-plane, and over the *R* and *Z* coordinates.

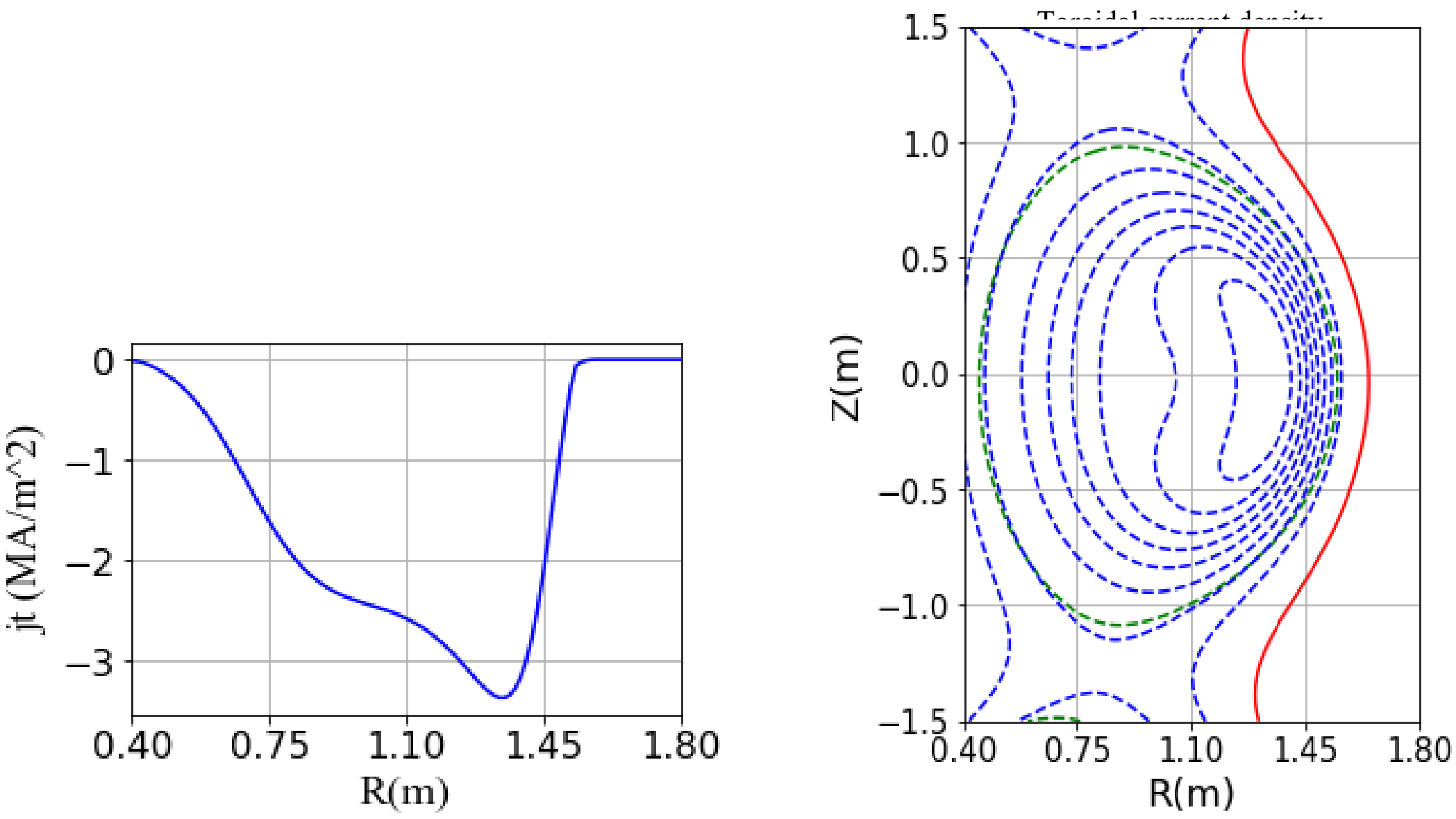

Figure 42. Toroidal current density across *R* on the mid-plane, and over the *R* and *Z* coordinates.

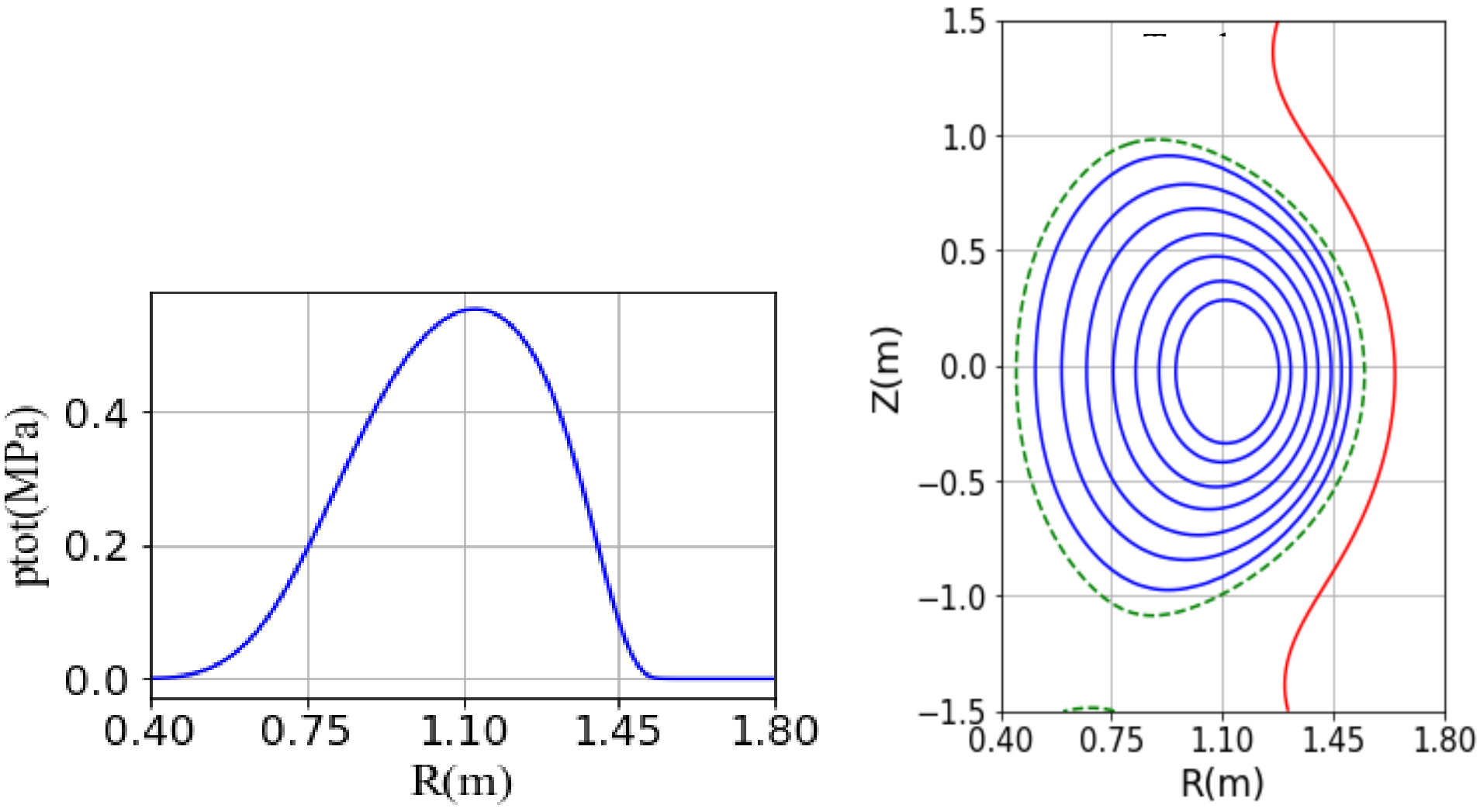

Figure 43. Plasma pressure across *R* on the mid-plane, and over the *R* and *Z* coordinates.

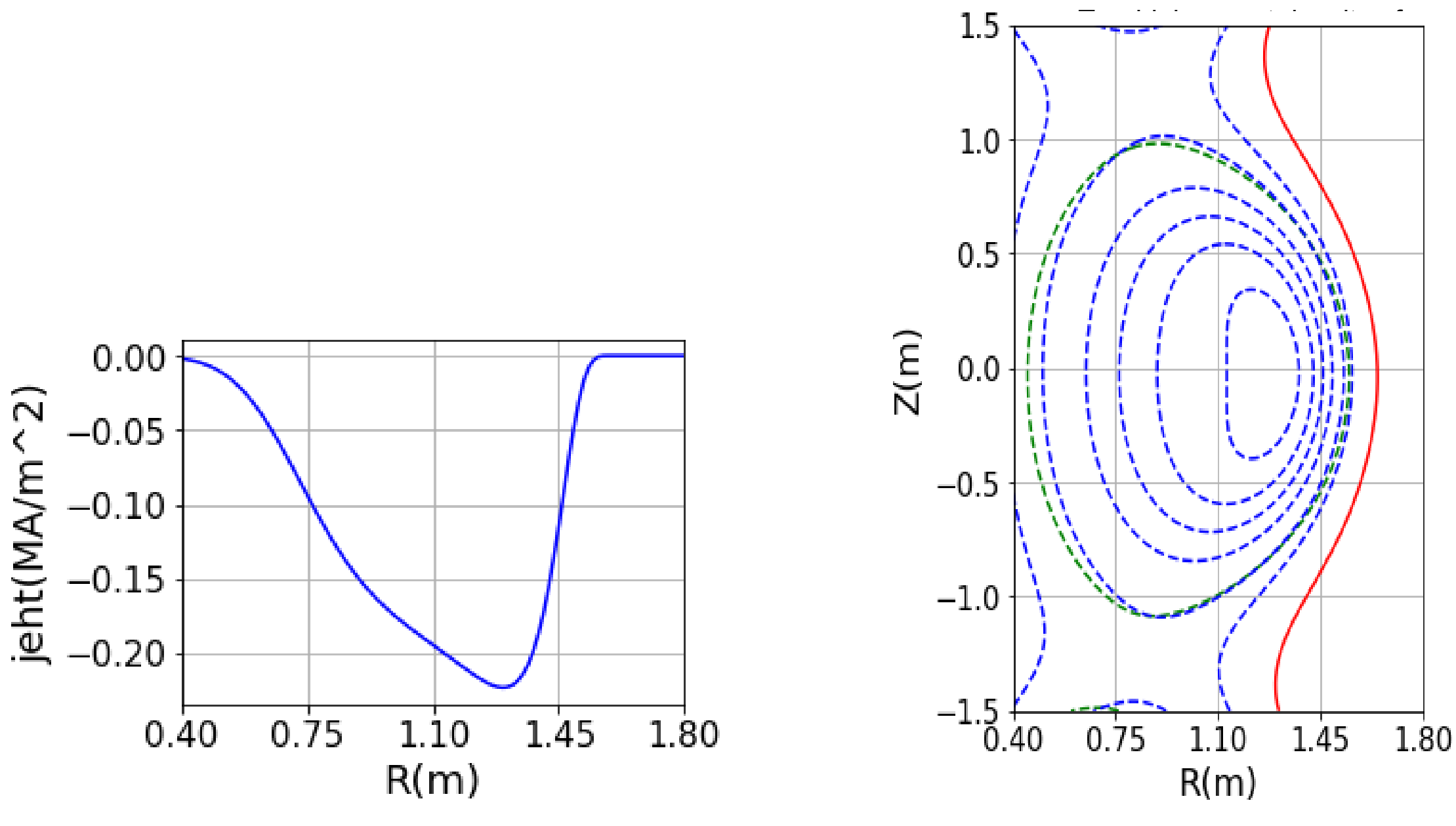

Figure 44. Toroidal current of the suprathermal electron fluid (eh) across *R* on the mid-plane, and over the *R* and *Z* coordinates.

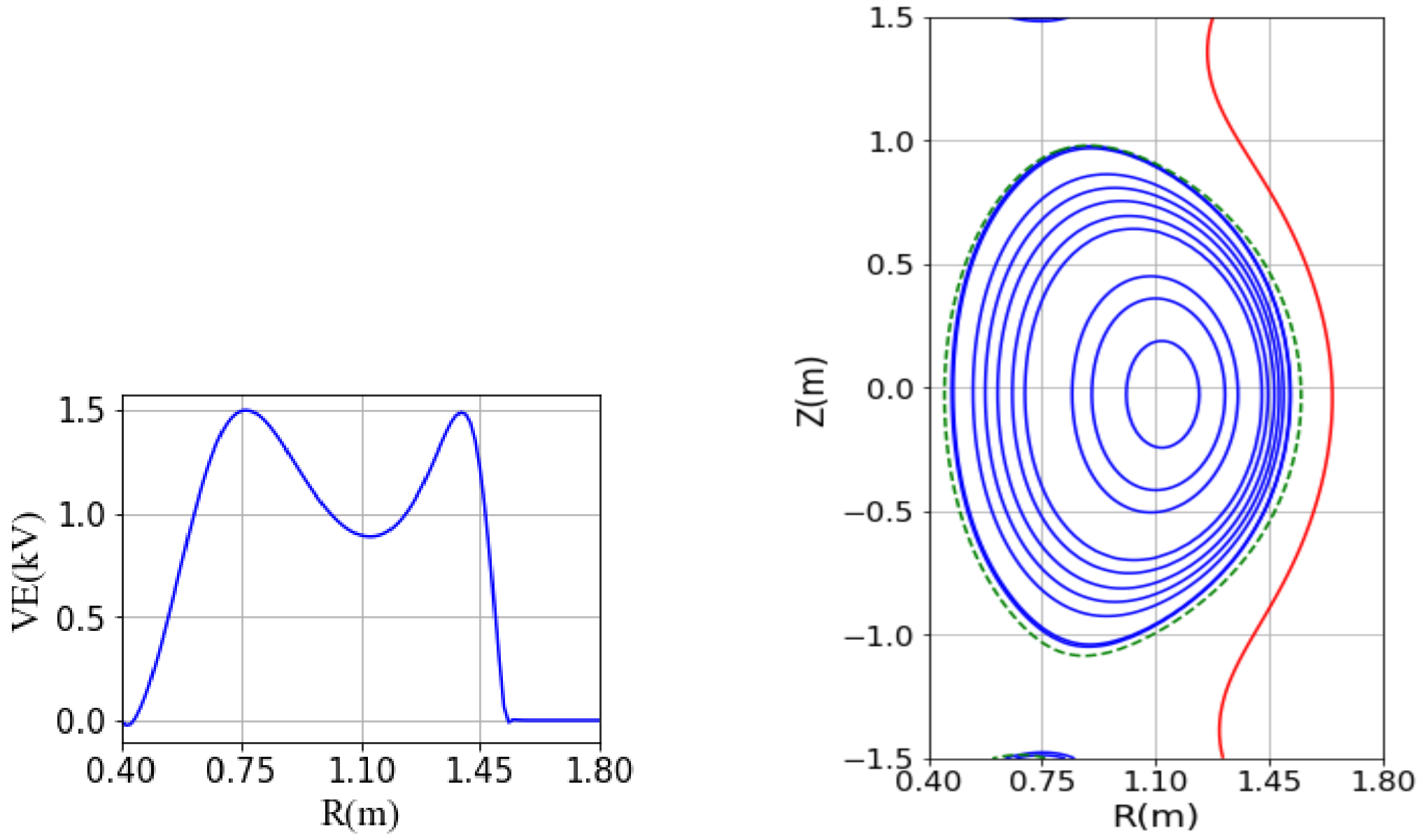

Figure 45. Electrostatic potential $V_E$ across $R$ on the mid-plane, and over the $R$ and $Z$ coordinates.

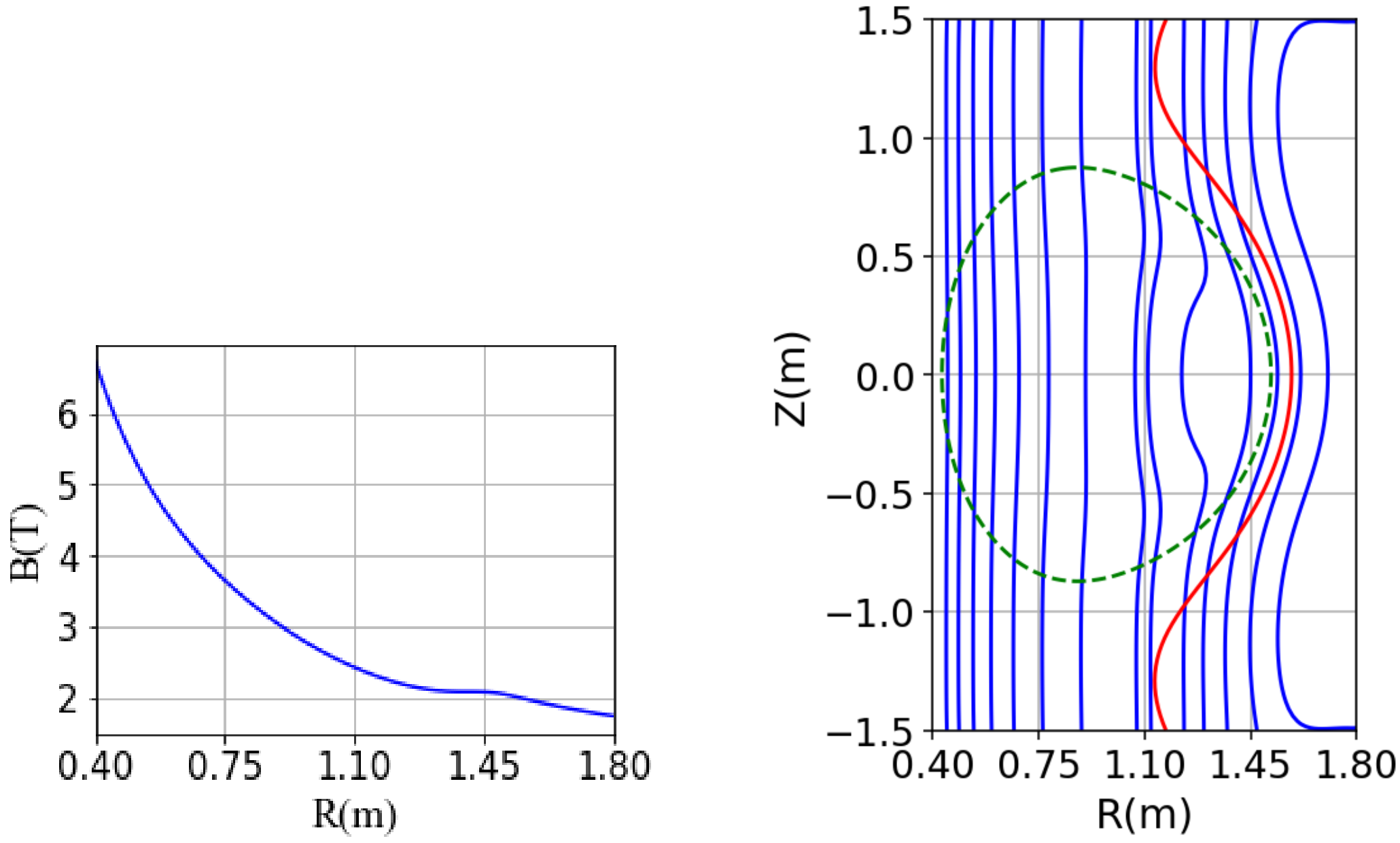

Figure 46. Magnetic field across $R$ on the mid-plane, and over the $R$ and $Z$ coordinates.

Figs. 47 and 48 show the mid-plane densities and temperatures of the six fluid components.

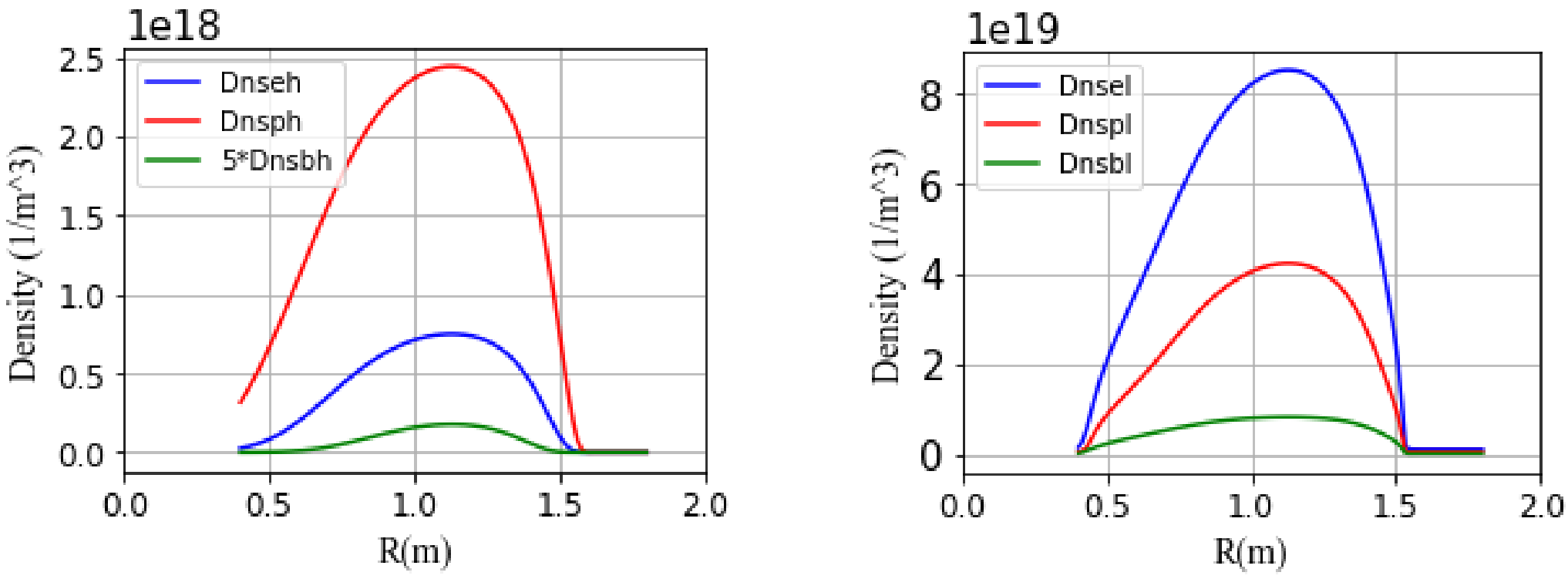

Figure 47. Densities of six fluid components.

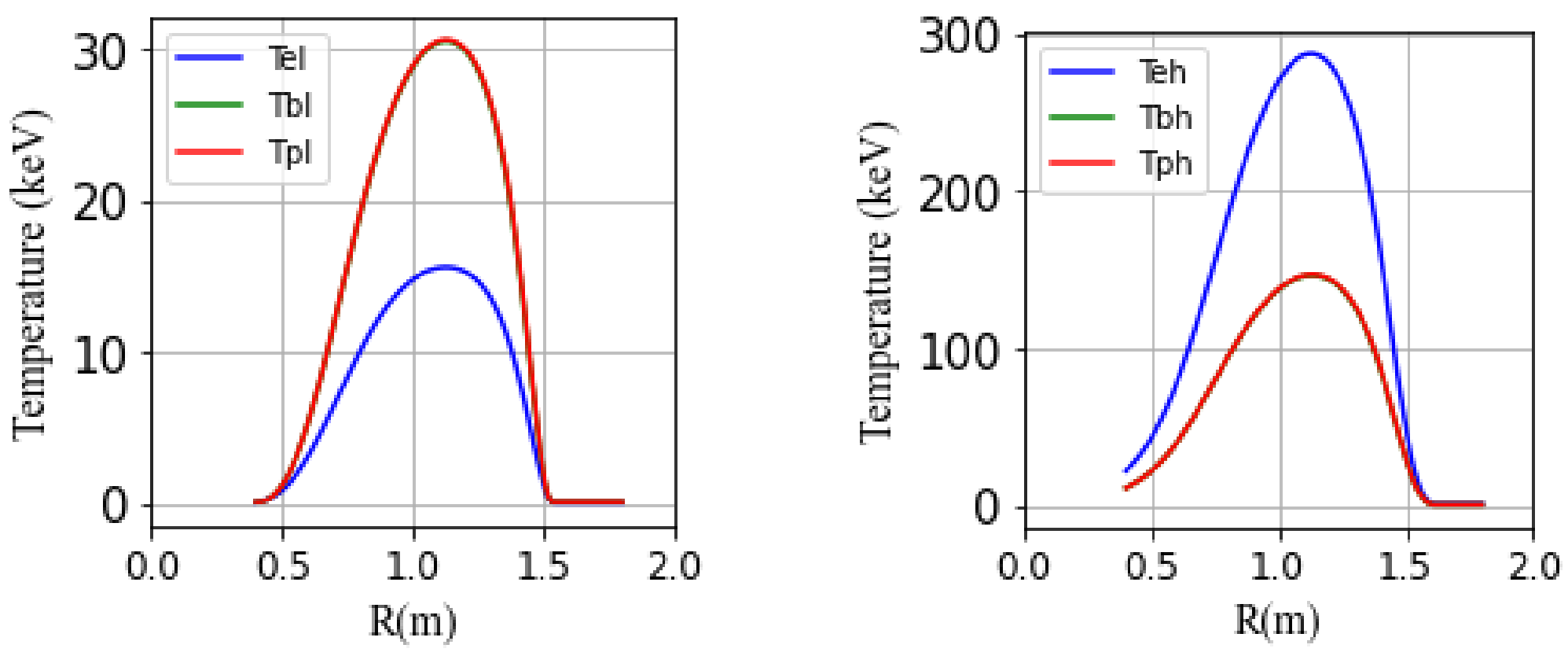

Figure 48. Temperatures of six fluids with Tpl = Tbl.

Fig. 49 shows the *h* functions, which represent the relativistic effects due to thermal motion, for the suprathermal electron fluid with temperature $T_{eh}$, and the thermal electron fluid with temperature $T_{el}$.

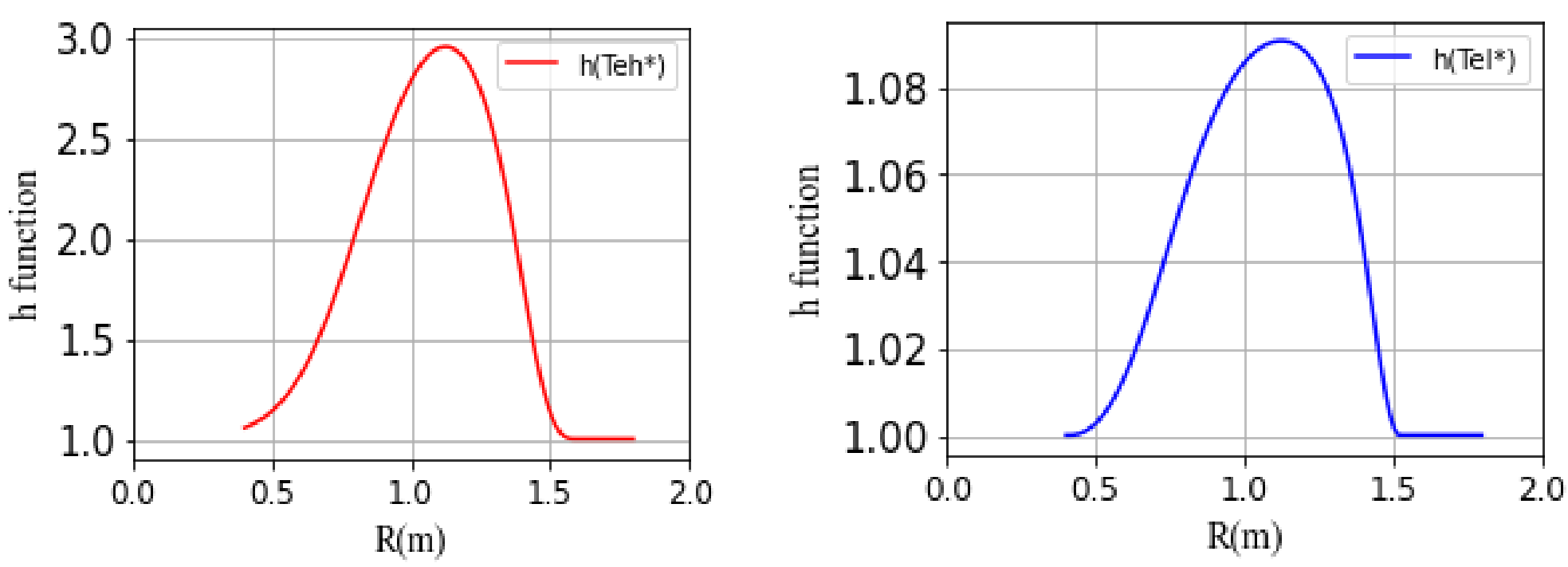

Figure 49. The *h* functions for the suprathermal electron and the thermal electron fluids.

Fig. 50 shows the mid-plane cyclotron frequency of the el-fluid, $eB/m_e h(T_{el}^*)$, and of the eh-fluid, $eB/m_e h(T_{eh}^*)$.

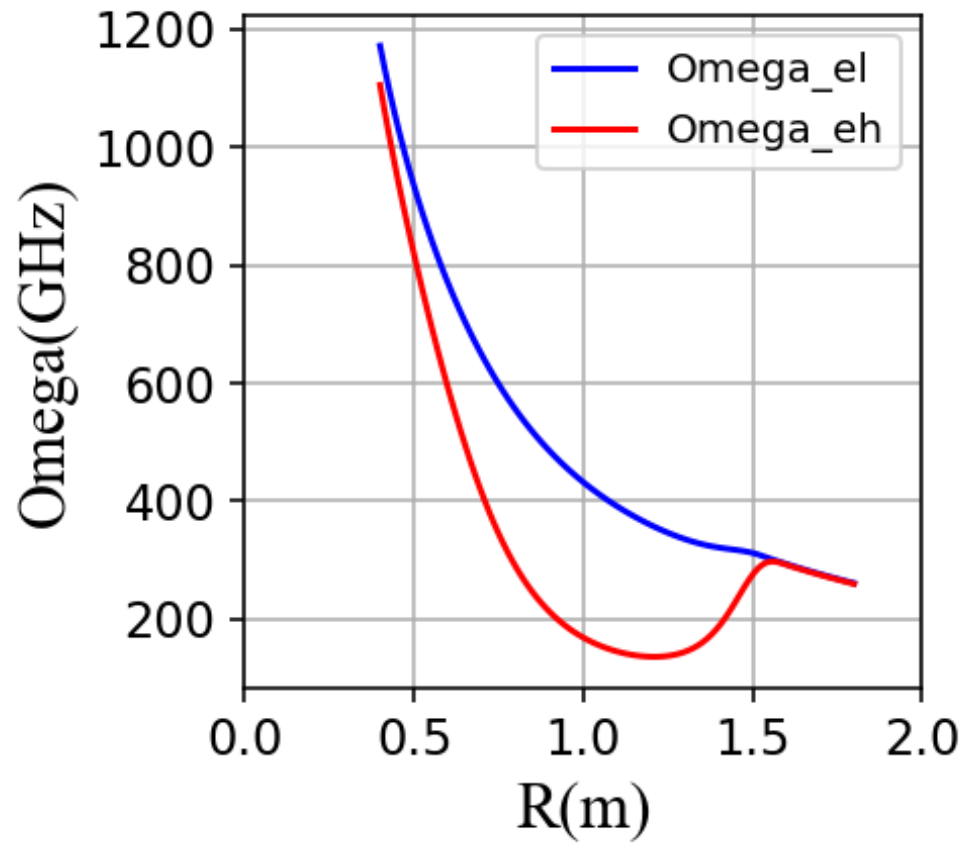

Fig. 50. The mid-plane cyclotron frequencies of the el-fluid and of the eh-fluid.

Figure 51 shows the mid-plane toroidal rotation velocities of the six fluid components.

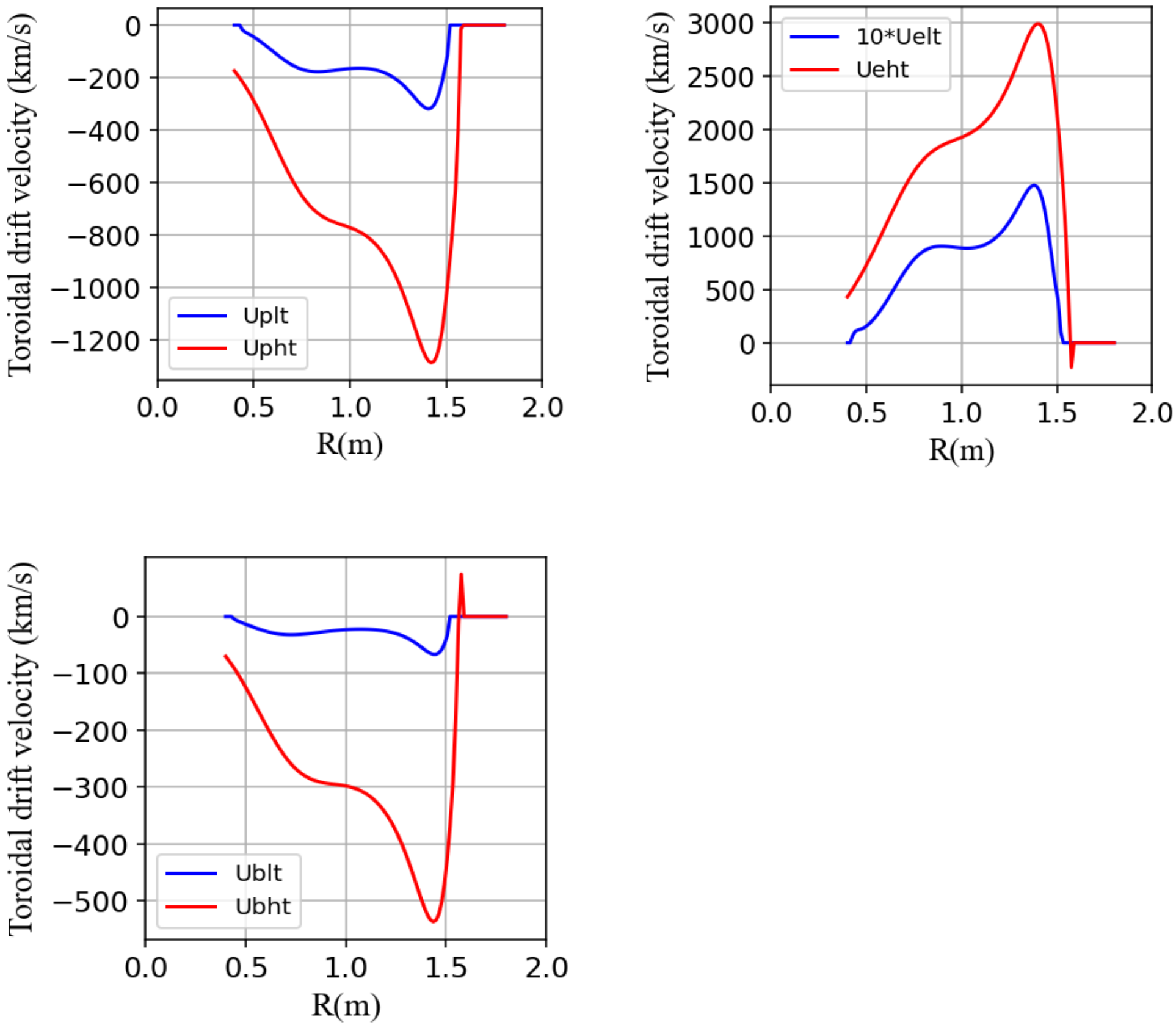

Fig. 51. The mid-plane toroidal rotation velocities of the six fluid components.

Table X. Plasma parameters of Example equilibrium 3.

| | | | |
|---|---|---|---|
| gfun_C1 | 0.004545 | CFpl1 | 40.9645 |
| gfun_nf | 2.6 | CTpl1 | 24 |
| gfun_nt | 1.8 | CFbl1 | -1.870366 |
| gfun_C2 | 0.09 | CTbl1 | 24 |
| gfun_C2l | 0.06 | CFel1 | 28 |
| CFpl0 | -0.000761 | CTel1 | 12 |
| CTpl0 | 0.000251 | CFph1 | -166.94308 |
| CFbl0 | -0.001166 | CTph1 | 144 |
| CTbl0 | 0.000251 | CFbh1 | -674.366704 |
| CFel0 | -0.000294 | CTbh1 | 144 |
| CTel0 | 0.000126 | CFeh1 | -622.387233 |
| CFph0 | -0.014234 | CTeh1 | 288 |
| CTph0 | 0.001257 | | |
| CFbh0 | -0.016258 | Psi_crit1 | -0.018 |
| CTbh0 | 0.001257 | Psi_crit2 | 0.01 |
| CFeh0 | -0.026725 | Psi_crit3 | -0.016 |
| CTeh0 | 0.002515 | | |

Table XI. Plasma parameters of Example equilibrium 3.

| | | | |
|---|---|---|---|
| Rmajor/m | 0.979 | Tpl-axis/keV | 30.6 |
| Rminor,a/m | 0.548 | Tph-axis/keV | 146.84 |
| Iplasma/MA | 2.697 | Tbl-axis/keV | 30.5 |
| B at Rmajor/T | 2.722 | Tbh-axis/keV | 146.3 |
| kappa | 2.08 | Upl-peak/km/s | -272 |
| A | 1.79 | Uph-peak/km/s | -1112 |
| delta | 0.18 | Ubl-peak/km/s | -57 |
| q0 | 1.62 | Ubh-peak/km/s | -458 |
| q-LCFS | 9.93 | Uel-peak/km/s | 127 |
| q95 | 6.69 | Ueh-peak/km/s | 2611 |
| li (int.inductance) | 0.380 | h(Teh)-peak | 2.77 |
| beta-T-LCFS | 0.058 | h(Tel)-peak | 1.077 |
| beta-pol-LCFS | 1.112 | Ipl/kA | 999.346 |
| beta-axis | 0.24 | Ibl/kA | 204.418 |
| nel-axis/e20/m3 | 0.8523 | Iel/kA | 1024.686 |
| neh-axis/e18/m3 | 0.7485 | Iph/kA | 297.361 |
| npl-axis/e20/m3 | 0.4233 | Ibh/kA | 3.768 |
| nph-axis/e18/m3 | 2.422 | Ieh/kA | 167.757 |
| nbl-axis/e20/m3 | 0.08207 | Ipl+Ibl+Iel/kA | 2228 |
| nbh-axis/e18/m3 | 0.03527 | Iph+Ibh+Ieh/kA | 469 |
| Tel-axis/keV | 15.6 | | |
| Teh-axis/keV | 287.42 | | |

Fig. 52 shows the contours of the poloidal magnetic flux $\psi$ $(= \psi_{pfcoil} + \psi_{plasma})$ of the low-beta EHL-2 equilibrium. $\psi_{pfcoil}$ is computed using the coil currents shown in Table VII, and $\psi_{plasma}$ is computed using the plasma current distribution obtained in Equilibrium 3. The red solid line shows the flux boundary of the suprathermal ions and electrons, and the red dashed line shows Psi_lcfs.

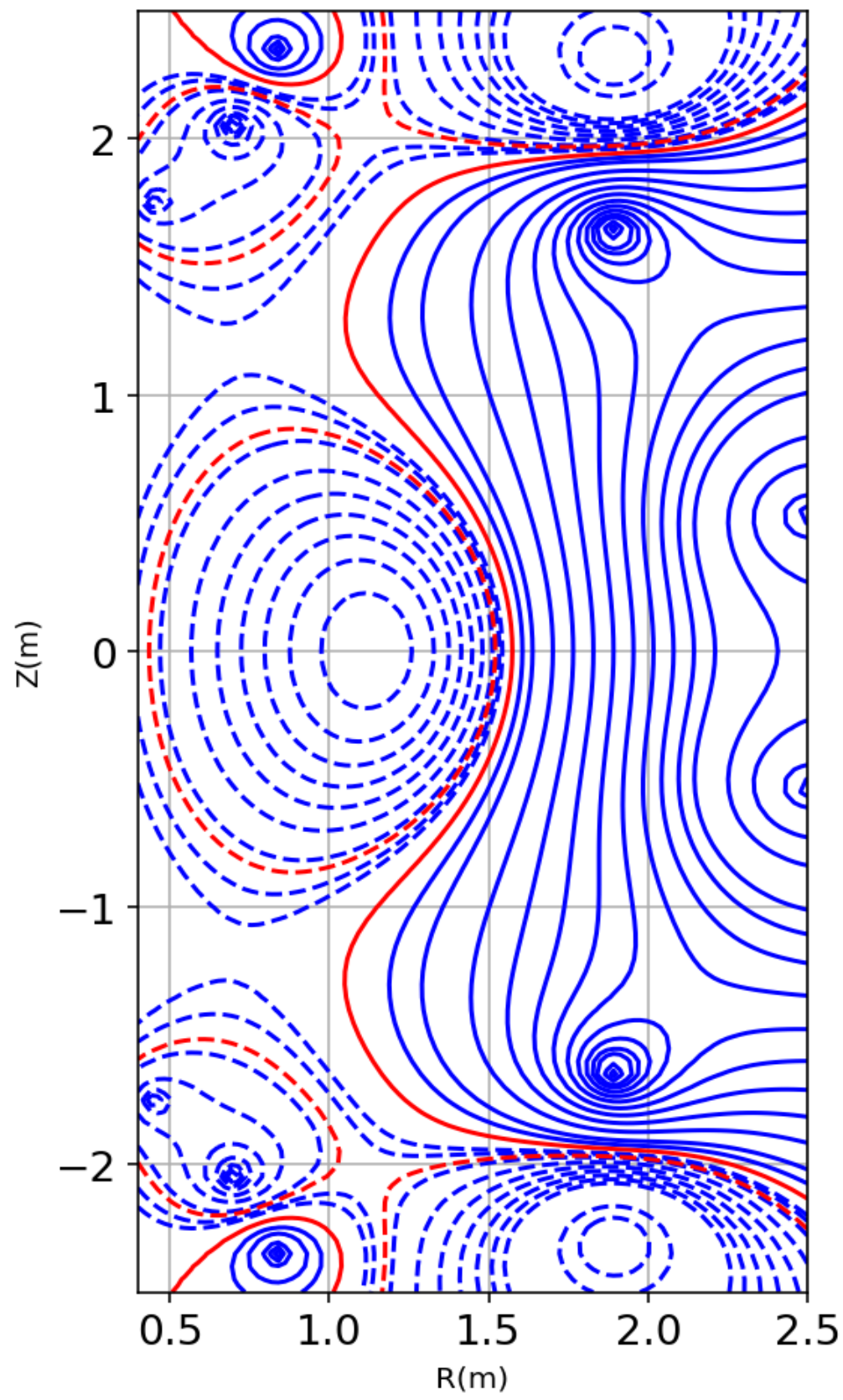

Figure 52. The contours of poloidal magnetic flux $\psi\ (= \psi_{pfcoil} + \psi_{plasma})$ of the low-beta EHL-2 equilibrium of Example equilibrium 3. The red solid line shows the flux boundary of the suprathermal ions and electrons, and the red dashed line shows Psi_lcfs.

### D. Example 4: High-$\beta_T$ fusion test EHL-2 equilibrium

The computational boundary is rectangular with $0.4 \le R(m) \le 1.8$ and $-1.5 \le Z(m) \le 1.5$. In Example 4 equilibrium, we adopt $L_{ref} = 1.78m$, $I_{ref} = 2683kA$ and $n_{ref} = 1.6 \times 10^{19} \frac{1}{m^3}$ as the primary scales. Therefore, $B_{ref} = 1.8941T$, $\psi_{ref} = 6.0014 \frac{Wb}{rad}$, $u_{ref} = \frac{10328.7km}{s}$, $T_{ref} = 1113.4424keV$, and $\varepsilon$ =0.03198. The poloidal field coil locations and currents used are provided in Table XII.

Table XII. Poloidal field coil locations and currents used for Example equilibrium 4.

| Poloidal field coil location and current (Equilibrium 4) | | | | | |
|---|---|---|---|---|---|
| | R(m) | Z(m) | Current(kA) | factor | net current (dimensionless) |
| PF1 | 0.45 | 1.756 | 720 | -1.6 | -0.429370 |
| PF2 | 0.45 | -1.756 | 720 | -1.6 | -0.429370 |
| PF3 | 0.7 | 2.05 | 782 | -1.6 | -0.466344 |
| PF4 | 0.7 | -2.05 | 782 | -1.6 | -0.466344 |
| PF5 | 0.835 | 2.35 | 761.6 | 1.6 | 0.454178 |
| PF6 | 0.835 | -2.35 | 761.6 | 1.6 | 0.454178 |
| PF7 | 1.89 | 2.31 | 1428 | -1.6 | -0.851584 |
| PF8 | 1.89 | -2.31 | 1428 | -1.6 | -0.851584 |
| PF9 | 1.89 | 1.64 | 952 | 1.7 | 0.603205 |
| PF10 | 1.89 | -1.64 | 952 | 1.7 | 0.603205 |
| PF11 | 2.5 | 0.53 | 989 | 1.6 | 0.589788 |
| PF12 | 2.5 | -0.53 | 989 | 1.6 | 0.589788 |

Ampere's equation, the electrostatic potential $V_E$, the poloidal magnetic flux $\psi$, and the radial force balance of each fluid in this equilibrium are solved within the relative errors shown below:

Ampere's equation is solved within an error of $\frac{average|GSop+RJt|}{average|RJt|} = 0.00106$.

The electrostatic potential VE is solved within an error of $\frac{averag|VEeh-VEel|}{average|VEel|}$ = 0.00772.

The quantity $(\psi - \psi_{plasma})$ is solved within an error of $\frac{average|\psi-\psi_{plasma}-\psi_{pfcoil}|}{average|\psi_{pfcoil}|} = 0.0277$.

The radial force balance of the pl-fluid is solved within an error of $\frac{average|balance|}{average|FplP|} = 0.0039.$

The radial force balance of the bl-fluid is solved within an error of $\frac{average|balance|}{average|FblP|} = 0.0026.$

The radial force balance of the el-fluid is solved within an error of $\frac{average|balance|}{average|FelP|} = 0.0049.$

The radial force balance of the ph-fluid is solved within an error of $\frac{average|balance|}{average|FphP|} = 0.0050.$

The radial force balance of the bh-fluid is solved within an error of $\frac{average|balance|}{average|FbhP|} = 0.0044.$

The radial force balance of the eh-fluid is solved within an error of $\frac{average|balance|}{average|FehP|} = 0.0052.$

Figs. 53 – 58 show the poloidal magnetic flux, the toroidal current density, the total pressure, the toroidal current density of the suprathermal electron fluid (eh), the electrostatic potential $V_E$, and the magnetic field strength, which shows a weak magnetic well, of a high-$\beta_T$ fusion test EHL-2 equilibrium. The green dotted lines in (a), (b), (c), (d), (e), and (f) represent the LCFS. The solid red line indicates the flux-surface boundary of the suprathermal components. This boundary is determined by the parameter Psi_crit2. The LCFS, indicated by green dashes, is determined via the convergent equilibrium. The $\beta_T$ and $\beta_p$ are computed within the LCFS

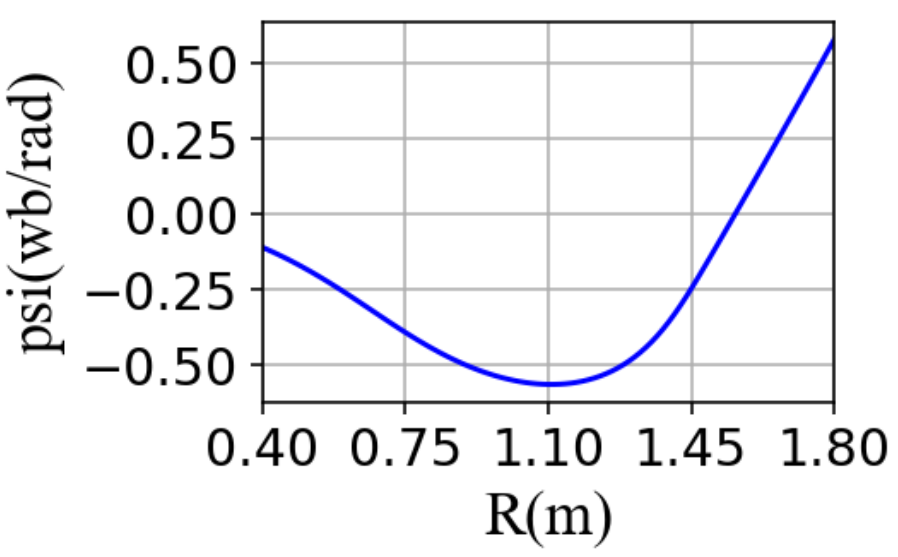

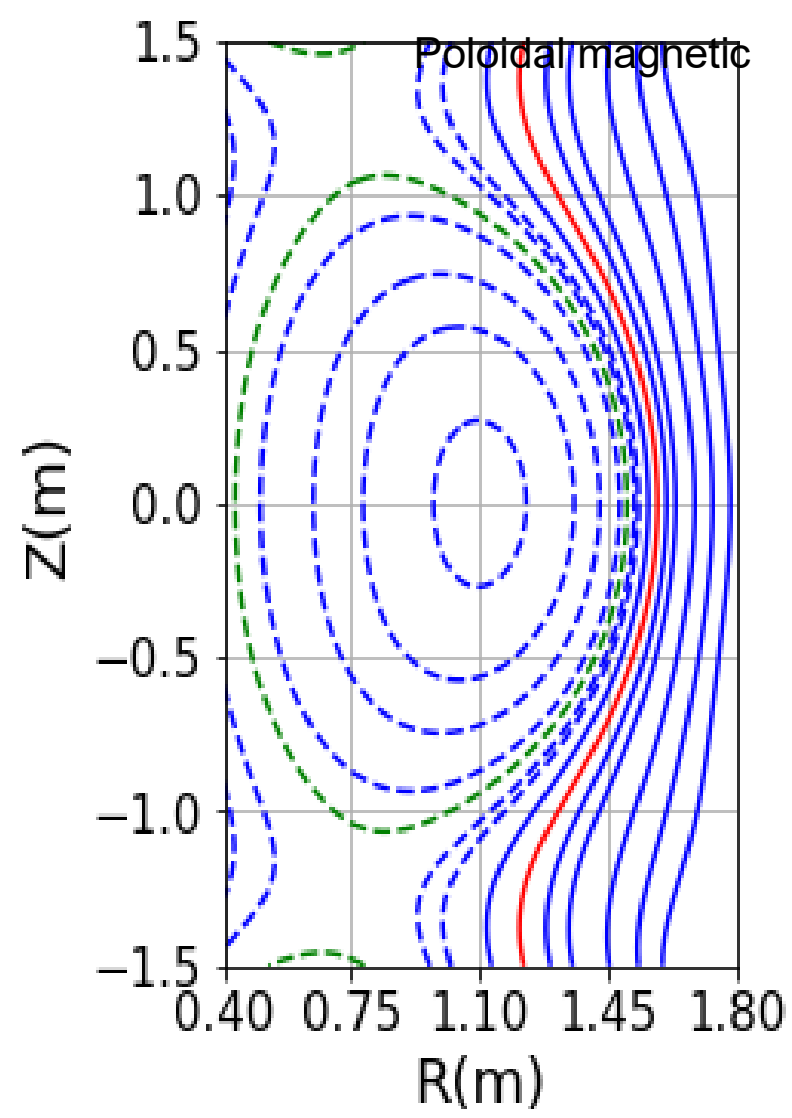

Figure 53. The poloidal magnetic flux across $R$ on the mid-plane, and over the $R$ and $Z$ coordinates.

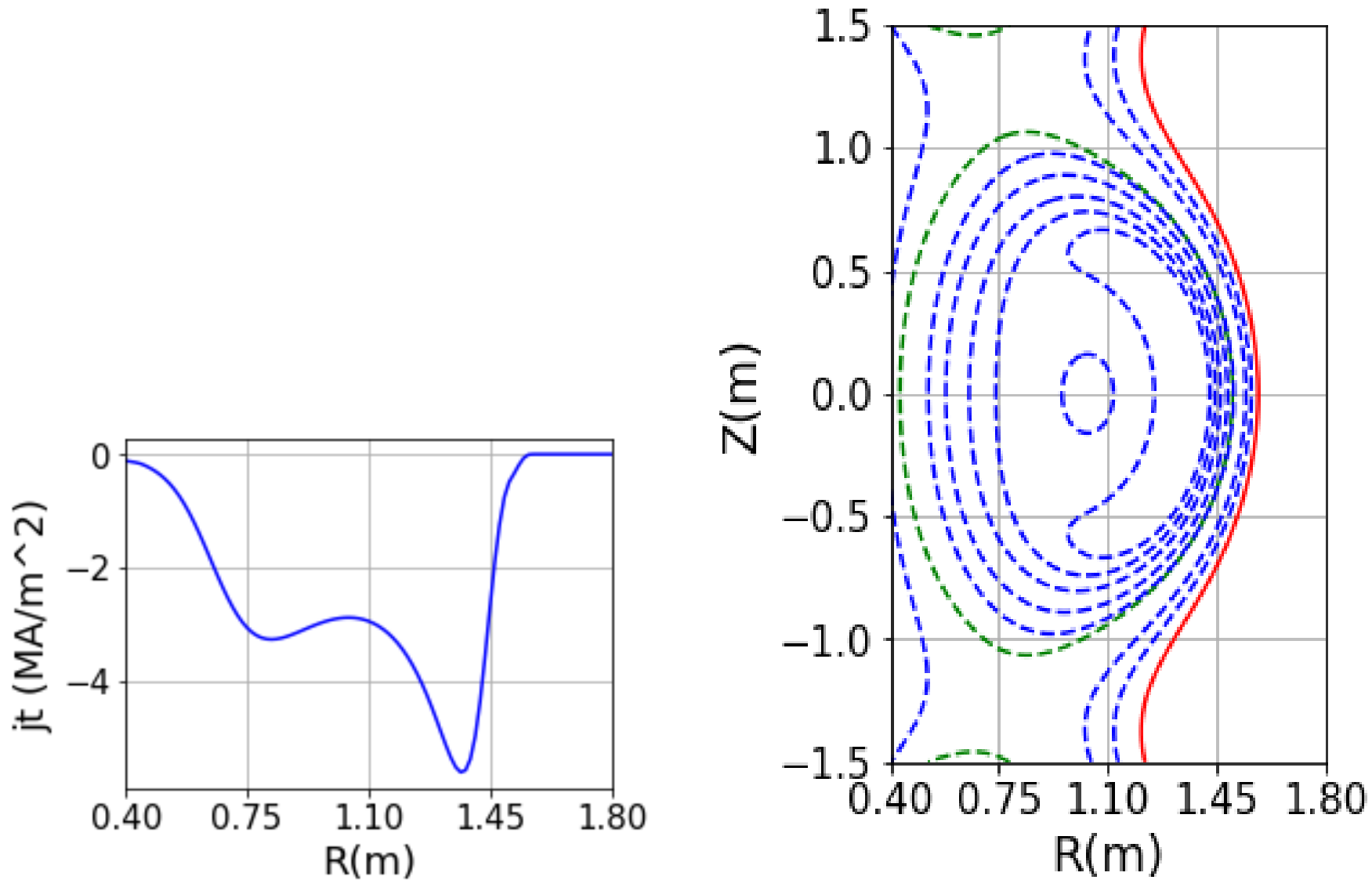

Figure 54. The toroidal current density across *R* on the mid-plane and over the *R* and *Z* coordinates.

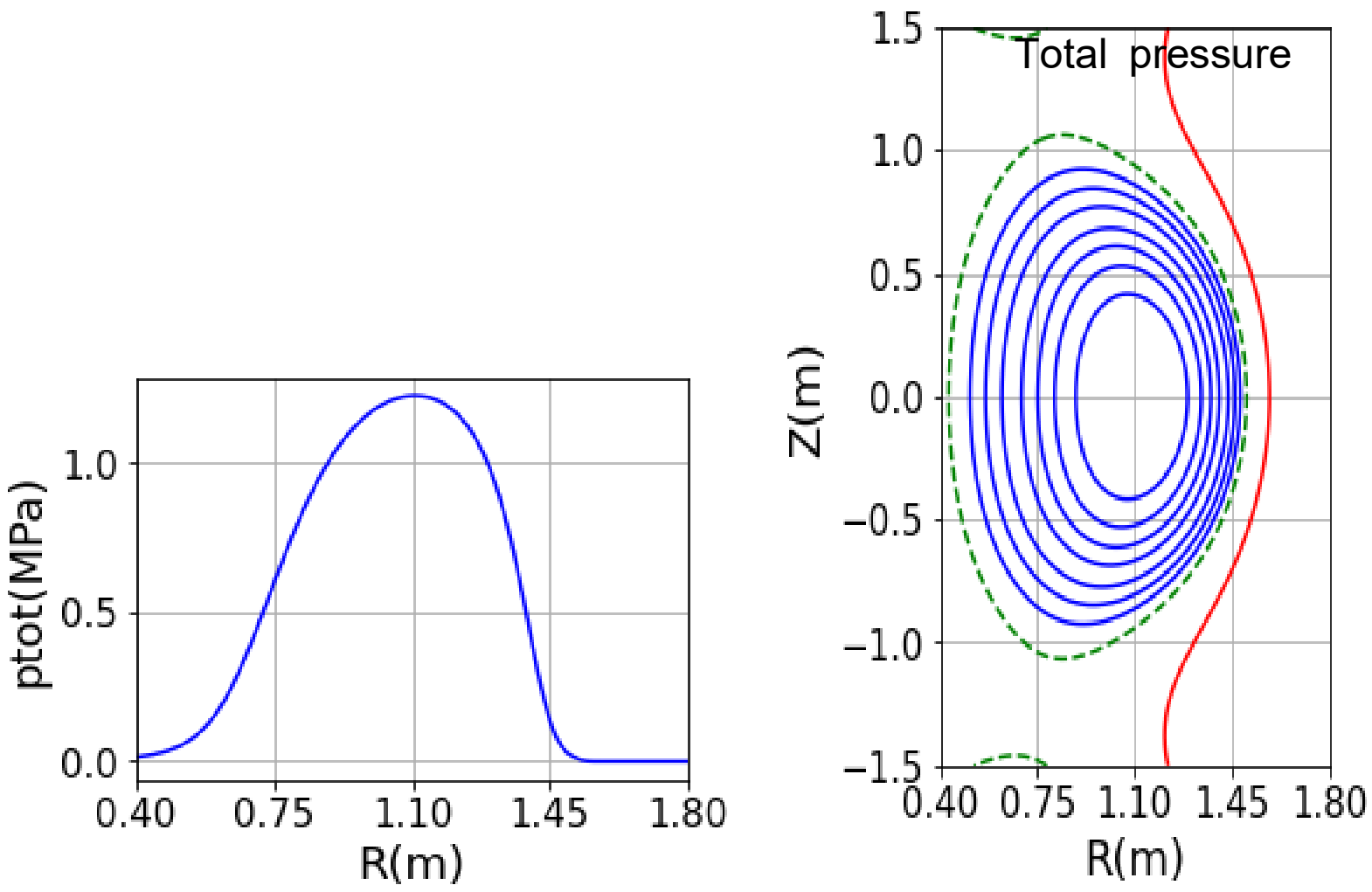

Figure 55. The total pressure across *R* on the mid-plane and over the *R* and *Z* coordinates.

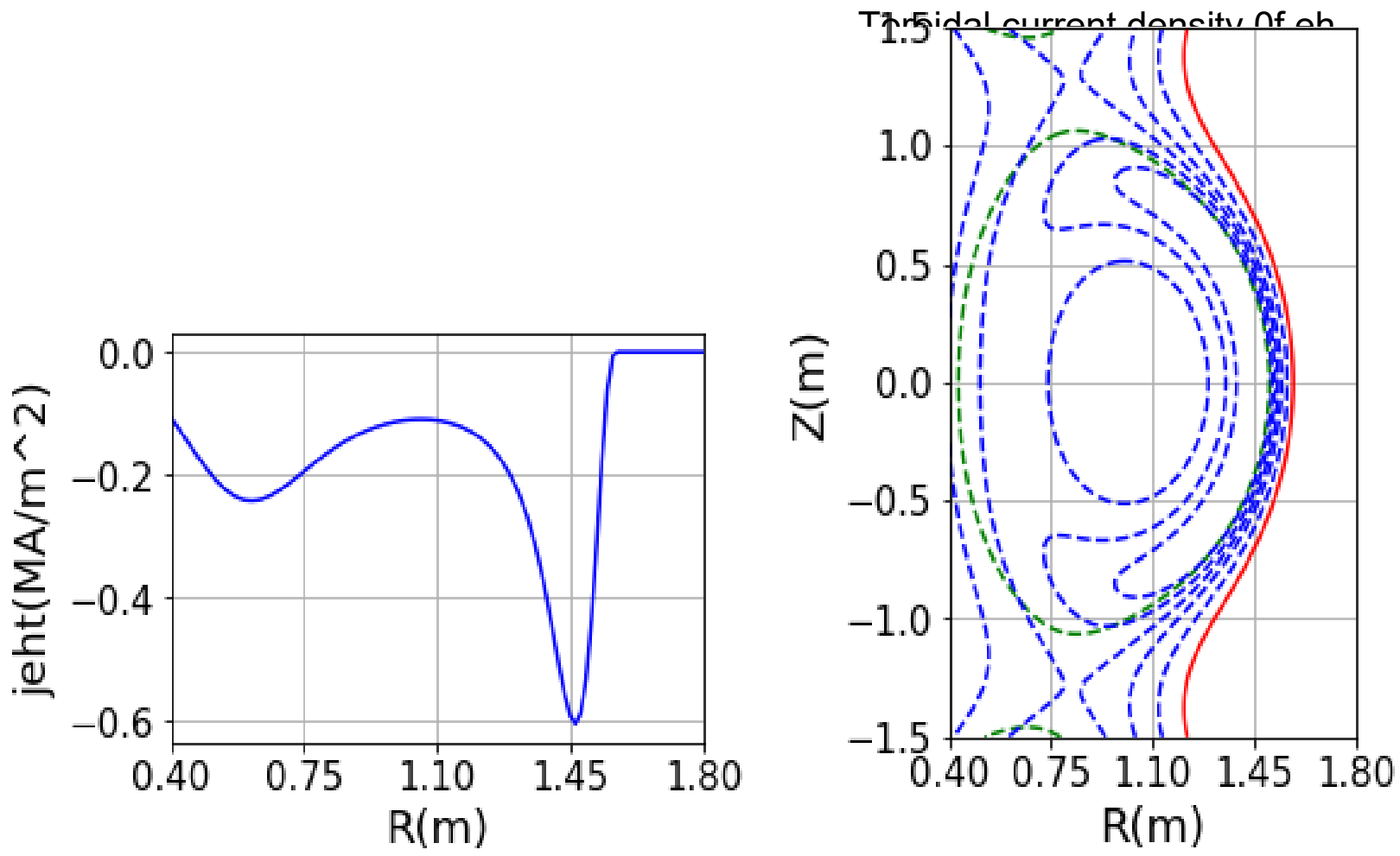

Figure 56. The toroidal current density of the suprathermal electron fluid (eh) across *R* on the mid-plane and over the *R* and *Z* coordinates.

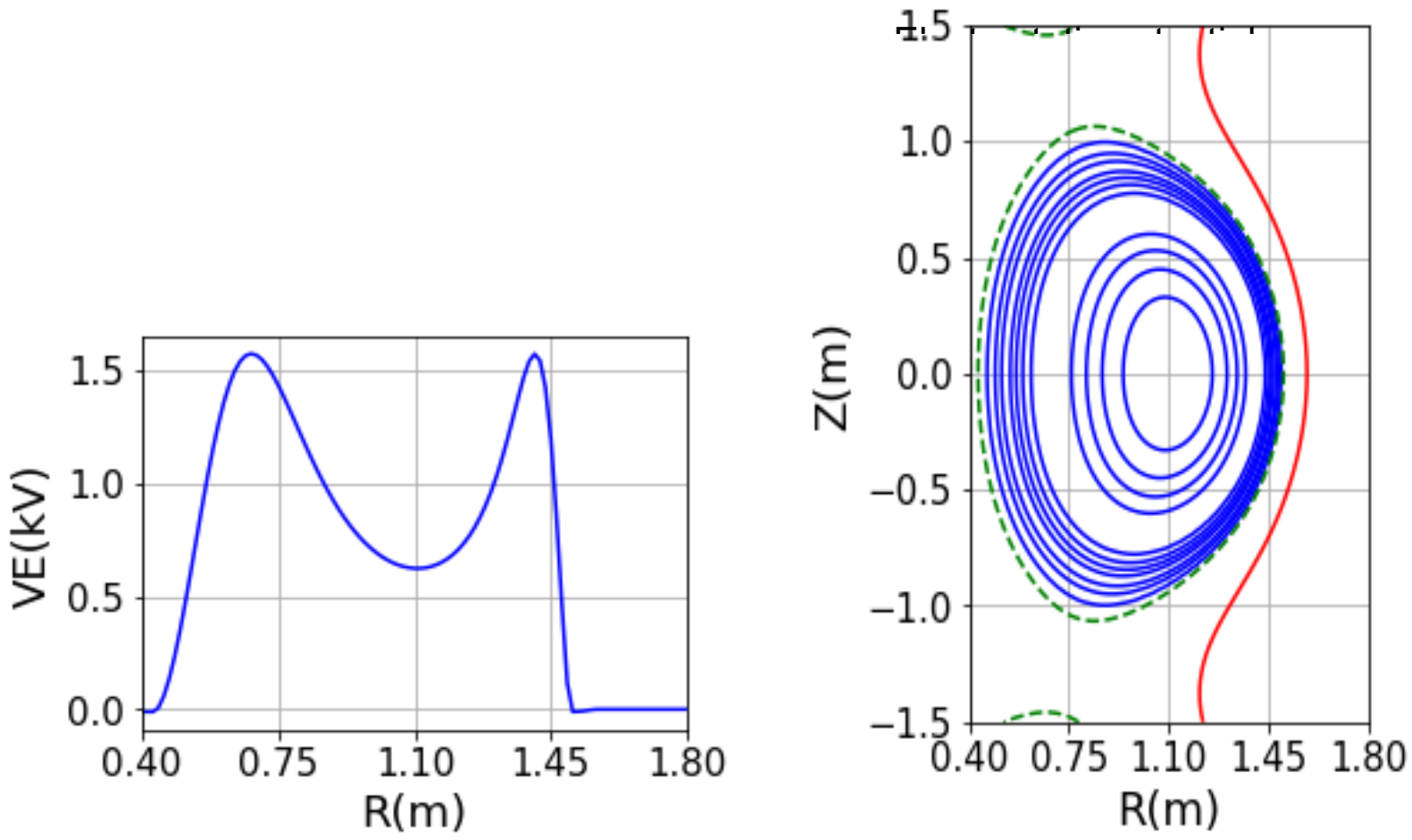

Figure 57. The electrostatic potential $V_E$ across *R* on the mid-plane and over the *R* and *Z* coordinates.

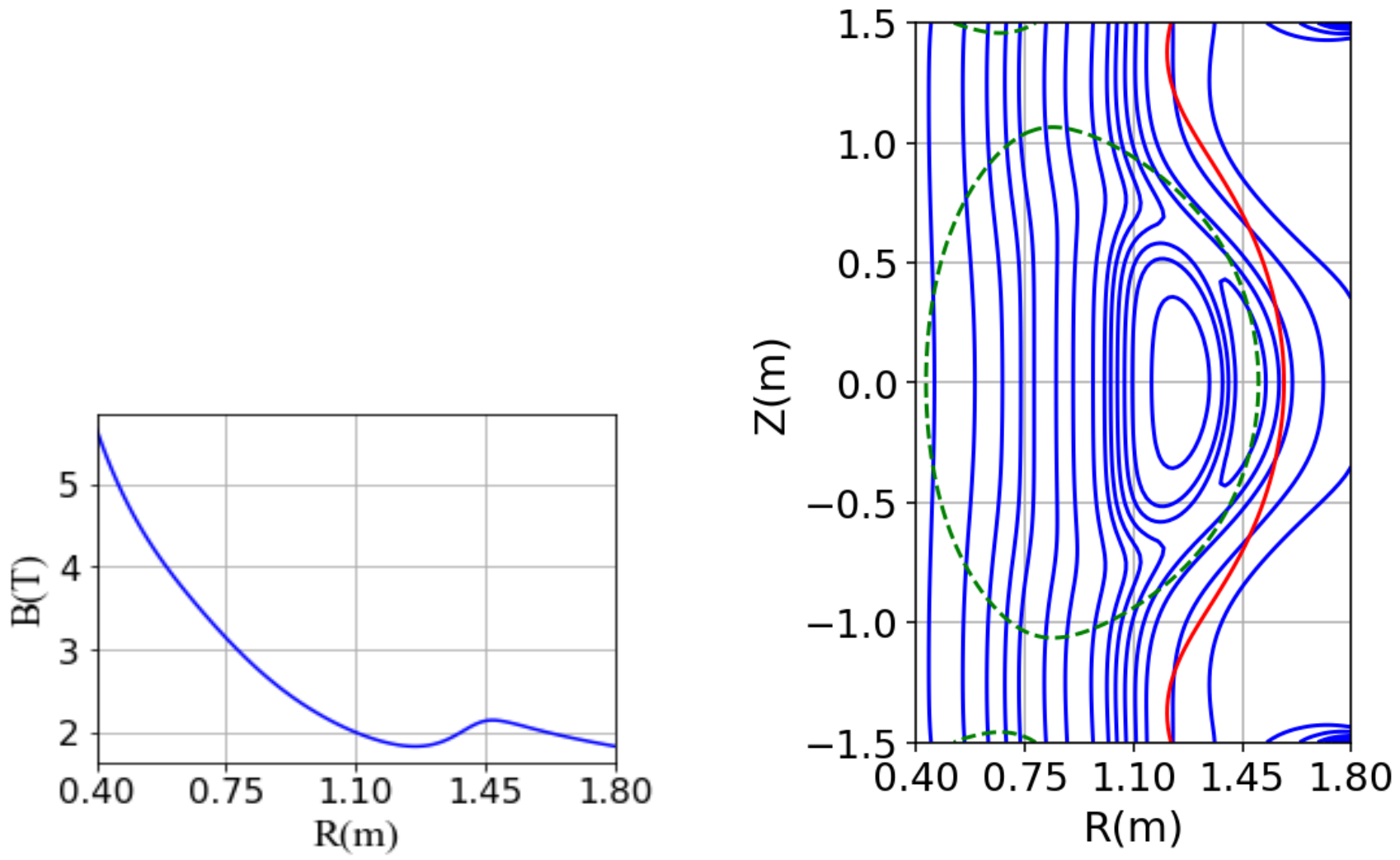

Figure 58. The magnetic field strength across *R* on the mid-plane and over the *R* and *Z* coordinates, showing a strong well over the outboard region of the poloidal plasma cross section.

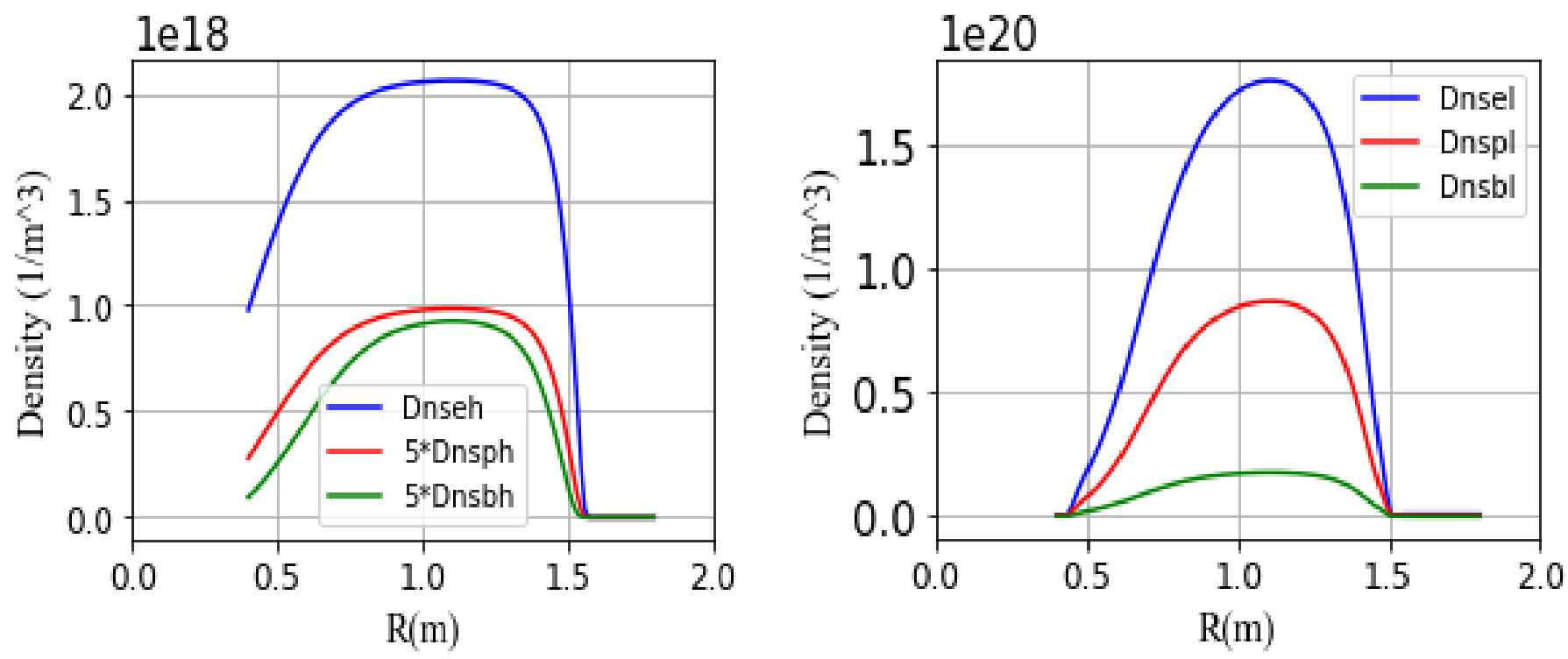

Fig. 59. The densities of the six fluid components.

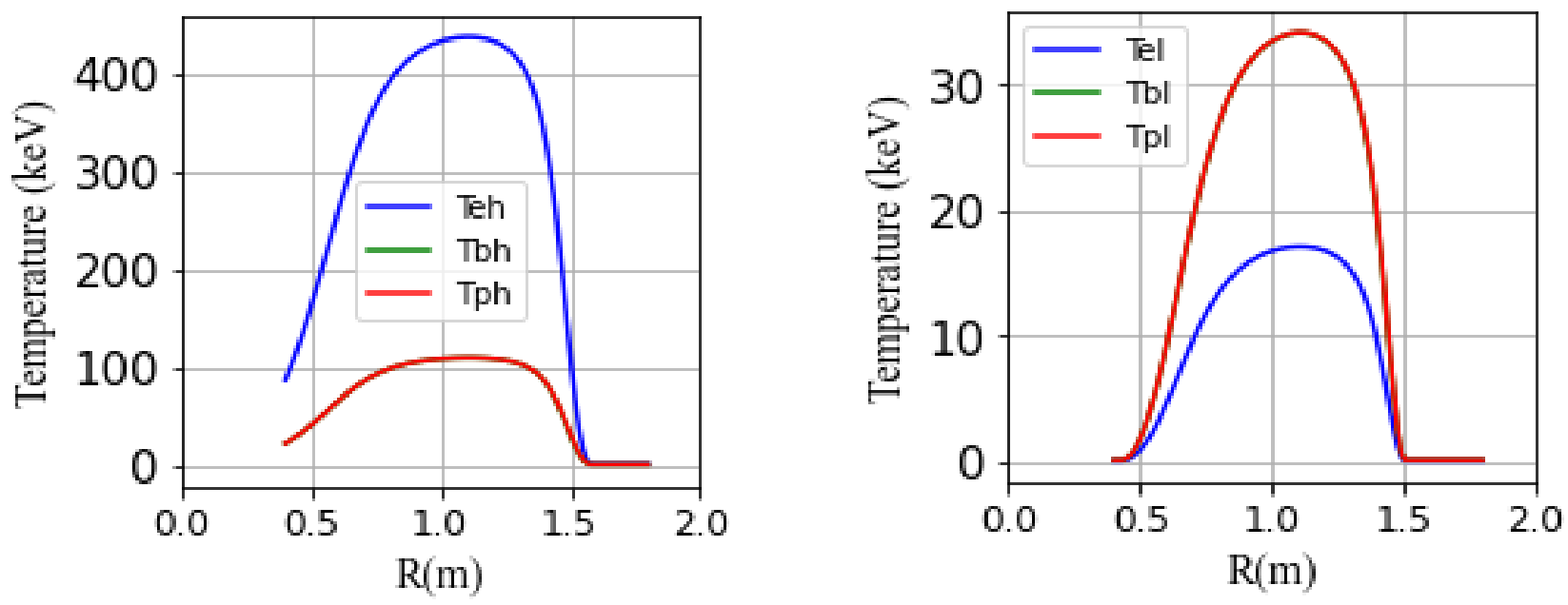

Fig. 60. The temperatures of the six fluid components with Tpl = Tbl.

Fig. 61 shows the $h$ functions representing relativistic effects due to the thermal motion of the suprathermal electron fluid at temperature Teh and the thermal electron fluid at temperature $T_{el}$.

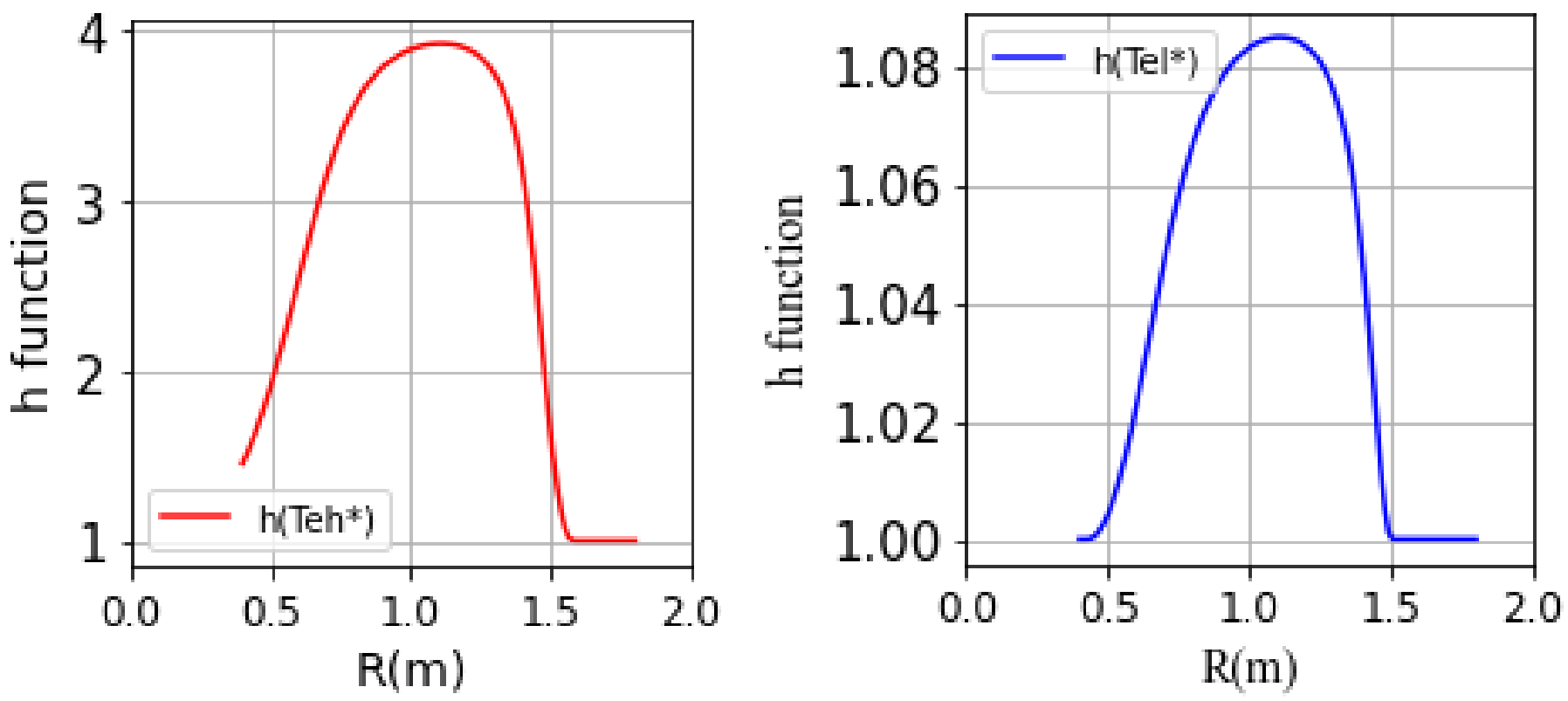

Figure 61. The $h$ functions of the suprathermal electron fluid and of the thermal electron fluid.

Fig. 62 shows the cyclotron frequencies of the el-fluid, $eB/m_e h(T_{el}^*)$, and the eh-fluid, $eB/m_e h(T_{eh}^*)$.

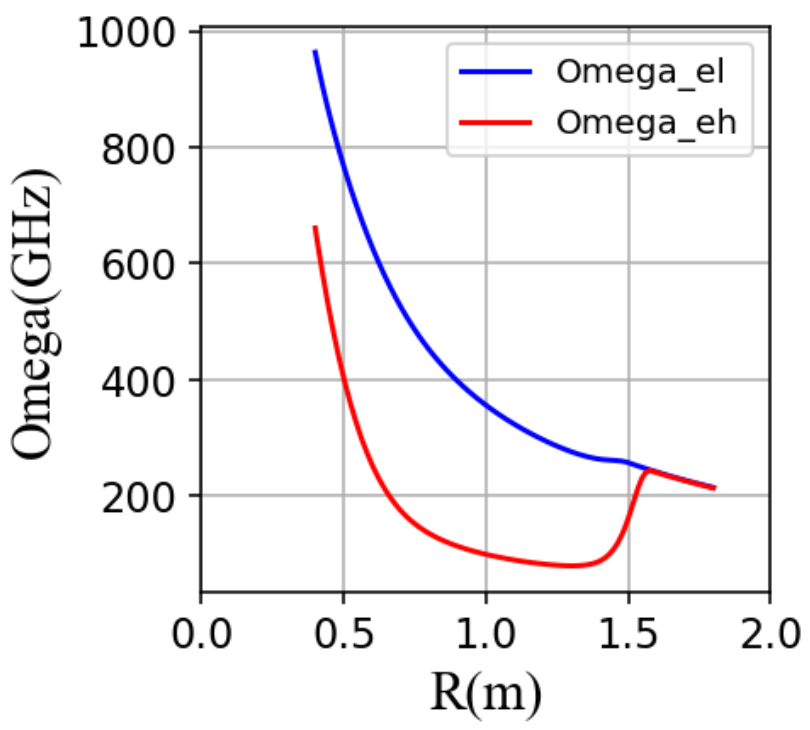

Figure 62. The cyclotron frequencies of the el-fluid and of the eh-fluid.

Fig. 63 shows the toroidal drift velocities of the six fluid components, showing a peak differential velocity of > 800 km/s between the suprathermal protons and the thermal boron ions.

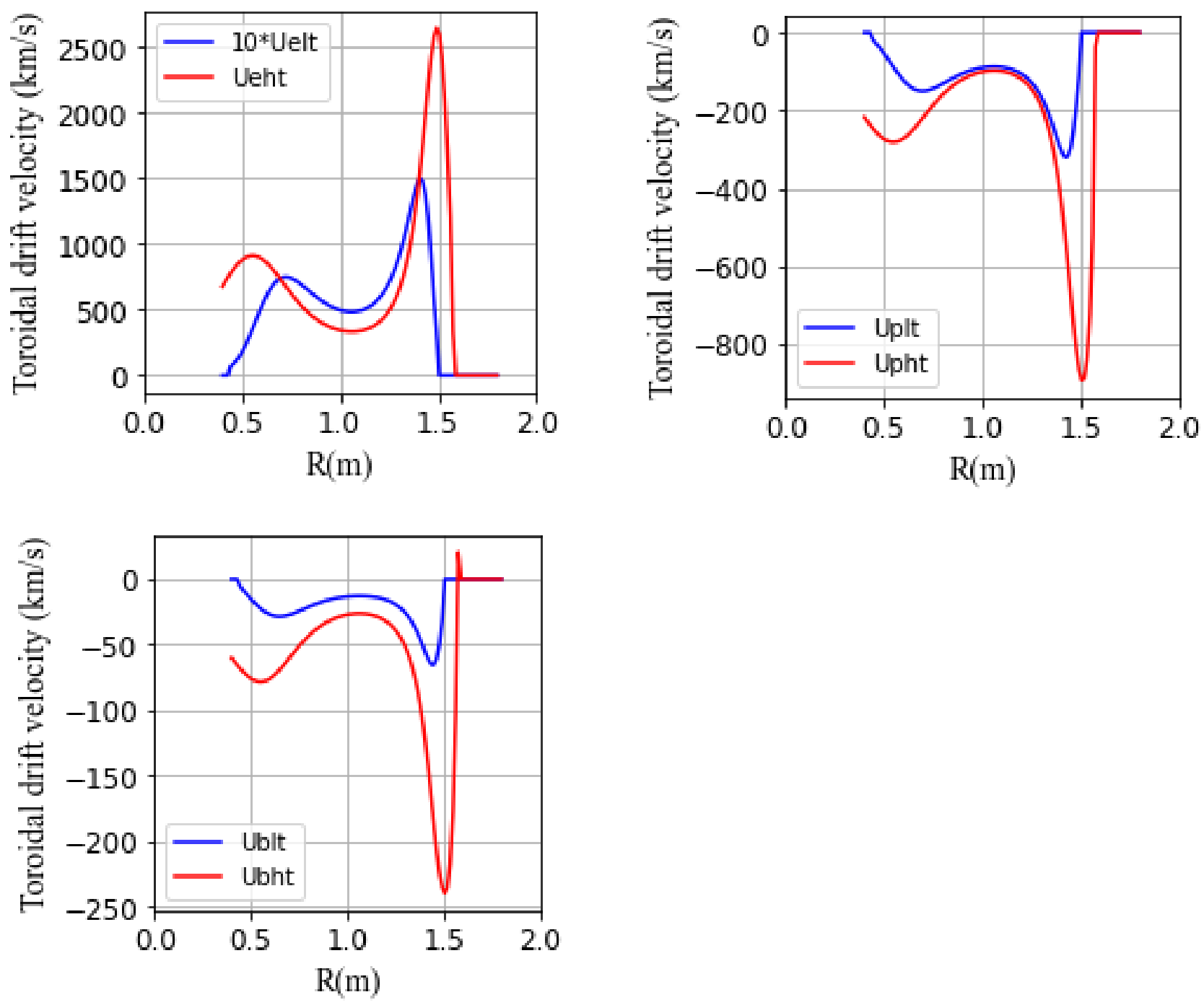

Figure 63. The toroidal drift velocities of the six fluid components.

The input coefficients of this high-$\beta_T$ fusion test EHL-2 equilibrium are provided in Table XIII. The plasma parameters are provided in Table XIV.

Table XIII. Input coefficients of Example equilibrium 4.

| | | | |
|---|---|---|---|
| gfun_C1 | 0.00147 | CFpl1 | 69.7645 |
| gfun_nf | 2.6 | CTpl1 | 24 |
| gfun_nt | 1.8 | CFbl1 | 31.137958 |
| gfun_C2 | 0.05 | CTbl1 | 24 |
| gfun_C2l | 0.045 | CFel1 | 43.2 |
| CFpl0 | -0.000221 | CTel1 | 12 |
| CTpl0 | 8.981E-05 | CFph1 | -122.27885 |
| CFbl0 | -0.000366 | CTph1 | 72 |
| CTbl0 | 8.981E-05 | CFbh1 | -238.15838 |
| CFel0 | -7.96E-05 | CTbh1 | 72 |
| CTel0 | 4.491E-05 | CFeh1 | -289.489014 |
| CFph0 | -0.004832 | CTeh1 | 288 |
| CTph0 | 0.0004491 | | |
| CFbh0 | -0.005554 | Psi_crit1 | -0.022 |
| CTbh0 | 0.0004491 | Psi_crit2 | -0.022 |
| CFeh0 | -0.009041 | Psi_crit3 | -0.022 |
| CTeh0 | 0.0008981 | | |

Table XIV. Plasma parameters of Example equilibrium 4.

| | | | |
|---|---|---|---|
| Rmajor/m | 0.967 | Tpl-axis/keV | 34.2 |
| Rminor,a/m | 0.535 | Tph-axis/keV | 109.44 |
| Iplasma/MA | 4.459 | Tbl-axis/keV | 34.2 |
| B at Rmajor/T | 2.267 | Tbh-axis/keV | 109.4 |
| kappa | 1.96 | Upl-peak/km/s | -320 |
| A | 1.81 | Uph-peak/km/s | -893 |
| delta | 0.26 | Ubl-peak/km/s | -65 |
| q0 | 1.29 | Ubh-peak/km/s | -240 |
| q-LCFS | 4.20 | Uel-peak/km/s | 150 |
| q95 | 3.26 | Ueh-peak/km/s | 2646 |
| li(int.inductance) | 0.32397 | h(Teh)-peak | 3.93 |
| beta-T-LCFS | 0.236 | h(Tel)-peak | 1.09 |
| beta-pol-LCFS | 1.10 | Ipl/kA | 1658.3 |
| beta-axis | 0.77 | Ibl/kA | 322.1 |
| nel-axis/e20/m3 | 1.763 | Iel/kA | 1749.2 |
| neh-axis/e18/m3 | 2.061 | Iph/kA | 73.6 |
| npl-axis/e20/m3 | 0.8711 | Ibh/kA | 12.6 |
| nph-axis/e18/m3 | 0.9854 | Ieh/kA | 643.2 |
| nbl-axis/e20/m3 | 0.1787 | Ipl+Ibl+Iel/kA | 3729.6 |
| nbh-axis/e18/m3 | 0.1852 | Iph+Ibh+Ieh/kA | 729.4 |
| Tel-axis/keV | 17.1 | | |
| Teh-axis/keV | 436.48 | | |

Fig. 64 shows the contours of poloidal magnetic flux $\psi = \psi_{pfcoil} + \psi_{plasma}$ of the high-beta EHL-2 equilibrium. $\psi_{pfcoil}$ is computed using the coil current shown in Table XII, and $\psi_{plasma}$ is computed using the plasma current distribution obtained in Example equilibrium 4. The red solid line shows the magnetic flux boundary for the suprathermal ions and electrons, and the red dashed line shows Psi_lcfs.

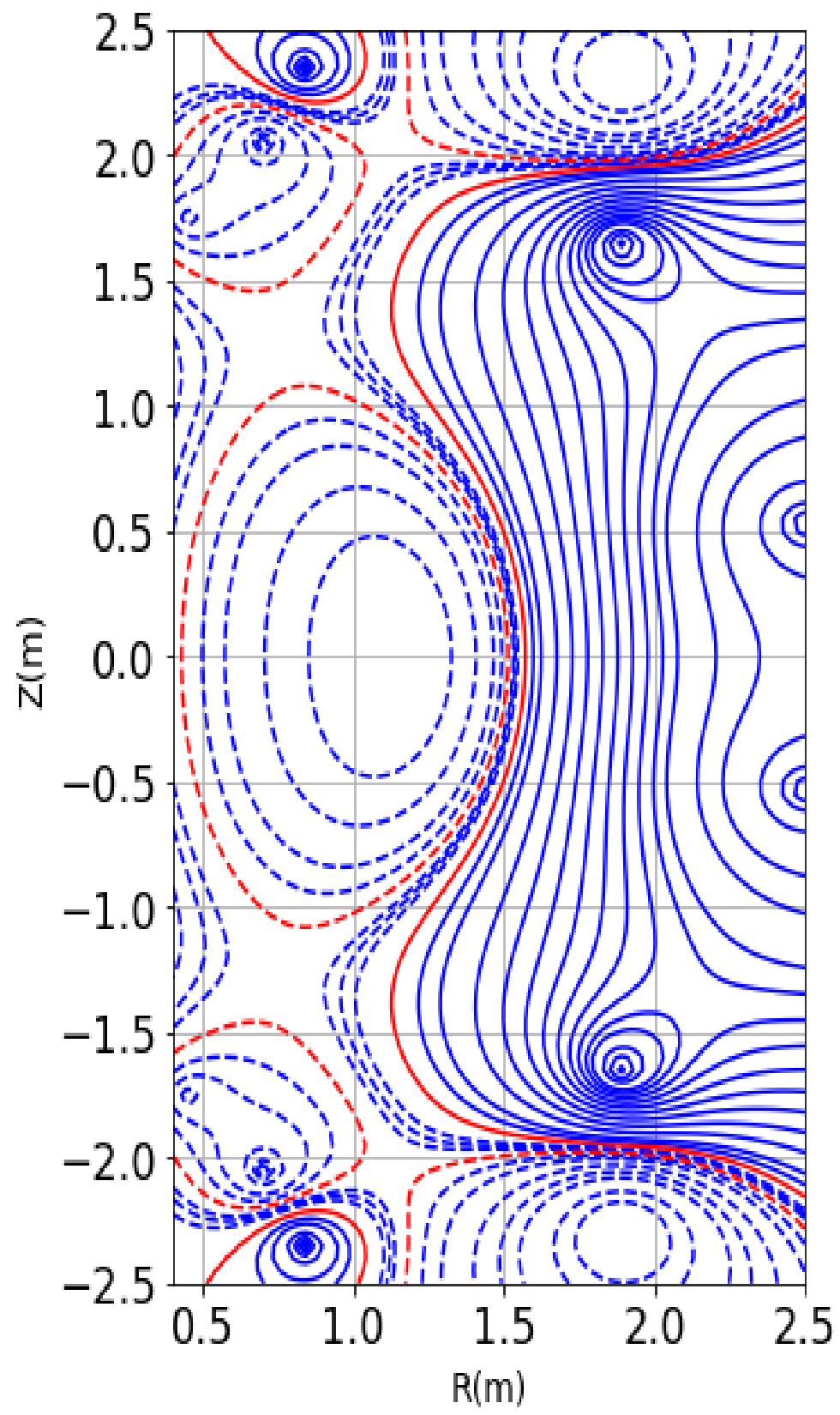

Figure 64. The contours of poloidal magnetic flux $\psi = \psi_{pfcoil} + \psi_{plasma}$ of the high-beta EHL-2 equilibrium. The red solid line shows the magnetic flux boundary for the suprathermal ions and electrons, and the red dashed line shows Psi_lcfs.

## DATA AVAILABILITY

The numerical output of the four six-fluid ST p-$^{11}$B fusion plasma equilibria is available upon request.